\documentclass{LMCS}

\usepackage[latin1]{inputenc}

\usepackage{amssymb}
\usepackage{amsthm}

\usepackage{logique}
\usepackage{enumerate,hyperref}

\newcommand{\NN}{\mathbb{N}}
\newcommand{\CC}{\mathbb{C}}

\newcommand{\nn}{\mbox{\sffamily n}}
\newcommand{\cp}{\mbox{\sffamily cp}}
\newcommand{\indi}{^{\mbox{\footnotesize int}}}
\newcommand{\inde}{^{\mbox{\footnotesize ent}}}
\newcommand{\ppt}{{\mbox{\small <}}}

\newcommand{\trl}{\triangleleft}

\newcommand{\forcec}{\;[\!]\hspace{-0.6em}-}

\newcommand{\sqle}{\sqsubseteq}
\newcommand{\et}{{\scriptstyle\land}}

\begin{document}

\title{Realizability algebras:\\
a program to well order $\mathbb{R}$}

\author[J.-L.~Krivine]{Jean-Louis Krivine}
\address{Université Paris VII, C.N.R.S.}
\email{krivine@pps.jussieu.fr}


\begin{abstract}\noindent
The theory of classical realizability is a framework in which we can develop the proof-program
correspondence. Using this framework, we show how to transform into programs the proofs in
classical analysis with dependent choice and the existence of a well ordering of the real line.
The principal tools are:\\
-~~The notion of \emph{realizability algebra}, which is a three-sorted variant of
the well known \emph{combinatory algebra} of Curry.\\
-~~An adaptation of the method of \emph{forcing} used in set theory to prove consistency
results. Here, it is used in another way, to obtain programs associated with a well ordering
of~$\mathbb{R}$ and the existence of a non trivial ultrafilter on $\NN$.
\end{abstract}

\def\doi{7 (3:02) 2011}
\lmcsheading%
{\doi}
{1--47}
{}
{}
{Jun.~14, 2010}
{Aug.~\phantom09, 2011}
{}

\maketitle\noindent

\section*{Introduction}\noindent
When we want to obtain programs from mathematical proofs, the main problem is, naturally, raised by the \emph{axioms}:
indeed, it has been a long time since we know how to transform a proof in pure (i.e. without axioms) intuitionistic logic, even at second order~\cite{curry,howard,girard}.\\
The very  first of these axioms is the \emph{excluded middle}, and it seemed completely
hopeless for decades. The solution, given by T. Griffin~\cite{griffin} in 1990, was absolutely surprising. It was an essential discovery in logic because, at this moment, it became clear that all other axioms will follow, as soon as we will work in a suitable framework.

\smallskip\noindent
The \emph{theory of classical realizability} is such a framework: it was developed in~\cite{krivine2,krivine3},
where we treat the axioms of \emph{Analysis} (second order arithmetic with dependent choice).\\
In~\cite{krivine5}, we attack a more difficult case of the general axiom of choice, which is the existence of a non trivial ultrafilter on~$\NN$~; the main tool is the notion of \emph{realizability structure}, in which the programs are written in $\lbd$-calculus.\\
In the present paper, we replace it with the notion of \emph{realizability algebra}, which has many advantages: it is simpler, first order and much more practical for implementation. It is a three-sorted variant of the usual notion of \emph{ combinatory algebra}. Thus, the programming language is no longer the $\lbd$-calculus, but a suitable set
of combinators~; remarkably enough, this is almost exactly the original set given by Curry. The $\lbd$-terms are now considered only as notations or abbreviations, very useful in fact: a $\lbd$-term is infinitely more readable than its translation into a sequence of combinators. The translation used here is new, as far as I know~;
its fundamental property is given in theorem~\ref{beta_red_gauche}.

\smallskip\noindent
The aim of this paper is to show how to transform into programs, the classical proofs which use dependent choice and:\\
i)~the existence of a non trivial ultrafilter on $\NN$~;\\
ii)~the existence of a well ordering on $\mathbb{R}$.\\
Of course, (ii) implies (i) but the method used for (i) is interesting, because it can give simpler programs.
This is an important point, because a new problem is appearing now, an important and very difficult problem: to understand the programs we obtain in this way, that is to explain their behavior. A fascinating, but probably long work.

\smallskip\noindent
The logical frame is given by \emph{classical second order logic}, in other words the (first order) theory of the comprehension scheme. However, since we use a binary membership relation on individuals, we work, in reality,
in at least third order logic. Moreover, this is indispensable since, although the axiom of dependent choice on $\mathbb{R}$ can be expressed as a second order scheme, axioms~(i) and~(ii) cannot be expressed in this way.\\
By using the method expounded in~\cite{krivine1}, we can obtain the same results in ZF\/.

\smallskip\noindent
It seems clear to me that, by developing the technology of classical realizability, we shall be able to treat all
``natural'' axioms introduced in set theory. It is already done for the \emph{continuum hypothesis}, which
will be the topic of a forthcoming paper. In my opinion, the axiom of choice and the generalized continuum
hypothesis in ZF do not pose serious issues, except this: it will be necessary to use the \emph{proper class forcing} of Easton~\cite{easton} inside the realizability model, and it will probably be very painful.\\
A very interesting open problem is posed by axioms such as the existence of measurable cardinals
or the determination axiom.

\smallskip\noindent
But the most important open problem is to understand what all these programs do and, in this way, to be able
to \emph{execute them}. I believe that big surprises are waiting for us here.\\
Indeed, when we realize usual axioms of mathematics, we need to introduce, one after the other, the very standard
tools in \emph{system programming}: for the law of Peirce, these are continuations (particularly useful for exceptions)~; for the axiom of dependent choice, these are the clock and the process numbering~; for the
ultrafilter axiom and the well ordering of $\mathbb{R}$, these are no less than read and write instructions on a
global memory, in other words \emph{assignment}.\\
It seems reasonable to conjecture that such tools are introduced for some worthwhile purpose, and therefore that the very complex programs we obtain by means of this formalization work, perform interesting and useful tasks. The question is: which ones~?

\smallskip\noindent
{\small{\bfseries Remark.}\\
The problem of obtaining a program from a proof which uses a given axiom, must be set correctly from the
point of view of computer science. As an example, consider a proof of a theorem of arithmetic, which uses
a well ordering of ${\mathcal P}(\NN)$: if you restrict this proof to the class of constructible sets, you easily
get a new proof of the same theorem, which does not use this well ordering any more. Thus, it looks like you
simply have to transform this new proof into a program.\\
But this program would be extracted from a proof which is \emph{deeply different from (and dramatically
more complicated than) the original one}. Moreover, with this method, it is impossible to associate a program with the well ordering axiom itself.
From the point of view of computer science, this is an unacceptable lack of \emph{modularity}: since we cannot put
the well ordering axiom in a \emph{program library}, we need to undertake again the programming work with each
new proof.\\
With the method which is explained below, we only use the $\lbd$-term \emph{extracted from the original proof}. Therefore, this term contains an unknown instruction for the well ordering axiom on ${\mathcal P}(\NN)$, which is not
yet implemented. Then, by means of a suitable compilation, we transform this term into a true program which \emph{realizes} the initial theorem.\\
As a corollary of this technology, we obtain a program which is associated with the well ordering axiom, which
we can put in a library for later use.}

\section{Realizability algebras}\noindent
A \emph{realizability algebra} is composed of three sets:
$\LLbd$ (the set of \emph{terms}), $\PPi$ (the set of \emph{stacks}),
$\LLbd\star\PPi$ (the set of \emph{processes}) with the following operations:

\smallskip\noindent
$(\xi,\eta)\mapsto(\xi)\eta$ from $\LLbd^2$ into $\LLbd$ (\emph{application})~;\\
$(\xi,\pi)\mapsto\xi\ps\pi$ from $\LLbd\fois\PPi$ into $\PPi$ (\emph{push})~;\\
$(\xi,\pi)\mapsto\xi\star\pi$ from $\LLbd\fois\PPi$ into $\LLbd\star\PPi$ (\emph{process})~;\\
$\pi\mapsto\kk_\pi$ from $\PPi$ into $\LLbd$ (\emph{continuation}).

\smallskip\noindent
We have, in $\LLbd$, the distinguished elements \ $B,C,E,I,K,W,\Ccc$, called \emph{elementary combinators} or \emph{instructions}.

\smallskip\noindent
{\bfseries Notation.}
The term $(\ldots(((\xi)\eta_1)\eta_2)\ldots)\eta_n$ will be also denoted by
$(\xi)\eta_1\eta_2\ldots\eta_n$ or even $\xi\eta_1\eta_2\ldots\eta_n$.
For example: \ $\xi\eta\zeta=(\xi)\eta\zeta=(\xi\eta)\zeta=((\xi)\eta)\zeta$. 

\smallskip\noindent
We define on $\LLbd\star\PPi$ a preorder relation, denoted by $\succ$. It is the least reflexive and transitive
relation such that we have, for any $\xi,\eta,\zeta\in\LLbd$ and $\pi,\varpi\in\PPi$:

\smallskip\noindent
$(\xi)\eta\star\pi\succ\xi\star\eta\ps\pi$.\\
$I\star\xi\ps\pi\succ\xi\star\pi$.\\
$K\star\xi\ps\eta\ps\pi\succ\xi\star\pi$.\\
$E\star\xi\ps\eta\ps\pi\succ(\xi)\eta\star\pi$.\\
$W\star\xi\ps\eta\ps\pi\succ\xi\star\eta\ps\eta\ps\pi$.\\
$C\star\xi\ps\eta\ps\zeta\ps\pi\succ\xi\star\zeta\ps\eta\ps\pi$.\\
$B\star\xi\ps\eta\ps\zeta\ps\pi\succ(\xi)(\eta)\zeta\star\pi$.\\
$\Ccc\star\xi\ps\pi\succ\xi\star\kk_\pi\ps\pi$.\\
$\kk_\pi\star\xi\ps\varpi\succ\xi\star\pi$.

\smallskip\noindent
Finally, we are given a subset $\bbot$ of $\LLbd\star\PPi$ which is a terminal segment for
this preorder, which means that: \ $\p\in\bbot$, $\p'\succ\p$ $\Fl$ $\p'\in\bbot$.\\
In other words, we ask that $\bbot$ be such that:

\smallskip\noindent
$(\xi)\eta\star\pi\notin\bbot\Fl\xi\star\eta\ps\pi\notin\bbot$.\\
$I\star\xi\ps\pi\notin\bbot\Fl\xi\star\pi\notin\bbot$.\\
$K\star\xi\ps\eta\ps\pi\notin\bbot\Fl\xi\star\pi\notin\bbot$.\\
$E\star\xi\ps\eta\ps\pi\notin\bbot\Fl(\xi)\eta\star\pi\notin\bbot$.\\
$W\star\xi\ps\eta\ps\pi\notin\bbot\Fl\xi\star\eta\ps\eta\ps\pi\notin\bbot$.\\
$C\star\xi\ps\eta\ps\zeta\ps\pi\notin\bbot\Fl\xi\star\zeta\ps\eta\ps\pi\notin\bbot$.\\
$B\star\xi\ps\eta\ps\zeta\ps\pi\notin\bbot\Fl(\xi)(\eta)\zeta\star\pi\notin\bbot$.\\
$\Ccc\star\xi\ps\pi\notin\bbot\Fl\xi\star\kk_\pi\ps\pi\notin\bbot$.\\
$\kk_\pi\star\xi\ps\varpi\notin\bbot\Fl\xi\star\pi\notin\bbot$.

\subsection*{\Cc-terms and \texorpdfstring{$\lbd$}{lambda}-terms}\noindent
We call \ \emph{\Cc-term} \ a term which is built with variables, the elementary combinators
$B$, $C$, $E$, $I$, $K$, $W$, $\Ccc$ and the application (binary function). A \Cc-term is called \emph{closed} if it contains
no variable~; it will then also be called \emph{proof-like}~; a proof-like term has a value in $\LLbd$.

\smallskip\noindent
Given a \Cc-term $t$ and a variable $x$, we define inductively on $t$, a new \Cc-term denoted by $\lbd x\,t$.
To this aim, we apply the first possible case in the following list:

\smallskip\noindent
1.~$\lbd x\,t=(K)t$\label{def_lbd} if $t$ does not contain $x$.\\
2.~$\lbd x\,x=I$.\\
3.~$\lbd x\,tu=(C\lbd x(E)t)u$ if $u$ does not contain $x$.\\
4.~$\lbd x\,tx=(E)t$ if $t$ does not contain $x$.\\
5.~$\lbd x\,tx=(W)\lbd x(E)t$ (if $t$ contains $x$).\\
6.~$\lbd x(t)(u)v=\lbd x(B)tuv$ (if $uv$ contains $x$).

\smallskip\noindent
We easily see that this rewriting is finite, for any given $\Cc$-term $t$: indeed, during the rewriting, no combinator is introduced inside $t$, but only in front of it. Moreover, the only changes in~$t$ are: moving
parentheses and erasing occurrences of $x$. Now, rules 1~to~5 strictly decrease the part of $t$ which remains under $\lbd x$, and rule~6 can
be applied consecutively only finitely many times.

\smallskip\noindent
The \emph{$\lbd$-terms} are defined as usual. But, in this paper, we consider $\lbd$-terms only as a notation
for particular \Cc-terms, by means of the above translation. This notation is essential, because almost
every $\Cc$-term we shall use, will be given as a $\lbd$-term. Theorem~\ref{beta_red_gauche} gives the
fundamental property of this translation.

\smallskip\noindent
{\small{\bfseries Remark.} We cannot use the well known $KS$-translation of $\lbd$-calculus,
because it does not satisfy Theorem~\ref{beta_red_gauche}.}

\begin{lem}\label{brg1}
If $t$ is a \ \Cc-term with the only variables $x,y_1,\ldots,y_n$, and if \
$\xi,\eta_1,\ldots,\eta_n\in\LLbd$, then: \
$(\lbd x\,t)[\eta_1/y_1,\ldots,\eta_n/y_n]\star\xi\ps\pi\succ t[\xi/x,\eta_1/y_1,\ldots,\eta_n/y_n]\star\pi$.
\end{lem}

\proof
To lighten the notation, \ let us put \ $u^*=u[\eta_1/y_1,\ldots,\eta_n/y_n]$ for each \Cc-term $u$~; \
thus, we have:\\
$u^*[\xi/x]=u[\xi/x,\eta_1/y_1,\ldots,\eta_n/y_n]$.\\
The proof is done by induction on the number of rules~1 to~6 used to translate the term $\lbd x\,t$.
Consider the rule used first.\\
If it is rule~1, then we have \ $(\lbd x\,t)^*\star\xi\ps\pi\equiv(K)t^*\star\xi\ps\pi\succ t^*\star\pi\\
\equiv t[\xi/x,\eta_1/y_1,\ldots,\eta_n/y_n]\star\pi$ since $x$ is not in $t$.\\
If it is rule~2, we have $t=x$ and \ $(\lbd x\,t)^*\star\xi\ps\pi\equiv I\star\xi\ps\pi\succ\xi\star\pi
\equiv t[\xi/x,\eta_1/y_1,\ldots,\eta_n/y_n]\star\pi$.\\
If it is rule~3, we have $t=uv$ \ and \ $(\lbd x\,t)^*\star\xi\ps\pi\equiv(C\lbd x(E)u)^*v^*\star\xi\ps\pi\\
\succ C\star(\lbd x(E)u)^*\ps v^*\ps\xi\ps\pi
\succ(\lbd x(E)u)^*\star\xi\ps v^*\ps\pi\succ(E)u^*[\xi/x]\star v^*\ps\pi$ (by induction hypothesis)
$\succ E\star u^*[\xi/x]\ps v^*\ps\pi\succ(u^*[\xi/x])v^*\star\pi
\equiv t[\xi/x,\eta_1/y_1,\ldots,\eta_n/y_n]\star\pi$ since $x$ is not in~$v$.\\
If it is rule~4, we have $t=ux$ \ and \ $(\lbd x\,t)^*\star\xi\ps\pi\equiv(E)u^*\star\xi\ps\pi\succ
E\star u^*\ps\xi\ps\pi\succ u^*\xi\star\pi\\
\equiv t[\xi/x,\eta_1/y_1,\ldots,\eta_n/y_n]\star\pi$ since $u$ does not contain $x$.\\
If it is rule~5, we have $t=ux$ \ and \ $(\lbd x\,t)^*\star\xi\ps\pi\equiv(W\lbd x(E)u)^*\star\xi\ps\pi\succ
W\star(\lbd x(E)u)^*\ps\xi\ps\pi\\
\succ(\lbd x(E)u)^*\star\xi\ps\xi\ps\pi\succ(E)u^*[\xi/x]\star\xi\ps\pi$ (by induction hypothesis)\\
$\succ E\star u^*[\xi/x]\ps\xi\ps\pi\succ(u^*[\xi/x])\xi\star\pi
\equiv t[\xi/x,\eta_1/y_1,\ldots,\eta_n/y_n]\star\pi$.\\
If it is rule~6, we have $t=(u)(v)w$ \ and \ $(\lbd x\,t)^*\star\xi\ps\pi\equiv(\lbd x(B)uvw)^*\star\xi\ps\pi\\
\succ(B)u^*[\xi/x]v^*[\xi/x]w^*[\xi/x]\star\pi$  (by induction hypothesis)\\
$\succ B\star u^*[\xi/x]\ps v^*[\xi/x]\ps w^*[\xi/x]\ps\pi\succ
(u^*[\xi/x])(v^*[\xi/x])w^*[\xi/x]\star\pi\\
\equiv t[\xi/x,\eta_1/y_1,\ldots,\eta_n/y_n]\star\pi$.
\qed

\begin{thm}\label{beta_red_gauche}
If $t$ is a \ \Cc-term with the only variables $x_1,\ldots,x_n$, and if
$\xi_1,\ldots,\xi_n\in\LLbd$, then $\lbd x_1\ldots\lbd x_n\,t\star\xi_1\ps\ldots\ps\xi_n\ps\pi
\succ t[\xi_1/x_1,\ldots,\xi_n/x_n]\star\pi$.
\end{thm}

\proof  By induction on $n$~; the case $n=0$ is trivial.\\
We have \ $\lbd x_1\ldots\lbd x_{n-1}\lbd x_n\,t\star\xi_1\ps\ldots\ps\xi_{n-1}\ps\xi_n\ps\pi\succ
(\lbd x_nt)[\xi_1/x_1,\ldots,\xi_{n-1}/x_{n-1}]\star\xi_n\ps\pi$\\
(by induction hypothesis) \ $\succ t[\xi_1/x_1,\ldots,\xi_{n-1}/x_{n-1},\xi_n/x_n]\star\pi$
by lemma~\ref{brg1}.
\qed

\subsection*{Natural deduction}\noindent
Before giving the formal language that we shall use, it is perhaps useful to describe informally
the structures (models) we have in mind. They are second order structures, with two types of objects:
{\em individuals} also called {\em conditions} and \emph{predicates} (of various arity).
Since we remain at an intuitive level, we start with a \emph{full model} which we call the
\emph{ground model}. Such a model consists of:\\
$\bullet$~~an infinite set $P$ (the set of individuals or conditions).\\
$\bullet$~~the set of $k$-ary predicates is ${\mathcal P}(P^k)$ (full model).\\
$\bullet$~~some functions from $P^k$ into $P$.\\
In particular, there is an individual $0$ and a bijective function \ $s:P\to(P\setminus\{0\})$.
This enables us to define the set of integers $\ennl$ as the least set which contains $0$ and which is
closed for $s$.\\
There is also a particular condition denoted by $\1$ and an application denoted by $\et$ from $P^2$
into $P$.\\
$\bullet$~~some relations (fixed predicates) on $P$. In particular, we have the equality relation
on individuals and the subset \ $\C$ of \emph{non trivial conditions}.\\
$\C[p\et q]$ reads as: ``$p$ and $q$ are two \emph{compatible} conditions''.

\smallskip\noindent
We now come to the formal language, in order to write formulas and proofs about such
structures. It consists of:

\smallskip\noindent
$\bullet$~~\emph{individual variables} or \emph{variables of conditions} \ called $x,y,\ldots$ or $p,q,\ldots$\\
$\bullet$~~\emph{predicate variables} or \emph{second order variables} \  $X,Y,\ldots$~;
\ each predicate variable has an arity which is in $\ennl$.\\
$\bullet$~~\emph{function symbols on individuals} $f,g,\ldots$~; each one has an arity which is in
$\ennl$.\\
In particular, there is a function symbol of arity $k$ for each recursive function
$f:\ennl^k\to\ennl$. This symbol will also be written as $f$.\\
There is also a constant symbol $\1$ (which represents the greatest condition) and a binary function
symbol $\et$ (which represents the $\inf$ of two conditions).

\smallskip\noindent
The \emph{terms} are built in the usual way with variables and function symbols.

\smallskip\noindent
The \emph{atomic formulas} are the expressions $X(t_1,\ldots,t_n)$, where $X$ is an $n$-ary predicate variable, and $t_1,\ldots,t_n$ are terms.

\smallskip\noindent
\emph{Formulas} are built as usual, from atomic formulas, with the only logical symbols \ $\to,\pt$:\\
$\bullet$~~each atomic formula is a formula~;\\
$\bullet$~~if $A,B$ are formulas, then $A\to B$ is a formula~;\\
$\bullet$~~if $A$ is a formula, then $\pt x\,A$ and $\pt X\,A$ are formulas.

\smallskip\noindent
{\bfseries Notations.}\\
The formula $A_1\to(A_2\to(\ldots(A_n\to B)\ldots)$ will be written
$A_1,A_2,\ldots,A_n\to B$.\\
The usual logical symbols are defined as follows:\\
($X$ is a predicate variable of arity~$0$, also called \emph{propositional variable})\\
$\bot\equiv\pt X\,X$~; $\neg A\equiv A\to\bot$~; $A\lor B\equiv(A\to\bot),(B\to\bot)\to\bot$~;
$A\land B\equiv(A,B\to\bot)\to\bot$~;\\
$\ex\;\mbox{\bfseries\sffamily y}\,F\equiv\pt\mbox{\bfseries\sffamily y}(F\to\bot)\to\bot$
(where {\bfseries\sffamily y} is an individual or predicate variable).\\
More generally, we shall write \ $\ex\;\mbox{\bfseries\sffamily y}\{F_1,\ldots,F_k\}$ \ for \
$\pt\;\mbox{\bfseries\sffamily y}(F_1,\ldots,F_k\to\bot)\to\bot$.\\
We shall sometimes write \ $\vec{F}$ \ for a finite sequence of formulas \ $F_1,\ldots,F_k$.\\
Then, we shall also write \ $\ex\;\mbox{\bfseries\sffamily y}\{\vec{F}\}$ \ and \
$\pt\;\mbox{\bfseries\sffamily y}(\vec{F}\to\bot)\to\bot$.

\smallskip\noindent
$x=y$ is the formula $\pt Z(Zx\to Zy)$, where $Z$ is a unary predicate variable.

\smallskip\noindent
The rules of natural deduction are the following (the $A_i$'s are formulas, the $x_i$'s
are variables of \Cc-terms, $t,u$ are \Cc-terms):

\smallskip\noindent
1.~$x_1:A_1,\ldots,x_n:A_n\vdash x_i:A_i$.\\
2.~$x_1:A_1,\ldots,x_n:A_n\vdash t:A\to B$, \ \ $x_1:A_1,\ldots,x_n:A_n\vdash u:A$ \ \
$\Fl$ \ \ $x_1:A_1,\ldots,x_n:A_n\vdash tu:B$.\\
3.~$x_1:A_1,\ldots,x_n:A_n,x:A\vdash t:B$ \ \ $\Fl$ \ \
$x_1:A_1,\ldots,x_n:A_n\vdash\lbd x\,t:A\to B$.\\
4.~$x_1:A_1,\ldots,x_n:A_n\vdash t:A$ \ \ $\Fl$ \ \
$x_1:A_1,\ldots,x_n:A_n\vdash t:\pt${\bfseries\sffamily x}$\,A$ \ \
for every variable {\bfseries\sffamily x} (individual or predicate) which does not appear in
$A_1,\ldots,A_n$.\\
5.~$x_1:A_1,\ldots,x_n:A_n\vdash t:\pt x\,A$ \ \ $\Fl$ \ \
$x_1:A_1,\ldots,x_n:A_n\vdash t:A[\tau/x]$ \ \ where $x$ is an individual variable and
$\tau$ is a term.\\
6.~$x_1:A_1,\ldots,x_n:A_n\vdash t:\pt X\,A$ \ \ $\Fl$ \ \
$x_1:A_1,\ldots,x_n:A_n\vdash t:A[F/Xy_1\ldots y_k]$ \ \
where $X$ is a predicate variable of arity $k$ \ and $F$ an arbitrary formula.

\smallskip\noindent
{\small{\bfseries Remark.}\\
In the notation $A[F/Xy_1\ldots y_k]$, the variables \ $y_1,\ldots,y_k$ are bound.
A more usual notation is:\\
$A[\lbd\!y_1\ldots\lbd\!y_k\,F/X]$. I prefer this one, to avoid confusion with the $\lbd$
defined for $\Cc$-terms.}

\subsection*{Realizability}\noindent
Given a realizability algebra ${\mathcal A}=(\LLbd,\PPi,\LLbd\star\PPi,\bbot)$,
a \emph{${\mathcal A}$-model} ${\mathcal M}$ consists of the following data:\\
$\bullet$~~An infinite set $P$ which is the domain of variation of individual variables.\\
$\bullet$~~The domain of variation of $k$-ary predicate variables is ${\mathcal P}(\PPi)^{P^k}$.\\
$\bullet$~~We associate with each $k$-ary function symbol $f$, a function from
$P^k$ into $P$, denoted by $\ov{f}$ or even $f$ if there is no ambiguity.\\
In particular, there is a distinguished element $0$ in $P$ and a function $s:P\to P$
(which is the interpretation of the symbol $s$). We suppose that $s$ is a bijection from $P$ onto $P\setminus\{0\}$.Then, we can identify $s^n0\in P$ with the integer $n$, and therefore, we have
$\ennl\subset P$.\\
Each recursive function $f:\ennl^k\to\ennl$ is, by hypothesis, a function symbol. Of course,
we assume that its interpretation $\ov{f}:P^k\to P$ takes the same values as $f$ on $\ennl^k$.\\
Finally, we have also a condition $\1\in P$ and a binary function $\et$ from $P^2$ into $P$.

\smallskip\noindent
A \emph{closed term} (resp. a \emph{closed formula}) \emph{with parameters in the model ${\mathcal M}$}
is, by definition, a term (resp. a formula) in which all free occurrences of each variable have been
replaced with a \emph{parameter}, i.e. an object of the same type in the model ${\mathcal M}$:
a condition for an individual variable, an application from $P^k$ into ${\mathcal P}(\PPi)$ for a
$k$-ary predicate variable.\\
Each closed term $t$, with parameters in ${\mathcal M}$ has a value $\ov{t}\in P$.

\smallskip\noindent
An {\em interpretation} $\mathcal I$ is an application which associates an individual (condition)
with each individual variable and a parameter of arity $k$ with each second order $k$-ary variable.\\
${\mathcal I}[x\lf p]$ (resp. ${\mathcal I}[X\lf{\mathcal X}]$) is, by definition, the interpretation obtained by
changing, in ${\mathcal I}$, the value of the variable $x$ (resp. $X$) and giving to it the value $p\in P$
(resp. ${\mathcal X}\in{\mathcal P}(\PPi)^{P^k}$).\\
For each formula $F$ (resp. term $t$), we denote by $F^{\mathcal I}$ (resp. $t^{\mathcal I})$
the closed formula (resp. term) with parameters obtained by replacing each free variable with the value
given by~$\mathcal I$.

\smallskip\noindent
For each closed formula $F^{\mathcal I}$ with parameters in ${\mathcal M}$, we define two truth values:\\
$\|F^{\mathcal I}\|\subset\PPi$ and $|F^{\mathcal I}|\subset\LLbd$.\\
$|F^{\mathcal I}|$ is defined as follows: \ $\xi\in|F^{\mathcal I}|$ \ $\Dbfl$ \
$(\pt\pi\in\|F^{\mathcal I}\|)\,\xi\star\pi\in\bbot$.\\
$\|F^{\mathcal I}\|$ is defined by recurrence on $F$:\\
$\bullet$~~$F$ is atomic: then $F^{\mathcal I}$ has the form ${\mathcal X}(t_1,\ldots,t_k)$ where
${\mathcal X}:P^k\to{\mathcal P}(\PPi)$ and the $t_i$'s are closed terms with parameters in ${\mathcal M}$.
We set \ $\|{\mathcal X}(t_1,\ldots,t_k)\|={\mathcal X}(\ov{t}_1,\ldots,\ov{t}_k)$.\\
$\bullet$~~$F\equiv A\to B$: we set \
$\|F^{\mathcal I}\|=\{\xi\ps\pi~;\;\xi\in|A^{\mathcal I}|,\pi\in\|B^{\mathcal I}\|\}$.\\
$\bullet$~~$F\equiv\pt x\,A$: we set \ $\|F^{\mathcal I}\|=\bigcup\{\|A^{{\mathcal I}[x\lf p]}\|~;\;p\in P\}$.\\
$\bullet$~~$F\equiv\pt X\,A$: we set \
$\|F^{\mathcal I}\|=\bigcup\{\|A^{{\mathcal I}[X\lf{\mathcal X}]}\|~;\;{\mathcal X}\in{\mathcal P}(\PPi)^{P^k}\}$
if $X$ is a $k$-ary predicate variable.

\smallskip\noindent
{\bfseries Notation.} \ We shall write \ $\xi\force F$ \ for \ $\xi\in|F|$.

\begin{thm}[Adequacy lemma]\label{adequat}\ \\
If \ $x_1:A_1,\ldots,x_k:A_k\vdash t:A$ \ and if \
$\xi_1\force A_1^{\mathcal I},\ldots,\xi_k\force A_k^{\mathcal I}$, where ${\mathcal I}$ is an interpretation,
then \ $t[\xi_1/x_1,\ldots,\xi_k/x_k]\force A^{\mathcal I}$.\\
In particular, if $A$ is closed and if \ $\vdash t:A$, then $t\force A$.
\end{thm}

\proof By recurrence on the length of the derivation of \ $x_1:A_1,\ldots,x_n:A_n\vdash t:A$.\\
We consider the last used rule.

\smallskip\noindent
1. We have $t=x_i,A\equiv A_i$. Now, we have assumed that $\xi_i\force A_i^{\mathcal I}$~; and it is
the desired result.

\smallskip\noindent
2. We have $t=uv$ and we already obtained:\\
$x_1:A_1,\ldots,x_k:A_k\vdash u:B\to A$ \ and \ $x_1:A_1,\ldots,x_k:A_k\vdash v:B$.\\
Given $\pi\in\|A^{\mathcal I}\|$, we must show \ $(uv)[\xi_1/x_1,\ldots,\xi_k/x_k]\star\pi\in\bbot$.\\
By hypothesis on $\bbot$, it suffices to show \
$u[\xi_1/x_1,\ldots,\xi_k/x_k]\star v[\xi_1/x_1,\ldots,\xi_k/x_k]\ps\pi\in\bbot$.\\
By the induction hypothesis, we have \ $v[\xi_1/x_1,\ldots,\xi_k/x_k]\force B^{\mathcal I}$ and therefore:\\
$v[\xi_1/x_1,\ldots,\xi_k/x_k]\ps\pi\in\|B^{\mathcal I}\to A^{\mathcal I}\|$.\\
But, by the induction hypothesis, we have also \ $u[\xi_1/x_1,\ldots,\xi_k/x_k]\force B^{\mathcal I}\to A^{\mathcal I}$, hence the result.

\smallskip\noindent
3. We have $A=B\to C$, $t=\lbd x\,u$. We must show \
$\lbd x\,u[\xi_1/x_1,\ldots,\xi_k/x_k]\force B^{\mathcal I}\to C^{\mathcal I}$~;
thus, we suppose $\xi\force B^{\mathcal I}$, $\pi\in\| C^{\mathcal I}\|$ and we have to show
$\lbd x\,u[\xi_1/x_1,\ldots,\xi_k/x_k]\star\xi\ps\pi\in\bbot$. By hypothesis on $\bbot$ and
lemma~\ref{brg1}, it suffices to show \ $u[\xi/x,\xi_1/x_1,\ldots,\xi_k/x_k]\star\pi\in\bbot$.\\
This follows from the induction hypothesis applied to \ $x_1:A_1,\ldots,x_n:A_n,x:B\vdash u:C$.

\smallskip\noindent
4. We have $A\equiv\pt X\,B$, and $X$ is not free in $A_1,\ldots,A_n$. We must show:\\
$t[\xi_1/x_1,\ldots,\xi_k/x_k]\force(\pt X\,B)^{\mathcal I}$, \ i.e. \
$t[\xi_1/x_1,\ldots,\xi_k/x_k]\force B^{\mathcal J}$ with
${\mathcal J}={\mathcal I}[X\leftarrow{\mathcal X}]$. But, by hypothesis, $\xi_i\force A_i^{\mathcal I}$
therefore $\xi_i\force A_i^{\mathcal J}$: indeed, since $X$ is not free in $A_i$, we have:\\
$\|A_i^{\mathcal I}\|=\|A_i^{\mathcal J}\|$. Then, the induction hypothesis gives the result.

\smallskip\noindent
6. We have $A=B[F/Xy_1\ldots y_n]$ and we must show:\\
$t[\xi_1/x_1,\ldots,\xi_k/x_k]\force B[F/Xy_1\ldots y_n]^{\mathcal I}$
assuming that $t[\xi_1/x_1,\ldots,\xi_k/x_k]\force(\pt X\,B)^{\mathcal I}$.\\
This follows from lemma~\ref{BF/Xy} below.
\qed

\begin{lem}\label{BF/Xy}
$\|B[F/Xy_1\ldots y_n]^{\mathcal I}\|=\|B^{{\mathcal I}[X\lf{\mathcal X}]}\|$ where
${\mathcal X}:P^n\to{\mathcal P}(\PPi)$ is defined by:\\
${\mathcal X}(p_1,\ldots,p_n)=\| F^{{\mathcal I}[y_1\lf p_1,\ldots,y_n\lf p_n]}\|$.
\end{lem}

\proof
The proof is by induction on $B$. That is trivial if $X$ is not free in $B$. Indeed, the only
non trivial case of the induction is \ $B=\pt Y\,C$~; and then, we have \ $Y\ne X$ \ and:\\
$\|B[F/Xy_1\ldots y_n]^{\mathcal I}\|=\|(\pt Y\,C[F/Xy_1\ldots y_n])^{\mathcal I}\|=
\bigcup_{\mathcal Y}\|C[F/Xy_1\ldots y_n]^{{\mathcal I}[Y\leftarrow{\mathcal Y}]}\|$.\\
By induction hypothesis, this gives \
$\bigcup_{\mathcal Y}\| C^{{\mathcal I}[Y\leftarrow{\mathcal Y}][X\leftarrow{\mathcal X}]}\|$, \ that is \
$\bigcup_{\mathcal Y}\| C^{{\mathcal I}[X\leftarrow{\mathcal X}][Y\leftarrow{\mathcal Y}]}\|$ \ i.e.
$\|(\pt Y\,C)^{{\mathcal I}[X\leftarrow{\mathcal X}]}\|$.
\qed

\begin{lem}\label{continuation}
Let ${\mathcal X,Y}\subset\PPi$ be truth values. If $\pi\in{\mathcal X}$, \ then \
$\kk_\pi\force{\mathcal X}\to{\mathcal Y}$.
\end{lem}

\proof
Suppose \ $\xi\force{\mathcal X}$ \ and \ $\rho\in{\mathcal Y}$~; we must show \
$\kk_\pi\star\xi\ps\rho\in\bbot$, that is $\xi\star\pi\in\bbot$, which is clear.
\qed

\begin{prop}[Law of Peirce]
$\Ccc\force\pt X\pt Y(((X\to Y)\to X)\to X)$.
\end{prop}

\proof
We want to show that \ $\Ccc\force(({\mathcal X}\to{\mathcal Y})\to{\mathcal X})\to{\mathcal X}$.
Thus, we take \ $\xi\force({\mathcal X}\to{\mathcal Y})\to{\mathcal X}$ and $\pi\in{\mathcal X}$~;
we must show that \ $\Ccc\star\xi\ps\pi\in\bbot$, that is  $\xi\star\kk_\pi\ps\pi\in\bbot$.
By hypothesis on $\xi$ and $\pi$, it is sufficient to show that $\kk_\pi\force{\mathcal X}\to{\mathcal Y}$,
which results from lemma~\ref{continuation}.
\qed

\begin{prop}\label{eta_exp}\ \\
\emph{\phantom ii)} If \ $\xi\force A\to B$, \ then \ $\pt\eta(\eta\force A\Fl\xi\eta\force B)$.\\
\emph{ii)} If \ $\pt\eta(\eta\force A\Fl\xi\eta\force B)$, \ then \ $(E)\xi\force A\to B$.
\end{prop}
\proof{\ }\ \\
\phantom ii) From \ $\xi\eta\star\pi\succ\xi\star\eta\ps\pi$.\\
ii) From \ $(E)\xi\star\eta\ps\pi\succ\xi\eta\star\pi$.
\qed

\smallskip\noindent
{\small{\bfseries Remark.} Proposition~\ref{eta_exp} shows that \ $\xi\force A\to B$ \ is ``almost'' equivalent
(i.e. up to an $\eta$-expansion of $\xi$) to  \ $\pt\eta(\eta\force A\Fl\xi\eta\force B)$.}

\subsubsection*{Predicate symbols}\noindent
In the following, we shall use \emph{extended formulas} which contain
\emph{predicate symbols (or predicate constants) \ {\sf R,S}, \ldots \ on individuals}.
Each one has an arity, which is an integer.\\
In particular, we have a unary predicate symbol $\C$ (which represents the set of non trivial conditions).\\
We have to add some rules of construction of formulas:

\smallskip\noindent
$\bullet$~~If $F$ is a formula, {\sf R} is a $n$-ary predicate constant and $t_1,\ldots,t_n$ are terms, then\\
${\sf R}(t_1,\ldots,t_n)\to F$ \ and \ ${\sf R}(t_1,\ldots,t_n)\mapsto F$ are formulas.\\
$\bullet$~~$\top$ is an atomic formula.

\smallskip\noindent
In the definition of a ${\mathcal A}$-model ${\mathcal M}$, we add the following clause:

\smallskip\noindent
$\bullet$~~With each relation symbol {\sf R} of arity $n$, we associate an application, denoted by
$\ov{\sf R}_{\mathcal M}$ or $\ov{\sf R}$, from $P^n$ into ${\mathcal P}(\LLbd)$.
We shall also write \ $|{\sf R}(p_1,\ldots,p_n)|$, \ instead of \ $\ov{\sf R}(p_1,\ldots,p_n)$, for
$p_1,\ldots,p_n\in P$.\\
In particular, we have an application $\ov{\sf C}:P\to{\mathcal P}(\LLbd)$, which we denote as \ $|\C[p]|$.

\smallskip\noindent
We define as follows the truth value in ${\mathcal M}$ of an extended formula:

\smallskip\noindent
$\|\top\|=\vide$.\\
$\|({\sf R}(t_1,\ldots,t_n)\to F)^{\mathcal I}\|=
\{t\ps\pi;\;t\in|{\sf R}(t_1^{\mathcal I},\ldots,t_n^{\mathcal I})|,\pi\in\|F^{\mathcal I}\|\}$.\\
$\|({\sf R}(t_1,\ldots,t_n)\mapsto F)^{\mathcal I}\|=\|F^{\mathcal I}\|$ if
$I\in|{\sf R}(t_1^{\mathcal I},\ldots,t_n^{\mathcal I})|$~;\\
$\|({\sf R}(t_1,\ldots,t_n)\mapsto F)^{\mathcal I}\|=\vide$ otherwise.

\begin{prop}\label{R_simple}\ \\
\emph{\phantom ii)}~$\lbd x(x)I\force\pt X\pt x_1\ldots\pt x_n[({\sf R}(x_1,\ldots,x_n)\to X)\to
({\sf R}(x_1,\ldots,x_n)\mapsto X)]$.\\
\emph{ii)}~If we have \ $|{\sf R}(p_1,\ldots,p_n)|\ne\vide$ $\Fl$ $I\in|{\sf R}(p_1,\ldots,p_n)|$ \
for every $p_1,\ldots,p_n\in P$, then:\\
$K\force\pt X\pt x_1\ldots\pt x_n[({\sf R}(x_1,\ldots,x_n)\mapsto X)\to
({\sf R}(x_1,\ldots,x_n)\to X)]$.
\end{prop}

\proof
Trivial.
\qed

\smallskip\noindent
{\small{\bfseries Remark.} By means of proposition~\ref{R_simple}, we see that, if the application
$\ov{\sf R}:P^n\to{\mathcal P}(\LLbd)$ takes only the values $\{I\}$ and $\vide$, we can
replace \ ${\sf R}(t_1,\ldots,t_n)\to F$ \ with \ ${\sf R}(t_1,\ldots,t_n)\mapsto F$.}

\smallskip\noindent
We define the binary predicate $\simeq$ by putting $|p\simeq q|=\{I\}$ if $p=q$ \ and \
$|p\simeq q|=\vide$ if $p\ne q$.\\
By the above remark, we can replace \ $p\simeq q\to F$ with $p\simeq q\mapsto F$.
Proposition~\ref{egal_simple} shows that we can also replace \ $p=q\to F$ with $p\simeq q\mapsto F$.

\smallskip\noindent
{\bfseries Notations.} We shall write \ $p=q\mapsto F$ \ instead of \ $p\simeq q\mapsto F$.
Thus, we have:\\
$\|p=q\mapsto F\|=\|F\|$ \ if $p=q$~; \ $\|p=q\mapsto F\|=\vide$ \ if $p\ne q$.\\
We shall write \ $p\ne q$ \ for \ $p=q\mapsto\bot$. Thus, we have:\\
$\|p\ne q\|=\PPi$ if $p=q$ and \ $\|p\ne q\|=\vide$ if $p\ne q$.

\smallskip\noindent
Using \ $p=q\mapsto F$ \ instead of \ $p=q\to F$, and \ $p\ne q$ \ instead of $p=q\to\bot$,
greatly simplifies the computation of the truth value of a formula which contains the symbol $=$.

\begin{prop}\label{egal_simple}\ \\
\emph{\phantom ii)}~$\lbd x\,xI\force\pt X\pt x\pt y((x=y\to X)\to(x=y\mapsto X))$~;\\
\emph{ii)}~$\lbd x\lbd y\,yx\force\pt X\pt x\pt y((x=y\mapsto X),x=y\to X)$.
\end{prop}
\proof{\ }\ \\
\phantom ii)~Let $a,b\in P\,,\;{\mathcal X}\subset\PPi,\xi\force a=b\to{\mathcal X}$
and $\pi\in\|a=b\mapsto{\mathcal X}\|$.\\
Then, we have $a=b$, thus $I\force a=b$, therefore $\xi\star I\ps\pi\in\bbot$, thus
$\lbd x\,xI\star\xi\ps\pi\in\bbot$.\\
ii)~Now let $\eta\force(a=b\mapsto{\mathcal X}),\;\zeta\force a=b$ and $\rho\in\|{\mathcal X}\|$.\\
We show that \ $\lbd x\lbd y\,yx\star\eta\ps\zeta\ps\rho\in\bbot$
in other words \ $\zeta\star\eta\ps\rho\in\bbot$.\\
If $a=b$, then $\eta\force{\mathcal X}$, $\zeta\force\pt Y(Y\to Y)$.
We have $\eta\ps\rho\in\|{\mathcal X}\to{\mathcal X}\|$, thus $\zeta\star\eta\ps\rho\in\bbot$.\\
If $a\ne b$, then $\zeta\force\top\to\bot$, thus $\zeta\star\eta\ps\rho\in\bbot$.\\
In both cases, we get the desired result.
\qed

\smallskip\noindent
{\small{\bfseries Remark.}\\
Let $R$ be a subset of $P^k$ and $1_R:P^k\to\{0,1\}$ its characteristic function, defined as follows:\\
$1_R(p_1,\ldots,p_n)=1$ (resp. $=0$) if $(p_1,\ldots,p_n)\in R$ (resp. $(p_1,\ldots,p_n)\notin R$).\\
Let us define the predicate $R$ in the model ${\mathcal M}$ by putting:\\
$|R(p_1,\ldots,p_n)|=\{I\}$ (resp. $=\vide$) if $(p_1,\ldots,p_n)\in R$ (resp. $(p_1,\ldots,p_n)\notin R$).\\
By propositions~\ref{R_simple} and~\ref{egal_simple}, we see that $R(x_1,\ldots,x_n)$ and
$1_R(x_1,\ldots,x_n)=1$ are interchangeable. More precisely, we have: \ \
$I\force\pt X\pt x_1\ldots\pt x_n((R(x_1,\ldots,x_n)\mapsto X)\dbfl(1_R(x_1,\ldots,x_n)=1\mapsto X))$.}

\smallskip\noindent
For each formula $A[x_1,\ldots,x_k]$, we can define the $k$-ary predicate symbol $N_A$, by putting
$|N_A(p_1,\ldots,p_k)|=\{\kk_\pi;\;\pi\in\|A[p_1,\ldots,p_k]\|\}$. Proposition~\ref{N_A_negA} below shows
that $N_A$ and $\neg A$ are interchangeable~; this may simplify truth value computations.

\begin{prop}\label{N_A_negA}\ \\
\emph{\phantom ii)}~~$I\force\pt x_1\ldots\pt x_k(N_A(x_1,\ldots,x_k)\to\neg A(x_1,\ldots,x_k))$~;\\
\emph{ii)}~~$\Ccc\force\pt x_1\ldots\pt x_k((N_A(x_1,\ldots,x_k)\to\bot)\to A(x_1,\ldots,x_k))$.
\end{prop}
\proof{\ }\ \\
\phantom ii)~~Let $p_1,\ldots,p_k\in P$, $\pi\in\|A(p_1,\ldots,p_k)\|$, $\xi\force A(p_1,\ldots,p_k)$ and $\rho\in\PPi$. We must show:\\
$I\star\kk_\pi\ps\xi\ps\rho\in\bbot$, that is $\xi\star\pi\in\bbot$, which is obvious.\\
ii)~~Let $\eta\force N_A(p_1,\ldots,p_k)\to\bot$ \ and \ $\pi\in\|A(p_1,\ldots,p_k)\|$. We must show:\\
$\Ccc\star\eta\ps\pi\in\bbot$, i.e. $\eta\star\kk_\pi\ps\pi\in\bbot$, \ which is clear, since \
$\kk_\pi\in|N_A(p_1,\ldots,p_k)|$.
\qed

\subsubsection*{Fixed point combinator}
\begin{thm}\label{bien_fonde}
Let \ $\Y=AA$ \ with $A=\lbd a\lbd f(f)(a)af$. Then, we have \ $\Y\star\xi\ps\pi\succ\xi\star\Y\xi\ps\pi$.\\
Let $f:P^2\to P$ such that $f(x,y)=1$ is a well founded relation on $P$. Then:\\
\phantom ii)~~$\Y\force\pt X\{\pt x[\pt y(f(y,x)=1\mapsto Xy)\to Xx]\to\pt x\,Xx\}$.\\
ii)~~$\Y\force\pt X_1\ldots\pt X_k\\
\hspace*{\fill}\{\pt x[\pt y(X_1y,\ldots,X_ky\to f(y,x)\ne1),X_1x,\ldots,X_kx\to\bot]
\to\pt x(X_1x,\ldots,X_kx\to\bot)\}$.
\end{thm}

\proof
The property \ $\Y\star\xi\ps\pi\succ\xi\star\Y\xi\ps\pi$ \ is immediate, from
theorem~\ref{beta_red_gauche}.\\
\phantom ii)~~We take ${\mathcal X}:P\to{\mathcal P}(\PPi)$, $p\in P$ and \
$\xi\force\pt x[\pt y(f(y,x)=1\mapsto{\mathcal X}y)\to{\mathcal X}x]$. We show, by induction on
the well founded relation $f(x,y)=1$, that $\Y\star\xi\ps\pi\in\bbot$ for every $\pi\in{\mathcal X}p$.\\
Let $\pi\in{\mathcal X}p$~; from (i), we get \ $\Y\star\xi\ps\pi\succ\xi\star\Y\xi\ps\pi$ \ and thus,
it is sufficient to prove that \ $\xi\star\Y\xi\ps\pi\in\bbot$.
By hypothesis, we have $\xi\force\pt y(f(y,p)=1\mapsto{\mathcal X}y)\to{\mathcal X}p$~; thus, it suffices
to show that $\Y\xi\force f(q,p)=1\mapsto{\mathcal X}q$ for every $q\in P$.
This is clear if $f(q,p)\ne1$, by definition of~~$\mapsto$.\\
If $f(q,p)=1$, we must show $\Y\xi\force{\mathcal X}q$, i.e. $\Y\star\xi\ps\rho\in\bbot$ for every
$\rho\in{\mathcal X}q$. But this follows from the induction hypothesis.

\smallskip\noindent
ii)~~The proof is almost the same: take ${\mathcal X}_1,\ldots,{\mathcal X}_k:P\to{\mathcal P}(\PPi)$, $p\in P$ and\\
$\xi\force\pt x[\pt y({\mathcal X}_1y,\ldots,{\mathcal X}_ky\to f(y,x)\ne1),{\mathcal X}_1x,\ldots,{\mathcal X}_kx\to\bot]$.
We show, by induction on the well founded relation $f(x,y)=1$, that $\Y\star\xi\ps\pi\in\bbot$
for every $\pi\in\|{\mathcal X}_1p,\ldots,{\mathcal X}_kp\to\bot\|$.\\
As before, we have to show that: \
$\Y\xi\force{\mathcal X}_1q,\ldots,{\mathcal X}_kq\to f(q,p)\ne1$ for all $q\in P$~;\\
this is obvious if $f(q,p)\ne1$. If \ $f(q,p)=1$, we must show \
$\Y\xi\force{\mathcal X}_1q,\ldots,{\mathcal X}_kq\to\bot$, or else:\\
$\Y\star\xi\ps\rho\in\bbot$ for every $\rho\in\|{\mathcal X}_1q,\ldots,{\mathcal X}_kq\to\bot\|$.
But this follows from the induction hypothesis.
\qed

\subsection*{Integers, storage and recursive functions}\noindent
Recall that we have a constant symbol $0$ and a unary function symbol  $s$ which is interpreted,
in the model ${\mathcal M}$ by a  bijective function $s:P\to(P\setminus\{0\})$.\\
And also, that we have identified $s^n0$ with  the integer $n$~; thus, we suppose  $\ennl\subset P$.

\smallskip\noindent
We denote by \ int$(x)$ the formula $\pt X(\pt y(Xy\to Xsy),X0\to Xx)$.

\smallskip\noindent
Let $u=(u_n)_{n\in\NN}$ be a sequence of elements of $\LLbd$.
We define the unary predicate symbol $e_u$ by putting: \ $|e_u(s^n0)|=\{u_n\}$~; \
$|e_u(p)|=\vide$ if $p\notin\ennl$.

\begin{thm}\label{memoire_gen}
Let $T_u,S_u\in\LLbd$ be such that \ $S_u\force(\top\to\bot),\top\to\bot$ \ and:\\
$T_u\star\phi\ps\nu\ps\pi\succ\nu\star S_u\ps\phi\ps u_0\ps\pi$~; \
$S_u\star\psi\ps u_n\ps\pi\succ\psi\star u_{n+1}\ps\pi$\\
for every $\nu,\phi,\psi\in\LLbd$ and $\pi\in\PPi$. Then:\\
$T_u\force\pt X\pt x[(e_u(x)\to X),$ int$(x)\to X]$.\\
$T_u$ is called a storage operator.
\end{thm}

\proof
Let \ $p\in P$, $\phi\force e_u(p)\to X$, \ $\nu\force$int$(p)$ \ and \ $\pi\in\|X\|$.
We must show \ $T_u\star\phi\ps\nu\ps\pi\in\bbot$ \ i.e. \ $\nu\star S_u\ps\phi\ps u_0\ps\pi\in\bbot$.

\smallskip\noindent
$\bullet$~~If $p\notin\NN$, we define the unary predicate $Y$ by putting:\\
$Y(q)\equiv\top$ if $q\in\NN$~; \ $Y(q)\equiv\top\to\bot$ if $q\notin\NN$.\\
Thus, we have obviously \ $\phi\force Y(0)$ and $u_0\ps\pi\in\|Y(p)\|$.\\
But, by hypothesis on $\nu$, we have \ $\nu\force\pt y(Yy\to Ysy),Y0\to Yp$.\\
 Thus, it is sufficient to show that:\\
$S_u\force\pt y(Yy\to Ysy)$, i.e. \ $S_u\force Y(q)\to Y(sq)$ \ for every $q\in P$.\\
This is clear if $q\in\NN$, since we have \ $\|Y(sq)\|=\vide$.\\
If $q\notin\NN$, we must show \	$S_u\force(\top\to\bot),\top\to\bot$, which follows from the hypothesis.

\smallskip\noindent
$\bullet$~~If $p\in\NN$, we have \ $p=s^p0$~; we define the unary predicate $Y$ by putting:\\
$\|Ys^i0\|=\{u_{p-i}\ps\pi\}$ for $0\le i\le p$ and
$\|Yq\|=\vide$ if $q\notin\{s^i0;\;0\le i\le p\}$.\\
By hypothesis on $\nu,\phi,\pi$, we have:\\
$\nu\force\pt y(Yy\to Ysy),Y0\to Ys^p0$~; \ $\phi\force Y0$~; \ $u_0\ps\pi\in\|Ys^p0\|$.\\
Thus, it suffices to show that \ $S_u\force\pt y(Yy\to Ysy)$, i.e. \ $S_u\force Yq\to Ysq$ for every $q\in P$.\\
This is clear if $q\notin\{s^i0;\;0\le i<p\}$, since then \ $\|Ysq\|=\vide$.\\
If $q=s^i0$ with  $i<p$, let \ $\xi\force Yq$~;
we must show \ $S_u\star\xi\ps u_{p-i-1}\ps\pi\in\bbot$.\\
But we have \ $S_u\star\xi\ps u_{p-i-1}\ps\pi\succ\xi\star u_{p-i}\ps\pi$ \ which is in~$\bbot$,
by hypothesis on $\xi$.
\qed

\smallskip\noindent
{\bfseries Notation.} We define the closed $\Cc$-terms \ $\ul{0}=\lbd x\lbd y\,y$~; \
$\sig=\lbd n\lbd f\lbd x(f)(n)fx$~; and, for each $n\in\NN$, we put \ $\ul{n}=(\sig)^n\ul{0}$.
We define the unary predicate symbol ent(x) by putting:\\
$|$ent$(n)|=\{\ul{n}\}$ if $n\in\NN$~;\\
$|$ent$(p)|=\vide$ if $p\notin\NN$.\\
In other words, ent(x) is the predicate $e_u(x)$ when the sequence $u$ is $(\ul{n})_{n\in\NN}$.

\begin{thm}\label{mm1}\ \\
We put \ $T=\lbd f\lbd n(n)Sf\ul{0}$, \ with  \ $S=\lbd g\lbd x(g)(\sigma)x$. Then, we have:\\
\emph{\phantom ii)}~~$T\force\pt X\pt x(($ent$(x)\to X),$ int$(x)\to X)$.\\
\emph{ii)}~~$I\,\force\pt x(($ent$(x)\to$int$(x))$.
\end{thm}

\noindent Therefore, $T$ is a storage operator (theorem~\ref{memoire_gen}).

\proof{\ }\ \\
\phantom ii)~~We immediately have, by theorem~\ref{beta_red_gauche}:\\
$T\star\phi\ps\nu\ps\pi\succ\nu\star S\ps\phi\ps\ul{0}\ps\pi$~; \
$S\star\psi\ps(\sig)^n\ul{0}\ps\pi\succ\psi\star(\sig)^{n+1}\ul{0}\ps\pi$\\
for every $\nu,\phi,\psi\in\LLbd$ and $\pi\in\PPi$.\\
Now, we check that \ $S\force(\top\to\bot),\top\to\bot$: indeed, if $\xi\force\top\to\bot$, then
$S\star\xi\ps\eta\ps\pi\succ\xi\star\sig\eta\ps\pi\in\bbot$ for every $\eta\in\LLbd$ and $\pi\in\PPi$
(by theorem~\ref{beta_red_gauche}).\\
Then, the result follows immediately, from theorem~\ref{memoire_gen}.

\smallskip\noindent
ii)~~We must show \ $I\force$ ent$(p)\to$ int$(p)$ for every $p\in P$. We may suppose
$p\in\NN$ (otherwise ent$(p)=\vide$ and the result is trivial).
Then, we must show:\\
$I\star\sig^p\ul{0}\ps\rho\in\bbot$ knowing that $\rho\in\|$int$(s^p0)\|$.\\
Therefore, we can find a unary predicate $X:P\to{\mathcal P}(\PPi)$, $\phi\force\pt y(Xy\to Xsy)$,
$\omega\force X0$ and $\pi\in\|Xs^p0\|$ such that $\rho=\phi\ps\omega\ps\pi$.
We must show \ $(\sig)^p\ul{0}\star\phi\ps\omega\ps\pi\in\bbot$. In fact, we show by recurrence on~$p$,
that \ $(\sig)^p\ul{0}\star\phi\ps\omega\ps\pi\in\bbot$ \ \emph{for all \ $\pi\in\|Xs^p0\|$}.\\
If $p=0$, let $\pi\in\|X0\|$~; we must show \ $\ul{0}\star\phi\ps\omega\ps\pi\in\bbot$,
i.e. $\omega\star\pi\in\bbot$, which is clear, since $\omega\force X0$.\\
To move up from $p$ to $p+1$, let $\pi\in\|Xs^{p+1}0\|$. We have:\\
$\sig^{p+1}\ul{0}\star\phi\ps\omega\ps\pi\equiv
(\sig)(\sig)^p\ul{0}\star\phi\ps\omega\ps\pi\succ\sig\star\sig^p\ul{0}\ps\phi\ps\omega\ps\pi
\succ\phi\star(\sig^p\ul{0})\phi\omega\ps\pi$.\\
But, by induction hypothesis, we have $\sig^p\ul{0}\star\phi\ps\omega\ps\rho\in\bbot$ for every
$\rho\in\|Xs^p0\|$. It follows that $(\sig^p\ul{0})\phi\omega\force Xs^p0$. Since
$\phi\force Xs^p0\to Xs^{p+1}0$, we obtain \ $\phi\star(\sig^p\ul{0})\phi\omega\ps\pi\in\bbot$.
\qed

\smallskip\noindent
Theorem~\ref{mm1} shows that we can use the predicate ent$(x)$ instead of int$(x)$,
which greatly simplifies many computations. In particular, we define the
\emph{universal quantifier restricted to integers} \ $\pt x\indi$ by putting \
$\pt x{\indi}F\equiv\pt x($int$(x)\to F)$.\\
Thus, we can replace it with the \emph{universal quantifier restricted to ent$(x)$} defined as follows:\\
$\pt x\inde\,F\equiv\pt x($ent$(x)\to F)$. Then, we have \
$\|\pt x\inde\,F\|=\{\ul{n}\ps\pi;\;n\in\NN,\pi\in\|F[s^n0/x]\|\}$.\\
Therefore, the truth value of the formula \ $\pt x\inde\,F$ \ is much simpler than the one of the
formula \ $\pt x^{\indi}F$.

\begin{thm}\label{red_tte_f}
Let $\phi:\NN\to\NN$ be a recursive function. There exists a closed $\lbd$-term \ $\theta$ such that,
if $m\in\NN$, $n=\phi(m)$ and $f$ is a $\lbd$-variable, then $\theta\ul{m}f$ reduces into
$f\ul{n}$ by weak head reduction.
\end{thm}\noindent
This is a variant of the theorem of representation of recursive functions by $\lbd$-terms. It is
proved in~\cite{krivine3}.

\begin{thm}
Let $\phi:\NN^k\to\NN$ be a recursive function. We define, in ${\mathcal M}$, a function symbol $f$,
by putting $f(s^{m_1}0,\ldots,s^{m_k}0)=s^n0$ with  $n=\phi(m_1,\ldots,m_k)$~; we extend $f$
on $P^k\setminus\NN^k$ in an arbitrary way. Then, there exists a proof-like term \ $\theta$ such that:\\
$\theta\force\pt x_1\ldots\pt x_k[$int$(x_1),\ldots,\,$int$(x_k)\to$int$(f(x_1,\ldots,x_k))]$.
\end{thm}

\proof
For simplicity, we assume $k=1$.
By theorem~\ref{mm1}, it suffices to find a proof-like term $\theta$ such that \
$\theta\force\pt x[$ent$(x),($ent$(f(x))\to\bot)\to\bot]$. In other words:\\
$\theta\force $ent$(p),($ent$(f(p))\to\bot)\to\bot$ for every $p\in P$.\\
We can suppose that $p=s^m0$ (otherwise, \ |ent$(p)|=\vide$ and the result is trivial).\\
Thus, we have \ ent$(p)=\{\ul{m}\}$~; we must show:\\
$\theta\star\ul{m}\ps\xi\ps\pi\in\bbot$ for all $\pi\in\PPi$ and
$\xi\force $ent$(s^n0)\to\bot$, with  $n=\phi(m)$.\\
Take the $\lbd$-term \ $\theta$ given by theorem~\ref{red_tte_f}. From this theorem, we get:\\
$\theta\star\ul{m}\ps\xi\ps\pi\succ\xi\star\ul{n}\ps\pi$, which is in $\bbot$, by hypothesis on $\xi$.
\qed

\smallskip\noindent
{\small{\bfseries Remark.} We have now found proof-like terms which realize all the axioms of
second order arithmetic, with  a function symbol for each recursive function.}

\section{Standard realizability algebras}\noindent
A realizability algebra ${\mathcal A}$ is called \emph{standard} if its set of terms $\Lbd$ and
its set of stacks $\Pi$ are defined as follows:\\
We have a countable set $\Pi_0$ which is the set of \emph{stack constants}.\\
The terms and the stacks of ${\mathcal A}$ are finite sequences of elements of the set:\\
\centerline{$\Pi_0\cup\{B,C,E,I,K,W,\Ccc,\vsig,\chi,\chi',\kk,(,),[,],\ps\}$}

\noindent
which are obtained by the following rules:

\smallskip\noindent
$\bullet$~~$B,C,E,I,K,W,\Ccc,\vsig,\chi,\chi'$ are terms~;\\
$\bullet$~~each element of $\Pi_0$ is a stack~;\\
$\bullet$~~if $\xi,\eta$ are terms, then $(\xi)\eta$ is a term~;\\
$\bullet$~~if $\xi$ is a term and $\pi$ a stack, then $\xi\ps\pi$ is a stack~;\\
$\bullet$~~if $\pi$ is a stack, then $\kk[\pi]$ is a term.

\smallskip\noindent
A term of the form $\kk[\pi]$ is called \emph{continuation}. It will also be denoted as $\kk_\pi$.

\smallskip\noindent
The set of processes of the algebra ${\mathcal A}$ is $\Lbd\fois\Pi$.\\
If $\xi\in\Lbd$ and $\pi\in\Pi$, the ordered pair $(\xi,\pi)$ is denoted as $\xi\star\pi$.

\smallskip\noindent
Therefore, every stack has the form \ $\pi=\xi_1\ps\ldots\ps\xi_n\ps\pi_0$,
where $\xi_1,\ldots,\xi_n\in\Lbd$ and $\pi_0\in\Pi_0$ ($\pi_0$ is a stack constant).
Given a term $\tau$, we put:\\
\centerline{$\pi^\tau=\xi_1\ps\ldots\ps\xi_n\ps\tau\ps\pi_0$.}

\smallskip\noindent
We choose a recursive bijection from $\Pi$ onto $\NN$, which is written \ $\pi\mapsto\nn_\pi$.

\smallskip\noindent
We define a preorder relation \ $\succ$, \ on $\Lbd\star\Pi$. It is the least reflexive and transitive
relation such that, for all $\xi,\eta,\zeta\in\Lbd$ and $\pi,\varpi\in\Pi$, we have:

\smallskip\noindent
$(\xi)\eta\star\pi\succ\xi\star\eta\ps\pi$.\\
$I\star\xi\ps\pi\succ\xi\star\pi$.\\
$K\star\xi\ps\eta\ps\pi\succ\xi\star\pi$.\\
$E\star\xi\ps\eta\ps\pi\succ(\xi)\eta\star\pi$.\\
$W\star\xi\ps\eta\ps\pi\succ\xi\star\eta\ps\eta\ps\pi$.\\
$C\star\xi\ps\eta\ps\zeta\ps\pi\succ\xi\star\zeta\ps\eta\ps\pi$.\\
$B\star\xi\ps\eta\ps\zeta\ps\pi\succ(\xi)(\eta)\zeta\star\pi$.\\
$\Ccc\star\xi\ps\pi\succ\xi\star\kk_\pi\ps\pi$.\\
$\kk_\pi\star\xi\ps\varpi\succ\xi\star\pi$.\\
$\vsig\star\xi\ps\pi\succ\xi\star\ul{\nn}_\pi\ps\pi$.\\
$\chi\star\xi\ps\pi^\tau\succ\xi\star\tau\ps\pi$.\\
$\chi'\star\xi\ps\tau\ps\pi\succ\xi\star\pi^\tau$.

\smallskip\noindent
Finally, we have a subset $\bbot$ of $\Lbd\star\Pi$ which is a final segment for this
preorder, which means that: \ $\p\in\bbot$, $\p'\succ\p$ $\Fl$ $\p'\in\bbot$.\\
In other words, we ask that $\bbot$ has the following properties:

\smallskip\noindent
$(\xi)\eta\star\pi\notin\bbot\Fl\xi\star\eta\ps\pi\notin\bbot$.\\
$I\star\xi\ps\pi\notin\bbot\Fl\xi\star\pi\notin\bbot$.\\
$K\star\xi\ps\eta\ps\pi\notin\bbot\Fl\xi\star\pi\notin\bbot$.\\
$E\star\xi\ps\eta\ps\pi\notin\bbot\Fl(\xi)\eta\star\pi\notin\bbot$.\\
$W\star\xi\ps\eta\ps\pi\notin\bbot\Fl\xi\star\eta\ps\eta\ps\pi\notin\bbot$.\\
$C\star\xi\ps\eta\ps\zeta\ps\pi\notin\bbot\Fl\xi\star\zeta\ps\eta\ps\pi\notin\bbot$.\\
$B\star\xi\ps\eta\ps\zeta\ps\pi\notin\bbot\Fl(\xi)(\eta)\zeta\star\pi\notin\bbot$.\\
$\Ccc\star\xi\ps\pi\notin\bbot\Fl\xi\star\kk_\pi\ps\pi\notin\bbot$.\\
$\kk_\pi\star\xi\ps\varpi\notin\bbot\Fl\xi\star\pi\notin\bbot$.\\
$\vsig\star\xi\ps\pi\notin\bbot\Fl\xi\star \ul{\nn}_\pi\ps\pi\notin\bbot$.\\
$\chi\star\xi\ps\pi^\tau\notin\bbot\Fl\xi\star\tau\ps\pi\notin\bbot$.\\
$\chi'\star\xi\ps\tau\ps\pi\notin\bbot\Fl\xi\star\pi^\tau\notin\bbot$.

\smallskip\noindent
{\small{\bfseries Remark.} Thus, the only arbitrary elements in a standard realizability algebra are
the set $\Pi_0$ of stack constants and the set $\bbot$ of processes.}

\subsection*{The axiom of choice for individuals (ACI)}\noindent
Let ${\mathcal A}$ be a standard realizability algebra and ${\mathcal M}$ a ${\mathcal A}$-model, the set of
individuals of which is denoted as $P$. Then, we have:

\begin{thm}[ACI]\label{ACI}
For each closed formula \ $\pt x_1\ldots\pt x_m\pt y\,F$ \ with  parameters, there exists a function
$f:P^{m+1}\to P$ such that:\\
\emph{\phantom ii)}~~$\vsig\force\pt x_1\ldots\pt x_m(\pt x($ent$(x)\to F[f(x_1,\ldots,x_m,x)/y])\to\pt y\,F)$.\\
\emph{ii)}~~$\vsig\force\pt x_1\ldots\pt x_m(\pt x($int$(x)\to F[f(x_1,\ldots,x_m,x)/y])\to\pt y\,F)$.
\end{thm}

\proof
For $p_1,\ldots,p_m,k\in P$, we define $f(p_1,\ldots,p_m,k)$ in an arbitrary way if $k\notin\NN$.\\
If $k\in\NN$, we have $k=\nn_{\pi_k}$ for one and only one stack $\pi_k\in\Pi$.\\ 
We define the function $f(p_1,\ldots,p_m,k)$ by means of the axiom of choice, in such a way that,
if there exists \ $q\in P$ such that:\\
$\pi_k\in\|F[p_1,\ldots,p_m,q]\|$, then we have $\pi_k\in\|F[p_1,\ldots,p_m,f(p_1,\ldots,p_m,k)]\|$.

\smallskip\noindent
\phantom ii)~We must show \ $\vsig\force\pt x($ent$(x)\to F[p_1,\ldots,p_m,f(p_1,\ldots,p_m,x)])\to F[p_1,\ldots,p_m,q]$, for every $p_1,\ldots,p_m,q\in P$.\\
Thus, let \ $\xi\force\pt x($ent$(x)\to F[p_1,\ldots,p_n,f(p_1,\ldots,p_n,x)])$ and
$\pi\in\|F[p_1,\ldots,p_m,q]\|$~; we must show
$\vsig\star\xi\ps\pi\in\bbot$, that is \ $\xi\star\ul{\nn}_\pi\ps\pi\in\bbot$. But we have:\\
$\xi\force$ent$(\nn_\pi)\to F[p_1,\ldots,p_m,f(p_1,\ldots,p_m,\nn_\pi)]$ by hypothesis on $\xi$~;\\
$\ul{\nn}_\pi\in|$ent$(\nn_\pi)|$ by definition of ent~;\\
$\pi\in\|F[p_1,\ldots,p_m,f(p_1,\ldots,p_m,\nn_\pi)]\|$ by hypothesis on $\pi$ and by definition of $f$.

\smallskip\noindent
ii)~The proof is the same~; in fact, (ii) is weaker than (i) since $|\,$ent$(x)|\subset|\,$int$(x)|$.
\qed

\smallskip\noindent
{\small{\bfseries Remarks.}\\
1.~A seemingly simpler formulation of this axiom of choice is the existence of a function\\
${\phi:P^m\to P}$ such that $\pt x_1\ldots\pt x_m(F[\phi(x_1,\ldots,x_m)/y]\to\pt y\,F)$.\\
This clearly follows from theorem~\ref{ACI}: simply define $\phi(x_1,\ldots,x_m)$ as $f(x_1,\ldots,x_m,x)$
for the first integer $x$ such that $\neg F[f(x_1,\ldots,x_m,x)/y]$ if there is such an integer~;
otherwise, $\phi(x_1,\ldots,x_m)$ is arbitrary.\\
But this function $\phi$ is not a \emph{function symbol}, i.e. it cannot be defined in the ground
model. For this reason, we prefer to use this axiom in the form stated in theorem~\ref{ACI}, which is, after all, much simpler.\\
2~.The axiom of dependent choice DC is a trivial consequence of ACI~; therefore theorem~\ref{ACI} shows that DC is realized by a proof-like term.
Theorem~\ref{ACI} is also crucial to prove theorem~\ref{cons_reels} (see lemma~\ref{H_existe}).\\
3.~In the following, there will be individuals which represent \emph{sets of integers}
(proposition~\ref{rep_pred}), but \emph{extensionality is not realized}. That is why ACI is much weaker
than the usual axiom of choice. For instance, it does not imply well-ordering.}

\subsection*{Generic models}\noindent
Given a \emph{standard} realizability algebra ${\mathcal A}$ and a ${\mathcal A}$-model ${\mathcal M}$,
we now build a new realizability algebra ${\mathcal B}$ and a ${\mathcal B}$-model ${\mathcal N}$,
which is called \emph{generic} over ${\mathcal M}$.
Then, we shall define the notion of \emph{forcing}, which is a syntactic transformation on formulas~;
it is the essential tool in order to compute truth values in the generic model~${\mathcal N}$.

\smallskip\noindent
Thus, we consider a standard realizability algebra ${\mathcal A}$ and a ${\mathcal A}$-model ${\mathcal M}$, the set of
individuals of which is $P$.\\
We have a unary predicate $\C:P\to{\mathcal P}(\LLbd)$, a binary function $\et:P^2\to P$
and a distinguished individual \ $\1\in P$. We suppose that the data \  $\{\C,\et,\1\}$ constitute what we call
a \emph{forcing structure in ${\mathcal M}$}, which means that we have the following property:

\smallskip\noindent
There exist six proof-like terms $\alpha_0,\alpha_1,\alpha_2,\beta_0,\beta_1,\beta_2$ such that:

\smallskip\noindent
$\tau\in|\C[(p\et q)\et r]|$ $\Fl$ $\alpha_0\tau\in|\C[p\et(q\et r)]|$~;\\
$\tau\in|\C[p]|$ $\Fl$ $\alpha_1\tau\in|\C[p\et\1]|$~;\\
$\tau\in|\C[p\et q]|$ $\Fl$ $\alpha_2\tau\in|\C[q]|$~;\\
$\tau\in|\C[p]|$ $\Fl$ $\beta_0\tau\in|\C[p\et p]|$~;\\
$\tau\in|\C[p\et q]|$ $\Fl$ $\beta_1\tau\in|\C[q\et p]|$~;\\
$\tau\in|\C[((p\et q)\et r)\et s]|$ $\Fl$ $\beta_2\tau\in|\C[(p\et(q\et r))\et s]|$.

\smallskip\noindent
We shall call \emph{$\C$-expression} any finite sequence of symbols of the form $\gamma=(\delta_0)(\delta_1)\ldots(\delta_k)$ where each $\delta_i$ is one of the proof-like terms
$\alpha_0,\alpha_1,\alpha_2,\beta_0,\beta_1,\beta_2$.\\
Such an expression is not a $\Cc$-term, but $\gamma\tau$ is, for every $\Cc$-term $\tau$~;\\
the term \ $\gamma\tau=(\delta_0)(\delta_1)\ldots(\delta_k)\tau$ will also be written $(\gamma)\tau$.

\smallskip\noindent
{\bfseries Notation.}
A \ \emph{$\et$-term} is, by definition, a term which is written with the variables $p_1,\ldots,p_k$,
the constant~$\1$ and the binary function symbol $\et$. Let $t(p_1,\ldots,p_k),u(p_1,\ldots,p_k)$ be two
$\et$-terms. The notation:\\
\centerline{$\gamma::t(p_1,\ldots,p_k)\Fl u(p_1,\ldots,p_k)$}
means that $\gamma$ is a $\C$-expression such that \
$\tau\in|\C[t(p_1,\ldots,p_k)]|$ $\Fl$ $(\gamma)\tau\in|\C[u(p_1,\ldots,p_k)]|$.

\smallskip\noindent
Thus, with this notation, the above hypothesis can be written as follows:

\smallskip\noindent
$\alpha_0::(p\et q)\et r\Fl p\et(q\et r)$~; \ $\alpha_1::p\Fl p\et\1$~; \
$\alpha_2::p\et q\Fl q$~;\\
$\beta_0::p\Fl p\et p$~; $\beta_1::p\et q\Fl q\et p$~; 
$\beta_2::((p\et q)\et r)\et s\Fl(p\et(q\et r))\et s$.

\begin{lem}\label{et_calc}
There exist \ $\C$-expressions $\beta'_0,\beta'_1,\beta'_2,\beta_3,\beta'_3$ such that:\\
$\beta'_0::p\et q\Fl (p\et q)\et q$~; \ $\beta'_1::(p\et q)\et r\Fl (q\et p)\et r$~; \
$\beta'_2::p\et(q\et r)\Fl (p\et q)\et r$~;\\
$\beta_3::p\et(q\et r)\Fl p\et(r\et q)$~; $\beta'_3::(p\et(q\et r))\et s\Fl(p\et(r\et q))\et s$.
\end{lem}

\proof
We write the sequence of transformations, with  the $\C$-expressions which perform them:

\smallskip\noindent
$\bullet$~~$\beta'_0=(\beta_1)(\alpha_2)(\alpha_0)(\beta_0)$.\\
$p\et q;\;\beta_0\;;\;(p\et q)\et(p\et q)\;;\;\alpha_0\;;p\et(q\et(p\et q))\;;\;\alpha_2\;;
q\et(p\et q)\;;\;\beta_1\;; \; (p\et q)\et q$.

\smallskip\noindent
$\bullet$~~$\beta'_2=(\beta_1)(\alpha_0)(\beta_1)(\alpha_0)(\beta_1)$.\\
$p\et(q\et r)\;;\beta_1\;;\;(q\et r)\et p\;;\;\alpha_0\;;\;q\et(r\et p)\;;\;\beta_1\;;\;
(r\et p)\et q\;;\;\alpha_0\;;\;r\et(p\et q)\;;\;\beta_1\;;\;(p\et q)\et r$.

\smallskip\noindent
$\bullet$~~$\beta'_1=(\alpha_2)(\alpha_0)(\beta_2)(\beta_1)(\alpha_0)(\alpha_2)(\beta_1)(\beta'_2)
(\beta'_0)(\beta_1)$.\\
$(p\et q)\et r\;;\;\beta_1\;;\;r\et(p\et q)\;;\;\beta'_0\;\;(r\et(p\et q))\et(p\et q)\;;\beta'_2\;;
((r\et(p\et q))\et p)\et q\;;\;\beta_1\;;\;q\et((r\et(p\et q))\et p)\;;\\
\alpha_2\;;\;(r\et(p\et q))\et p\;;\;\alpha_0\;;\;r\et((p\et q)\et p)\;;\beta_1\;;\;
((p\et q)\et p)\et r\;;\;\beta_2\;;\;(p\et(q\et p))\et r\;;\;\alpha_0\;;\\
p\et((q\et p)\et r)\;;\;\alpha_2\;;\;(q\et p)\et r$.

\smallskip\noindent
$\bullet$~~$\beta_3=(\beta_1)(\beta'_1)(\beta_1)$.\\
$p\et(q\et r)\;;\;\beta_1\;;\;(q\et r)\et p\;;\;\beta'_1\;;\;(r\et q)\et p\;;\;\beta_1\;;\;
p\et(r\et q)$.

\smallskip\noindent
$\bullet$~~$\beta'_3=(\beta'_1)(\beta'_2)(\beta'_1)(\alpha_0)(\beta'_1)$.\\
$(p\et(q\et r))\et s\;;\;\beta'_1\;;\;((q\et r)\et p)\et s\;;\;\alpha_0\;;\;(q\et r)\et(p\et s)\;;\;
\beta'_1\;;\;(r\et q)\et(p\et s)\;;\;\beta'_2\;;\;((r\et q)\et p)\et s\;;\\
\beta'_1\;;\;(p\et(r\et q))\et s$.
\qed

\begin{lem}\label{gamma_t_tp}
Let $t$ be a $\et$-term and $p$ a variable of \ $t$. Then, there exists a \ $\C$-expression \ $\gamma$ \ such that \
$\gamma::t\Fl t\et p$.
\end{lem}

\proof By induction on the number of symbols of $t$ which stand after the last occurrence of $p$.
If this number is $0$, then $t=p$ or $t=u\et p$. Then, we have $\gamma=\beta_0$ or $\beta'_0$
(lemma~\ref{et_calc}).\\
Otherwise, we have $t=u\et v$~; if the last occurrence of $p$ is in $u$, the recurrence hypothesis gives \
$\gamma'::v\et u\Fl(v\et u)\et p$. Then, we have \ $\gamma=(\beta'_1)(\gamma')(\beta_1)$.\\
If the last occurrence of $p$ is in $v$, we have \ $v=v_0\et v_1$. If this occurrence is in $v_0$,  
the recurrence hypothesis gives \ $\gamma'::u\et(v_1\et v_0)\Fl(u\et(v_1\et v_0))\et p$. We put \ $\gamma=(\beta'_3)(\gamma')(\beta_3)$ (lemma~\ref{et_calc}).\\
If this occurrence is in $v_1$, the recurrence hypothesis gives\\
$\gamma'::(u\et v_0)\et v_1\Fl((u\et v_0)\et v_1)\et p$. Then, we put $\gamma=(\beta_2)(\gamma')(\beta'_2)$.
\qed

\begin{lem}\label{gamma_t_tu}
Let $t,u$ be two $\et$-terms such that each variable of $u$ appears in $t$. Then, there exists a $\C$-expression $\gamma$ such that $\gamma::t\Fl t\et u$.
\end{lem}\noindent
Proof by recurrence on the length of $u$.\\
If $u=\1$, then $\gamma=\alpha_1$~; if $u$ is a variable, we apply lemma~\ref{gamma_t_tp}.\\
If $u=v\et w$, the recurrence hypothesis gives \ $\gamma'::t\Fl t\et v$ \ and also \
$\gamma''::t\et v\Fl (t\et v)\et w$. Then, we put \ $\gamma=(\alpha_0)(\gamma'')(\gamma')$.
\qed

\begin{thm}\label{gamma_t_u}
Let $t,u$ be two $\et$-terms such that each variable of $u$ appears in $t$. Then, there exists a \
$\C$-expression $\gamma$ such that $\gamma::t\Fl u$.
\end{thm}

\proof
By lemma~\ref{gamma_t_tu}, we have \ $\gamma'::t\Fl t\et u$. Thus, we can put $\gamma=(\alpha_2)(\gamma')$.
\qed

\begin{cor}\label{CpqCqp}
There exist \ $\C$-expressions $\gamma_I,\gamma_K,\gamma_E,\gamma_W,\gamma_C,\gamma_B,
\gamma_{\sf cc},\gamma_{\sf k}$ such that:\\
$\gamma_I::p\et q\Fl q$~; \ $\gamma_K::\1\et(p\et(q\et r))\Fl p\et r$~; \
$\gamma_E::\1\et(p\et(q\et r))\Fl (p\et q)\et r$~;\\
$\gamma_W::\1\et(p\et(q\et r))\Fl p\et(q\et(q\et r))$~; \
$\gamma_C::\1\et(p\et(q\et(r\et s)))\Fl p\et(r\et(q\et s))$~;\\
$\gamma_B::\1\et(p\et(q\et(r\et s)))\Fl (p\et(q\et r))\et s$~; \
$\gamma_{\sf cc}::\1\et(p\et q)\Fl p\et(q\et q)$~;\\
$\gamma_{\sf k}::p\et(q\et r)\Fl q\et p$.\qed
\end{cor}\noindent

\subsection*{The algebra \texorpdfstring{${\mathcal B}$}{B}}\noindent
We define now a new realizability algebra ${\mathcal B}=(\LLbd,\PPi,\LLbd\star\PPi,\bbbot)$: its set of terms is $\LLbd=\Lbd\fois P$, its set of stacks is \ $\PPi=\Pi\fois P$ \ and its set of processes is \
$\LLbd\star\PPi=(\Lbd\star\Pi)\fois P$.\\
The distinguished subset $\bbot_{\mathcal B}$ of $\LLbd\star\PPi$ is denoted by $\bbbot$. It is defined as follows:\\
$(\xi\star\pi,p)\in\bbbot$ \ $\Dbfl$ \ $(\pt\tau\in\C[p])\,\xi\star\pi^\tau\in\bbot$.

\smallskip\noindent
For $(\xi,p)\in\LLbd$ and $(\pi,q)\in\PPi$, we put:

\smallskip\noindent
$(\xi,p)\star(\pi,q)=(\xi\star\pi,p\et q)$~;\\
$(\xi,p)\ps(\pi,q)=(\xi\ps\pi,p\et q)$.

\smallskip\noindent
For $(\xi,p),(\eta,q)\in\LLbd$, we put:

\smallskip\noindent
$(\xi,p)(\eta,q)=(\ov{\alpha}_0\xi\eta,p\et q)$ \ with \
$\ov{\alpha}_0=\lbd x(\chi)\lbd y(\chi' x)(\alpha_0)y$.

\begin{lem}\label{A_gamma}
For each \ $\C$-expression $\gamma$, we put \ $\ov{\gamma}=\lbd x(\chi)\lbd y(\chi' x)(\gamma)y$.\\
Then, we have \ $\ov{\gamma}\star\xi\ps\pi^\tau\succ\xi\star\pi^{\gamma\tau}$.
\end{lem}

\proof
This is immediate, by means of theorem~\ref{beta_red_gauche}.\\ We could take also \
$\ov{\gamma}=(\chi)\lbd x\lbd y(\chi'y)(\gamma)x$.
\qed

\begin{prop}\label{gamma::tu}
If we have \ $\gamma::t(p_1,\ldots,p_k)\Fl u(p_1,\ldots,p_k)$, then:\\
$(\ov{\gamma}\star\xi\ps\pi,t(p_1,\ldots,p_k))\succ(\xi\star\pi,u(p_1,\ldots,p_k))$.
\end{prop}

\proof
Suppose that $(\ov{\gamma}\star\xi\ps\pi,t(p_1,\ldots,p_k))\notin\bbbot$. Thus, there exists
$\tau\in\C[t(p_1,\ldots,p_k)]$ such that:\\
$\ov{\gamma}\star\xi\ps\pi^\tau\notin\bbot$. Therefore, we have $\xi\star\pi^{\gamma\tau}\notin\bbot$
et $\gamma\tau\in\C[u(p_1,\ldots,p_k)]$. It follows that:\\
$(\xi\star\pi,u(p_1,\ldots,p_k))\notin\bbbot$.
\qed

\begin{lem}
We have \ $(\xi,p)(\eta,q)\star(\pi,r)\notin\bbbot$ \ $\Fl$ \ $(\xi,p)\star(\eta,q)\ps(\pi,r)\notin\bbbot$.
\end{lem}

\proof
By hypothesis, we have \ $(\ov{\alpha}_0\xi\eta\star\pi,(p\et q)\et r)\notin\bbbot$~; thus, there exists
$\tau\in\C[(p\et q)\et r]$ such that \
$\ov{\alpha}_0\xi\eta\star\pi^\tau\notin\bbot$. By lemma~\ref{A_gamma}, we have
$\xi\star\eta\ps\pi^{\alpha_0\tau}\notin\bbot$~; since $\alpha_0\tau\in\C[p\et(q\et r)]$, we have \ $(\xi\star\eta\ps\pi,p\et(q\et r))\notin\bbot$ and thus \ $(\xi,p)\star(\eta,q)\ps(\pi,r)\notin\bbbot$.
\qed

\smallskip\noindent
We define the elementary combinators {\bfseries B, C, E, I, K, W, {\sffamily cc}} of the algebra ${\mathcal B}$
by putting:

\smallskip\noindent
{\bf B} $=(B^*,\1)$~; {\bf C} $=(C^*,\1)$~; {\bf E} $=(E^*,\1)$~;
{\bf I} $=(I^*,\1)$~; {\bf K} $=(K^*,\1)$~; {\bf W} $=(W^*,\1)$~;\\
{\bf\sffamily cc} $=(\Ccc^*,\1)$\\
with $B^*=\lbd x\lbd y\lbd z(\ov{\gamma}_B)(\ov{\alpha}_0x)(\ov{\alpha}_0)yz$~; $C^*=\ov{\gamma}_CC$~;
$E^*=\lbd x\lbd y(\ov{\gamma}_E)(\ov{\alpha}_0)xy$~; $I^*=\ov{\gamma}_II$~;\\
$K^*=\ov{\gamma}_KK$~; $W^*=\ov{\gamma}_WW$~; \
$\Ccc^*=(\chi)\lbd x\lbd y(\Ccc)\lbd k((\chi'y)(\gamma_{\sf cc})x)(\chi)\lbd x\lbd y(k)(\chi'y)(\gamma_{\sf k})x$.

\smallskip\noindent
We put \ {\bf\sffamily k}$_{(\pi,p)}=(\kk^*_\pi,p)$ \ with  \
$\kk^*_\pi=(\chi)\lbd x\lbd y(\kk_\pi)(\chi'y)(\gamma_{\sf k})x$.

\begin{thm}\label{combi_compo}
For every \ $\tilde{\xi},\tilde{\eta},\tilde{\zeta}\in\LLbd$ and $\tilde{\pi},\tilde{\varpi}\in\PPi$,
we have:\\
$\mathbf{I}\star\tilde{\xi}\ps\tilde{\pi}\notin\bbbot$ \ $\Fl$ \
$\tilde{\xi}\star\tilde{\pi}\notin\bbbot$~;\\
$\mathbf{K}\star\tilde{\xi}\ps\tilde{\eta}\ps\tilde{\pi}\notin\bbbot$ \ $\Fl$ \ $\tilde{\xi}\star\tilde{\pi}\notin\bbbot$~;\\
$\mathbf{E}\star\tilde{\xi}\ps\tilde{\eta}\ps\tilde{\pi}\notin\bbbot$ \ $\Fl$ \ $(\tilde{\xi})\tilde{\eta}\star\tilde{\pi}\notin\bbbot$~;\\
$\mathbf{W}\star\tilde{\xi}\ps\tilde{\eta}\ps\tilde{\pi}\notin\bbbot$ \ $\Fl$ \ $\tilde{\xi}\star\tilde{\eta}\ps\tilde{\eta}\ps\tilde{\pi}\notin\bbbot$.\\
$\mathbf{B}\star\tilde{\xi}\ps\tilde{\eta}\ps\tilde{\zeta}\ps\tilde{\pi}\notin\bbbot$ \ $\Fl$ \ $(\tilde{\xi})(\tilde{\eta})\tilde{\zeta}\star\tilde{\pi}\notin\bbbot$~;\\
$\mathbf{C}\star\tilde{\xi}\ps\tilde{\eta}\ps\tilde{\zeta}\ps\tilde{\pi}\notin\bbbot$ \ $\Fl$ \ $\tilde{\xi}\star\tilde{\zeta}\ps\tilde{\eta}\ps\tilde{\pi}\notin\bbbot$.\\
{\bf\sffamily cc} $\star\,\tilde{\xi}\ps\tilde{\pi}\notin\bbbot$ \ $\Fl$ \
$\tilde{\xi}\;\star$ {\bf\sffamily k}$_{\tilde{\pi}}\ps\tilde{\pi}\notin\bbbot$.\\
{\bf\sffamily k}$_{\tilde{\pi}}\star\tilde{\xi}\ps\tilde{\varpi}\notin\bbbot$ \ $\Fl$ \
$\tilde{\xi}\star\tilde{\pi}\notin\bbbot$.
\end{thm}

\proof
We shall prove only the cases \ {\bfseries W, B, {\sffamily k}}$_{\tilde{\pi}}$,
{\bf\sffamily cc}.\\
We put \ $\tilde{\xi}=(\xi,p),\tilde{\eta}=(\eta,q),\tilde{\zeta}=(\zeta,r),
\tilde{\pi}=(\pi,s),\tilde{\varpi}=(\varpi,q)$.

\smallskip\noindent
Suppose \ $\mathbf{W}\star\tilde{\xi}\ps\tilde{\eta}\ps\tilde{\pi}\notin\bbbot$, and therefore \
$(\ov{\gamma}_WW\star\xi\ps\eta\ps\pi,\1\et(p\et(q\et s)))\notin\bbbot$.\\
Thus, there exists $\tau\in\C[\1\et(p\et(q\et s))]$ \ such that \
$\ov{\gamma}_WW\star\xi\ps\eta\ps\pi^\tau\notin\bbot$.\\
Since $\ov{\gamma}_WW\star\xi\ps\eta\ps\pi^\tau\succ\xi\star\eta\ps\eta\ps\pi^{\gamma_W\tau}$,
we have \ $\xi\star\eta\ps\eta\ps\pi^{\gamma_W\tau}\notin\bbot$.\\
But\ $\gamma_W\tau\in\C[p\et(q\et(q\et s))]$ and it follows that \
$\tilde{\xi}\star\tilde{\eta}\ps\tilde{\eta}\ps\tilde{\pi}\notin\bbbot$.

\smallskip\noindent
Suppose \ $\mathbf{B}\star\tilde{\xi}\ps\tilde{\eta}\ps\tilde{\zeta}\ps\tilde{\pi}\notin\bbbot$, that is \
$(B^*\star\xi\ps\eta\ps\zeta\ps\pi,\1\et(p\et(q\et(r\et s))))\notin\bbbot$.\\
Thus, there exists $\tau\in\C[\1\et(p\et(q\et(r\et s)))]$ \ such that \
$B^*\star\xi\ps\eta\ps\zeta\ps\pi^\tau\notin\bbot$.\\
But, we have \
$B^*\star\xi\ps\eta\ps\zeta\ps\pi^\tau\succ(\ov{\gamma}_B)(\ov{\alpha}_0\xi)(\ov{\alpha}_0)\eta\zeta\star\pi^\tau$
(by theorem~\ref{beta_red_gauche})\\
$\succ(\ov{\alpha}_0\xi)(\ov{\alpha}_0)\eta\zeta\star\pi^{\gamma_B\tau}$ (by lemma~\ref{A_gamma}).
Therefore, we have \ $(\ov{\alpha}_0\xi)(\ov{\alpha}_0)\eta\zeta\star\pi^{\gamma_B\tau}\notin\bbot$.\\
But \ $\gamma_B\tau\in\C[(p\et(q\et r))\et s]$ and thus, we have:\\
$((\ov{\alpha}_0\xi)(\ov{\alpha}_0)\eta\zeta\star\pi,(p\et(q\et r))\et s)\notin\bbbot$,
in other words $(\tilde{\xi})(\tilde{\eta})\tilde{\zeta}\star\tilde{\pi}\notin\bbbot$.

\smallskip\noindent
Suppose \ {\bf\sffamily k}$_{\tilde{\pi}}\star\tilde{\xi}\ps\tilde{\varpi}\notin\bbbot$, \ that is \
$(\kk^*_\pi\star\xi\ps\varpi,s\et(p\et q))\notin\bbbot$.\\
Thus, there exists $\tau\in\C[s\et(p\et q)]$
such that \ $\kk^*_\pi\star\xi\ps\varpi^\tau\notin\bbot$. But we have:\\
$\kk^*_\pi\star\xi\ps\varpi^\tau\succ
\lbd x\lbd y(\kk_\pi)(\chi'y)(\gamma_{\sf k})x\star\tau\ps\xi\ps\varpi\succ
(\kk_\pi)(\chi'\xi)(\gamma_{\sf k})\tau\star\varpi$ (by theorem~\ref{beta_red_gauche})\\
$\succ(\chi'\xi)(\gamma_{\sf k})\tau\star\pi\succ\chi'\star\xi\ps\gamma_{\sf k}\tau\ps\pi\succ
\xi\star\pi^{\gamma_{\sf k}\tau}$.\\
Thus, we have \ $\xi\star\pi^{\gamma_{\sf k}\tau}\notin\bbot$~; but, since \ $\gamma_{\sf k}\tau\in\C[p\et s]$,
we get \ $\tilde{\xi}\star\tilde{\pi}\notin\bbbot$.

\smallskip\noindent
Suppose \ {\bf\sffamily cc} $\star\,\tilde{\xi}\ps\tilde{\pi}\notin\bbbot$, \ that is \
$(\Ccc^*\star\xi\ps\pi,\1\et(p\et s))\notin\bbbot$.\\
Thus, there exists \ $\tau\in\C[\1\et(p\et s)]$ such that \ $\Ccc^*\star\xi\ps\pi^\tau\notin\bbot$.
But we have:\\
$\Ccc^*\star\xi\ps\pi^\tau\succ
\lbd x\lbd y(\Ccc)\lbd k((\chi'y)(\gamma_{\sf cc})x)(\chi)\lbd x\lbd y(k)(\chi'y)(\gamma_{\sf k})x\star
\tau\ps\xi\ps\pi\\
\succ(\Ccc)\lbd k((\chi'\xi)(\gamma_{\sf cc})\tau)(\chi)\lbd x\lbd y(k)(\chi'y)(\gamma_{\sf k})x\star\pi\\
\succ((\chi'\xi)(\gamma_{\sf cc})\tau)(\chi)\lbd x\lbd y(\kk_\pi)(\chi'y)(\gamma_{\sf k})x\star\pi
\succ\chi'\star\xi\ps\gamma_{\sf cc}\tau\ps(\chi)\lbd x\lbd y(\kk_\pi)(\chi'y)(\gamma_{\sf k})x\ps\pi\\
\succ\xi\star(\chi)\lbd x\lbd y(\kk_\pi)(\chi'y)(\gamma_{\sf k})x\ps\pi^{\gamma_{\sf cc}\tau}
\equiv\xi\star\kk^*_\pi\ps\pi^{\gamma_{\sf cc}\tau}$.\\
It follows that \ $\xi\star\kk^*_\pi\ps\pi^{\gamma_{\sf cc}\tau}\notin\bbot$. But we have \
$\gamma_{\sf cc}\tau\in\C[p\et(s\et s)]$ and it follows that we have \
$(\xi,p)\star(\kk^*_\pi,s)\ps(\pi,s)\notin\bbbot$, \ that is \
$\tilde{\xi}\star$ {\bf\sffamily k}$_{\tilde{\pi}}\ps\tilde{\pi}\notin\bbbot$.
\qed

\smallskip\noindent
We have now completely defined the realizability algebra ${\mathcal B}$.

\smallskip\noindent
For each closed $\Cc$-term $t$ (proof-like term), let us denote by $t_{\mathcal B}$ its value in the algebra
${\mathcal B}$ (its value in the standard algebra ${\mathcal A}$ is $t$ itself). Thus, we have
$t_{\mathcal B}=(t^*,\1_t)$, where $t^*$ is a proof-like term and $\1_t$ a condition written with  $\1$,
$\et$ and parentheses, which are obtained as follows, by recurrence on $t$:

\smallskip\noindent
$\bullet$~~If $t$ is an elementary combinator $B,C,E,I,K,W,{\sf cc}$, then $t^*$ is already defined~;
$\1_t=\1$.\\
$\bullet$~~$(tu)^*=\ov{\alpha}_0t^*u^*$~; $\1_{tu}=\1_t\et\1_u$.

\subsection*{The model \texorpdfstring{${\mathcal N}$}{}}\noindent
The ${\mathcal B}$-model ${\mathcal N}$ has the same set $P$ of individuals and the same functions as ${\mathcal M}$.\\
By definition, the $k$-ary predicates of ${\mathcal N}$ are the applications from $P^k$ into ${\mathcal P}(\PPi)$.
But, since $\PPi=\Pi\fois P$, they are the same as the applications from $P^{k+1}$ into ${\mathcal P}(\Pi)$,
i.e. the $k+1$-ary predicates of the model ${\mathcal M}$.\\
Each predicate constant \ {\sf R}, of arity $k$, is interpreted, in the model ${\mathcal M}$,
by an application ${\sf R}_{\mathcal M}$ from $P^k$ into ${\mathcal P}(\Lbd)$.
In the model ${\mathcal N}$, this predicate constant is interpreted by the application \
${\sf R}_{\mathcal N}:P^k\to{\mathcal P}(\LLbd)$, where \
${\sf R}_{\mathcal N}(p_1,\ldots,p_k)={\sf R}_{\mathcal M}(p_1,\ldots,p_k)\fois\{\1\}$.

\smallskip\noindent
For each closed formula $F$, with parameters in ${\mathcal N}$, its truth value, which is a subset of~$\PPi$,
will be denoted by $\vv F\vv$. We shall write \ $(\xi,p)\fforce F$ to mean that $(\xi,p)\in\LLbd$
realizes $F$, in other words \
$(\pt\pi\in\Pi)(\pt q\in P)(((\pi,q)\in\vv F\vv)\Fl(\xi,p)\star(\pi,q)\in\bbbot)$.

\begin{thm}\label{adequat_B}\ \\
If we have \ $\vdash t:A$ in classical second order logic, where $A$ is a closed formula, then\\
$t_{\mathcal B}=(t^*,\1_t)\fforce A$.
\end{thm}

\proof
Immediate application of theorem~\ref{adequat} (adequacy lemma) in the ${\mathcal B}$-model ${\mathcal N}$.
\qed

\begin{prop}\label{1_to_p}\ \\
\emph{\phantom ii)} If \ $(\xi,\1)\fforce F$, then \ $(\ov{\gamma}\xi,p)\fforce F$ for each $p\in P$, with  \
$\gamma::p\et q\Fl\1\et q$.\\
\emph{ii)} Let $\xi,\eta\in\Lbd$ be such that $\xi\star\pi\succ\eta\star\pi$ for each $\pi\in\Pi$. Then, we have:\\
$(\xi\star\pi,p)\notin\bbbot\Fl(\eta\star\pi,p)\notin\bbbot$ for every $\pi\in\Pi$ and $p\in P$~;\\
$(\eta,p)\fforce F$ \ $\Fl$ \ $(\xi,p)\fforce F$ for every closed formula $F$.
\end{prop}
\proof{\ }\ \\
\phantom ii)~~We must show that, for each $(\pi,q)\in\vv F\vv$, we have \
$(\ov{\gamma}\xi,p)\star(\pi,q)\in\bbbot$, that is:\\
$(\ov{\gamma}\xi\star\pi,p\et q)\in\bbbot$. Thus, let \ $\tau\in\C[p\et q]$, so that $\gamma\tau\in\C[\1\et q]$.\\
Since we have, by hypothesis, $(\xi\star\pi,\1\et q)\in\bbbot$, it follows that \
$\xi\star\pi^{\gamma\tau}\in\bbot$ and therefore $\ov{\gamma}\xi\star\pi^\tau\in\bbot$.\\
ii)~~By hypothesis, there exists \ $\tau\in\C[p]$ such that $\xi\star\pi^\tau\notin\bbot$. Thus, we have \ $\eta\star\pi^\tau\notin\bbot$, so that \ $(\eta\star\pi,p)\notin\bbbot$.\\
Let $(\pi,q)\in\vv F\vv$~; we have $(\eta,p)\star(\pi,q)\in\bbbot$, that is
$(\eta\star\pi,p\et q)\in\bbot$. From what we have just shown, it follows that
$(\xi\star\pi,p\et q)\in\bbot$, and therefore \ $(\xi,p)\star(\pi,q)\in\bbbot$.
\qed

\subsubsection*{The integers of the model ${\mathcal N}$}\noindent
Recall that we have put:\\
$\sig=\lbd n\lbd f\lbd x(f)(n)fx$, \ $\ul{0}=\lbd x\lbd y\,y$ \ and \
$\ul{n}=(\sig)^n\ul{0}$ \ for every integer $n$.\\  
Thus, we have \ $\sig_{\mathcal B}=(\sig^*,\1_\sig)$ \ and \
$\ul{n}_{\mathcal B}=((\sig)^n\ul{0})_{\mathcal B}=(\ul{n}^*,\1_{\ul{n}})$.\\
Therefore \ $\ul{0}_{\mathcal B}=(KI)_{\mathcal B}=(K^*,\1)(I^*,\1)$ \ and \
$\ul{n+1}_{\mathcal B}=\sig_{\mathcal B}\ul{n}_{\mathcal B}=(\sig^*,\1_\sig)(\ul{n}^*,\1_{\ul{n}})$.\\
Thus, the recursive definitions of $\ul{n}^*,\1_{\ul{n}}$ are the following:\\
$\ul{0}^*=\ov{\alpha}_0K^*I^*$~; $(\ul{n+1})^*=\ov{\alpha}_0\sig^*\ul{n}^*$~;\\
$\1_{\ul{0}}=\1\et\1$~; $\1_{\ul{n+1}}=\1_\sig\et\1_{\ul{n}}$.

\smallskip\noindent
We can define the unary predicate \ ent$(x)$ \ in the model ${\mathcal N}$ in two distinct ways:

\smallskip\noindent
\phantom ii)~~From the predicate \ ent$(x)$ \ of the model ${\mathcal M}$, by putting:\\
$|$ent$(s^n0)|=\{(\ul{n},\1)\}$~; $|$ent$(p)|=\vide$ if $p\notin\NN$.\\
ii)~~By using directly the definition of ent(x) in the model ${\mathcal N}$~; we denote this predicate by
ent$_{\mathcal N}(x)$. Therefore, we have:\\
$|$ent$_{\mathcal N}(s^n0)|=\{\ul{n}_{\mathcal B}\}$~; $|$ent$_{\mathcal N}(p)|=\vide$ if $p\notin\NN$.\\
From theorem~\ref{mm1}, applied in the model ${\mathcal N}$, we know that the predicates \ int$(x)$ \
and \ ent$_{\mathcal N}(x)$ are interchangeable. Theorem~\ref{entMN} shows that the
predicates int$(x)$ and ent$(x)$ are also interchangeable. Thus, we have three predicates which
define the integers in the model ${\mathcal N}$~; it is the predicate ent$(x)$ that we shall mostly use
in the sequel. In particular, we shall often replace the quantifier $\pt x\indi$ with $\pt x\inde$.

\begin{thm}\label{entMN}\ \\
There exist two proof-like terms $T,J$ such that:\\
\emph{\phantom ii)}~~$(T,\1)\fforce\pt X\pt x(($ent$(x)\to X),$ int$(x)\to X)$.\\
\emph{ii)}~~$(J,\1)\fforce\pt x($ent$(x)\to$int$(x))$.
\end{thm}
\proof{\ }\ \\
\phantom ii)~~We apply theorem~\ref{memoire_gen} to the sequence $u:\NN\to\LLbd$ defined by $u_n=(\ul{n},\1)$.\\
We are looking for two proof-like terms \ $T,S$ \ such that:\\
$(S,\1)\star(\psi,p)\ps(\ul{n},\1)\ps(\pi,r)\succ(\psi,p)\star(\ul{n+1},\1)\ps(\pi,r)$~; \
$(S,\1)\fforce\top\to\bot,\top\to\bot$.\\
$(T,\1)\star(\phi,p)\ps(\nu,q)\ps(\pi,r)\succ(\nu,q)\star(S,\1)\ps(\phi,p)\ps(\ul{0},\1)\ps(\pi,r)$.

\smallskip\noindent
Then theorem~\ref{memoire_gen} will give the desired result:\\
$(T,\1)\fforce\pt X\pt x(($ent$(x)\to X),$ int$(x)\to X)$.

\smallskip\noindent
We put \ $S=\lbd f\lbd x(\ov{\gamma}f)(\sig)x$, with  $\gamma::\1\et(p\et(q\et r))\Fl p\et(q\et r)$.

\smallskip\noindent
Then, we have \ $(S,\1)\star(\psi,p)\ps(\nu,q)\ps(\pi,r)\equiv
(S\star\psi\ps\nu\ps\pi,\1\et(p\et(q\et r)))\succ\\
(\ov{\gamma}\psi\star\sig\nu\ps\pi,\1\et(p\et(q\et r)))$
(theorem~\ref{beta_red_gauche} and proposition~\ref{1_to_p}(ii))\\
$\succ(\psi\star\sig\nu\ps\pi,p\et(q\et r))$ (proposition~\ref{gamma::tu})
$\equiv(\psi,p)\star(\sig\nu,q)\ps(\pi,r)$.\\
Suppose first that $(\psi,p)\fforce\top\to\bot$~; then, we have 
$(\psi,p)\star(\sig\nu,q)\ps(\pi,r)\in\bbbot$ and thus:\\
$(S,\1)\star(\psi,p)\ps(\nu,q)\ps(\pi,r)\in\bbbot$. This shows that
$(S,\1)\fforce\top\to\bot,\top\to\bot$.\\
Moreover, if we put \ $\nu=\ul{n}$, so that \ $\sig\nu=\ul{n+1}$, and $q=\1$, we have shown that:\\
$(S,\1)\star(\psi,p)\ps(\ul{n},\1)\ps(\pi,r)\succ(\psi,p)\star(\ul{n+1},\1)\ps(\pi,r)$.

\smallskip\noindent
Now, we put \ $T=\lbd f\lbd x(\ov{\gamma}'x)Sf\ul{0}$, with  \
$\gamma'::\1\et(p\et(q\et r))]\Fl q\et(\1\et(p\et(\1\et r)))$.

\smallskip\noindent
Then, we have \ $(T,\1)\star(\phi,p)\ps(\nu,q)\ps(\pi,r)\equiv
(T\star\phi\ps\nu\ps\pi,\1\et(p\et(q\et r)))\succ\\
(\ov{\gamma}'\nu\star S\ps\phi\ps\ul{0}\ps\pi,\1\et(p\et(q\et r)))$
(theorem~\ref{beta_red_gauche} and proposition~\ref{1_to_p}(ii))\\
$\succ(\nu\star S\ps\phi\ps\ul{0}\ps\pi,q\et(\1\et(p\et(\1\et r))))$
(proposition~\ref{gamma::tu})\\
$\equiv(\nu,q)\star(S,\1)\ps(\phi,p)\ps(\ul{0},\1)\ps(\pi,r)$ \
which is the desired result.

\smallskip\noindent
ii)~~We are looking for a proof-like term $J$ such that $(J,\1)\fforce\pt x($ent$(x)\to$int$(x))$. It is
sufficient to have \ $(J,\1)\fforce$ent$(s^n0)\to$int$(s^n0)$ for each $n\in\NN$,
since $|\,$ent$(p)|=\vide$ if $p\notin\NN$.\\
Let \ $(\pi,q)\in\vv$int$(n)\vv$~; we must have \ $(J,\1)\star(\ul{n},\1)\ps(\pi,q)\in\bbbot$, \
that is:\\
$(J\star\ul{n}\ps\pi,\1\et(\1\et q))\in\bbbot$.\\
But, we have \ $(\ul{n}^*,\1_{\ul{n}})=((\sig)^n\ul{0})_{\mathcal B}\fforce$int$(s^n0)$
(theorem~\ref{adequat}, applied in ${\mathcal B}$) and therefore:\\
$(\ul{n}^*,\1_{\ul{n}})\star(\pi,q)\in\bbbot$ \ or else \
$(\ul{n}^*\star\pi,\1_{\ul{n}}\et q)\in\bbbot$.

\smallskip\noindent
Thus, let \ $\tau\in\C[\1\et(\1\et q)]$~; we have then \ $(\gamma)^n(\gamma_0)\tau\in\C[\1_n\et q]$\\
where $\gamma_0$ and $\gamma$ are two $\C$-expressions such that:\\
$\gamma_0::\1\et(\1\et q)\Fl(\1\et\1)\et q$~; \ $\gamma::p\et q\Fl(\1_\sig\et p)\et q$.\\
Indeed, we have seen that $\1_{\ul{0}}=\1\et\1$ and $\1_{\ul{n+1}}=\1_\sig\et\1_{\ul{n}}$.
It follows that, if $\tau\in\C[\1\et(\1\et q)]$, then
$(\gamma_0)\tau\in\C[\1_{\ul{0}}\et q]$, \ and therefore \ $(\gamma)^n(\gamma_0)\tau\in\C[\1_n\et q]$.\\
Thus, we have \ $\ul{n}^*\star\pi^{(\gamma)^n(\gamma_0)\tau}\in\bbot$.\\
Now, we build below two proof-like terms $g,j$ such that, for each $n\in\NN$, we have:\\
a)~~$g\star\ul{n}\ps\xi\ps\pi^\tau\succ\xi\star\pi^{(\gamma)^n(\gamma_0)\tau}$~;\\
b)~~$j\star\ul{n}\ps\xi\ps\pi\succ\xi\star\ul{n}^*\ps\pi$.\\
Then, by putting \ $J=\lbd x(gx)(j)x$, we have \ $J\star\ul{n}\ps\pi^\tau\succ\ul{n}^*\star\pi^{(\gamma)^n(\gamma_0)\tau}\in\bbot$,
which is the desired result.

\smallskip\noindent
a)~~We put $g=\lbd k\lbd x(\ov{\gamma}_0)(k)\ov{\gamma}x$~; from theorem~\ref{beta_red_gauche}, we have:\\
$g\star\ul{n}\ps\xi\ps\pi^\tau\succ\ov{\gamma}_0\star(\ul{n})\ov{\gamma}\xi\ps\pi^\tau\succ
(\ul{n})\ov{\gamma}\xi\star\pi^{(\gamma_0)\tau}$.\\
Thus, it suffices to show that \ $(\ul{n})\ov{\gamma}\xi\star\pi^\tau\succ\xi\star\pi^{(\gamma)^n\tau}$ \
which we do by recurrence on~$n$.\\
If $n=0$, we have immediately \
$\ul{0}\star\ov{\gamma}\ps\xi\ps\pi^\tau\succ\xi\star\pi^\tau$ \ since $\ul{0}=\lbd x\lbd y\,y$.\\
Going from $n$ to $n+1$: we have \ $(\ul{n+1})\ov{\gamma}\xi\star\pi^\tau\equiv
(\sig\ul{n})\ov{\gamma}\xi\star\pi^\tau\succ\sig\star\ul{n}\ps\ov{\gamma}\ps\xi\ps\pi^\tau\\
\succ\ov{\gamma}\star(\ul{n})\ov{\gamma}\xi\ps\pi^\tau\succ
(\ul{n})\ov{\gamma}\xi\star\pi^{(\gamma)\tau}\succ
\xi\star\pi^{(\gamma)^{n+1}\tau}$ \ by induction hypothesis.

\smallskip\noindent
b)~~We put \ $\beta=\ov{\alpha}_0\sig^*$, $U=\lbd g\lbd y(g)(\beta)y$ \ and \ $j=\lbd k\lbd f(k)Uf\ul{0}^*$.\\
Therefore, we have \ $j\star\ul{n}\ps\xi\ps\pi\succ\ul{n}U\xi\star\ul{0}^*\ps\pi$. We show,
by recurrence on $n$, that:\\
$\ul{n}U\xi\star\ul{k}^*\ps\pi\succ\xi\star(\ul{n+k})^*\ps\pi$ for each integer $k$,
which gives the desired result with  $k=0$.\\
For $n=0$, we have \ $\ul{0}U\xi\star\ul{k}^*\ps\pi\succ\xi\star\ul{k}^*\ps\pi$
since $\ul{0}=\lbd x\lbd y\,y$.\\
Going from $n$ to $n+1$: we have \ $(\ul{n+1})\star U\ps\xi\ps\ul{k}^*\ps\pi\equiv
\sig\ul{n}\star U\ps\xi\ps\ul{k}^*\ps\pi\succ U\star\ul{n}U\xi\ps\ul{k}^*\ps\pi$\\
(since $\sig=\lbd n\lbd f\lbd x(f)(n)fx$) $\succ\ul{n}U\xi\star\beta\ul{k}^*\ps\pi
\equiv\ul{n}U\xi\star(\ul{k+1})^*\ps\pi\succ\xi\star(\ul{n+k+1})^*\ps\pi$\\
by induction hypothesis.
\qed

\section{Forcing}\noindent
{\em Forcing} is a method to compute truth values of formulas in the generic ${\mathcal B}$-model ${\mathcal N}$.\\
For each $k$-ary predicate variable $X$, we add to the language a new predicate variable, denoted by $X^+$,
which has arity $k+1$. In the ${\mathcal A}$-model ${\mathcal M}$, we use the variables $X$ and $X^+$~;
in the ${\mathcal B}$-model ${\mathcal N}$, only the variables $X$.

\smallskip\noindent
With each $k$-ary second order parameter ${\mathcal X}:P^k\to{\mathcal P}(\PPi)$ of the model ${\mathcal N}$, we associate
a $(k+1)$-ary second order parameter \ ${\mathcal X}^+:P^{k+1}\to{\mathcal P}(\Pi)$ of the model ${\mathcal M}$.
It is defined in an obvious way, since $\PPi=\Pi\fois P$~; we put:\\
${\mathcal X}^+(p,p_1,\ldots,p_k)=\{\pi\in\Pi;\;(\pi,p)\in{\mathcal X}(p_1,\ldots,p_k)\}$.

\smallskip\noindent
For each formula $F$ \emph{written without the variables $X^+$}, with  parameters in the model ${\mathcal N}$,
we define, by recurrence on $F$, a formula denoted by \ $p\forcec F$ (read ``~$p$ forces $F$~''),
with parameters in the model ${\mathcal A}$, \emph{written with the variables $X^+$} and a free condition
variable~$p$:

\smallskip\noindent
If $F$ is atomic of the form \ $X(t_1,\ldots,t_k)$, then $p\forcec F$ is \
$\pt q(\C[p\et q]\to X^+(q,t_1,\ldots,t_k))$.\\
If $F$ is atomic of the form \ ${\mathcal X}(t_1,\ldots,t_k)$, then $p\forcec F$ is \
$\pt q(\C[p\et q]\to{\mathcal X}^+(q,t_1,\ldots,t_k))$.\\
If $F\equiv(A\to B)$ where $A,B$ are formulas, then \ $p\forcec F$ is \
$\pt q(q\forcec A\to p\et q\forcec B)$.\\
If $F\equiv({\sf R}(t_1,\ldots,t_k)\to B)$, where ${\sf R}$ is a predicate constant, then:\\
$p\forcec F$ \ is \ $({\sf R}(t_1,\ldots,t_k)\to p\forcec B)$.\\
If $F\equiv(t_1=t_2\mapsto B)$, \ then \ $p\forcec F$ \ is \ $(t_1=t_2\mapsto p\forcec B)$.\\
If $F\equiv\pt x\,A$, \ then \ $p\forcec F$ \ is \ $\pt x(p\forcec A)$.\\
If $F\equiv\pt X\,A$, \ then \ $p\forcec F$ \ is \ $\pt X^+(p\forcec A)$.

\smallskip\noindent
Thus we have, in particular:\\
If $F\equiv\pt x\inde\,A\,$, \ then \ $p\forcec F$ \ is \ $\pt x\inde(p\forcec A)$.

\begin{lem}\label{forcecXX+}
Let $F$ be a formula the free variables of which are amongst $X_1,\ldots,X_k$ and let \
${\mathcal X}_1,\ldots,{\mathcal X}_k$ be second order parameters in the model ${\mathcal N}$, with corresponding arities.
Then, we have: \ $(p\forcec F)[{\mathcal X}^+_1/X^+_1,\ldots,{\mathcal X}^+_k/X^+_k]\equiv
(p\forcec F[{\mathcal X}_1/X_1,\ldots,{\mathcal X}_k/X_k])$.
\end{lem}

\proof
Immediate, by recurrence on $F$.
\qed

\begin{thm}\label{forcec-fforce}\ \\
For each closed formula $F$ with parameters in the model ${\mathcal N}$, there exist two proof-like terms $\chi_F,\chi'_F$, which only depend on the propositional structure of $F$, such that we have:\\
$\xi\force(p\forcec F)$ \ $\Fl$ \ $(\chi_F\xi,p)\fforce F$~;\\
$(\xi,p)\fforce F$ \ $\Fl$ \ $\chi'_F\xi\force(p\forcec F)$\\
for every \ $\xi\in\Lbd$ and \ $p\in P$.
\end{thm}\noindent
The \emph{propositional structure} of $F$ is the simple type built with only one atom $O$ and the connective~~$\to$, which is obtained from $F$ by deleting all quantifiers, all symbols $\mapsto$ with their hypothesis, and by identifying all atomic formulas with  $O$.\\
For instance, the propositional structure of the formula:\\
$\pt X(\pt x(\pt y(f(x,y)=0\mapsto Xy)\to Xx)\to\pt x\,Xx)$ \ is \ $(O\to O)\to O$.

\smallskip\noindent
\proof By recurrence on the length of $F$.\\
$\bullet$~~If $F$ is atomic, we have $F\equiv{\mathcal X}(t_1,\ldots,t_k)$~; we show that \ $\chi_F=\chi$
and $\chi'_F=\chi'$.\\
Indeed, we have: \
$\|p\forcec F\|=\|\pt q(\C[p\et q]\to{\mathcal X}^+(q,t_1,\ldots,t_k)\|\\
\hspace*{12em}=\bigcup_q\{\tau\ps\pi;\;\tau\in\C[p\et q],(\pi,q)\in\vv{\mathcal X}(t_1,\ldots,t_k)\vv\}$,\\
because, by definition of ${\mathcal X}^+$, we have \
$\pi\in\|{\mathcal X}^+(q,t_1,\ldots,t_k)\|$ $\Dbfl$ $(\pi,q)\in\vv{\mathcal X}(t_1,\ldots,t_k)\vv$.\\
Therefore, we have:\\
$(*)$\hspace{2em}$\xi\force(p\forcec F)$ \ $\Dbfl$\\
\hspace*{\fill}$(\pt q\in P)(\pt\tau\in\C[p\et q])(\pt\pi\in\Pi)
((\pi,q)\in\vv{\mathcal X}(t_1,\ldots,t_k)\vv\Fl\xi\star\tau\ps\pi\in\bbot)$.

\smallskip\noindent
Moreover, we have \ $(\xi,p)\fforce F$ $\Dbfl$ 
$(\pt q\in P)(\pt\pi\in\Pi)
((\pi,q)\in\vv F\vv\Fl(\xi,p)\star(\pi,q)\in\bbbot)$\\
$\Dbfl$ \ $(\pt q\in P)(\pt\pi\in\Pi)
((\pi,q)\in\vv F\vv\Fl(\xi\star\pi,p\et q)\in\bbbot)$ \ and finally, by definition of $\bbbot$:

\smallskip\noindent
$(**)$\hspace{2em}$(\xi,p)\fforce F$ \ $\Dbfl$ \ $(\pt q\in P)(\pt\tau\in\C[p\et q])(\pt\pi\in\Pi)
((\pi,q)\in\vv F\vv\Fl\xi\star\pi^\tau\in\bbot)$.

\smallskip\noindent
Suppose that \ $\xi\force(p\forcec F)$. Since \ $\chi\xi\star\pi^\tau\succ\xi\star\tau\ps\pi$, we have
from $(*)$:\\
$(\pt q\in P)(\pt\tau\in\C[p\et q])(\pt\pi\in\Pi)
((\pi,q)\in\vv{\mathcal X}(t_1,\ldots,t_k)\vv\Fl\chi\xi\star\tau\ps\pi\in\bbot)$\\
and therefore \ $(\chi\xi,p)\fforce F$ \ from $(**)$.\\
Conversely, suppose that  \ $(\xi,p)\fforce F$. By applying $(**)$ and
$\chi'\xi\star\tau\ps\pi\succ\xi\star\pi^\tau$, we obtain \
$(\pt q\in P)(\pt\tau\in\C[p\et q])(\pt\pi\in\Pi)
((\pi,q)\in\vv F\vv\Fl\chi'\xi\star\tau\ps\pi\in\bbot)$\\
and therefore \ $\chi'\xi\force(p\forcec F)$ \ from $(*)$.

\smallskip\noindent
$\bullet$~~If $F\equiv\pt X\,A$, then $p\forcec F\equiv\pt X^+(p\forcec A)$.\\
Therefore, we have \ $\xi\force(p\forcec F)\equiv\pt X^+(\xi\force(p\forcec A))$.\\
Moreover, we have \
$(\xi,p)\fforce F\equiv\pt X((\xi,p)\fforce A)$.\\
Let ${\mathcal X}:P^k\to{\mathcal P}(\PPi)$ be a second order parameter in the model ${\mathcal N}$, with the same arity
as $X$, and let ${\mathcal X}^+$ be the corresponding parameter of the model ${\mathcal M}$.\\
If \ $\xi\force(p\forcec F)$, then we have $(\xi\force(p\forcec A))[{\mathcal X}^+/X^+]$, thus 
$\xi\force(p\forcec A[{\mathcal X}/X])$, from lemma~\ref{forcecXX+}.\\
By the recurrence hypothesis, we have \ $(\chi_A\xi,p)\fforce A[{\mathcal X}/X]$. Since ${\mathcal X}$ is arbitrary,
it follows that \ $(\chi_A\xi,p)\fforce\pt X\,A$.\\
Conversely, if we have \ $(\xi,p)\fforce F$, then \ $(\xi,p)\fforce A[{\mathcal X}/X]$ for every ${\mathcal X}$.\\
By the recurrence hypothesis, we have \ $\chi'_A\xi\force(p\forcec A[{\mathcal X}/X])$, and therefore:\\
$\chi'_A\xi\force(p\forcec A)[{\mathcal X}^+/X^+])$, from lemma~\ref{forcecXX+}.
Since \ ${\mathcal X}^+$ is arbitrary, it follows that:\\
$\chi'_A\xi\force\pt X^+(p\forcec A)$, that is \ $\chi'_A\xi\force(p\forcec\pt X\,A)$.

\smallskip\noindent
$\bullet$~~If $F\equiv\pt x\,A$, then $p\forcec F\equiv\pt x(p\forcec A)$. Therefore
$\xi\force p\forcec F\equiv\pt x(\xi\force(p\forcec A))$.\\
Moreover, \ $(\xi,p)\fforce F\equiv\pt x((\xi,p)\fforce A)$.\\
The result is immediate, from the recurrence hypothesis.

\smallskip\noindent
$\bullet$~~If $F\equiv(t_1=t_2\mapsto A)$, then
$p\forcec F\equiv t_1=t_2\mapsto p\forcec A$. Therefore:\\
$\xi\force(p\forcec F)\equiv(t_1=t_2\mapsto\xi\force(p\forcec A))$.\\
Moreover, \ $(\xi,p)\fforce F\equiv(t_1=t_2\mapsto(\xi,p)\fforce A)$.\\
The result is immediate, from the recurrence hypothesis.

\smallskip\noindent
$\bullet$~~If $F\equiv A\to B$, we have \ $p\forcec F\equiv\pt q(q\forcec A\to p\et q\forcec B)$ and therefore:\\
$(*)$\hspace{2em}
$\xi\force(p\forcec F)$ $\Fl$ $\pt\eta\pt q(\eta\force(q\forcec A)\to\xi\eta\force(p\et q\forcec B))$.\\
Suppose that \ $\xi\force(p\forcec F)$ \ and put \ $\chi_F=\lbd x\lbd y(\ov{\gamma}_0)(\chi_B)(x)(\chi'_A)y$.\\
We must show \ $(\chi_F\xi,p)\fforce A\to B$~; thus, let \ $(\eta,q)\fforce A$ and $(\pi,r)\in\vv B\vv$.\\
We must show \ $(\chi_F\xi,p)\star(\eta,q)\ps(\pi,r)\in\bbbot$ \ that is \
$(\chi_F\xi\star\eta\ps\pi,p\et(q\et r))\in\bbbot$.\\
Thus, let \ $\tau\in\C[p\et(q\et r)]$~; we must show \ $\chi_F\xi\star\eta\ps\pi^\tau\in\bbot$ \
or else \ $\chi_F\star\xi\ps\eta\ps\pi^\tau\in\bbot$.

\smallskip\noindent
From the recurrence hypothesis applied to $(\eta,q)\fforce A$, we have \ $\chi'_A\eta\force(q\forcec A)$.\\
From $(*)$, we have therefore \ $(\xi)(\chi'_A)\eta\force(p\et q\forcec B)$.\\
Applying again the recurrence hypothesis, we get:\\
$((\chi_B)(\xi)(\chi'_A)\eta,p\et q)\fforce B$. But since $(\pi,r)\in\vv B\vv$, we have:\\
$((\chi_B)(\xi)(\chi'_A)\eta,p\et q)\star(\pi,r)\in\bbbot$, \ that is \
$((\chi_B)(\xi)(\chi'_A)\eta\star\pi,(p\et q)\et r)\in\bbbot$.\\
Since \ $\tau\in\C[p\et(q\et r)]$, we have \ $\gamma_0\tau\in\C[(p\et q)\et r]$ \ and therefore \
$(\chi_B)(\xi)(\chi'_A)\eta\star\pi^{\gamma_0\tau}\in\bbot$.\\
But, by definition of $\chi_F$, we have, from theorem~\ref{beta_red_gauche}:\\
$\chi_F\star\xi\ps\eta\ps\pi^\tau\succ(\chi_B)(\xi)(\chi'_A)\eta\star\pi^{\gamma_0\tau}$ \
which gives the desired result: \ $\chi_F\star\xi\ps\eta\ps\pi^\tau\in\bbot$.

\smallskip\noindent
Suppose now that \ $(\xi,p)\fforce A\to B$~; we put \
$\chi'_F=\lbd x\lbd y(\chi'_B)(\ov{\alpha}_0 x)(\chi_A)y$.\\
We must show \ $\chi'_F\xi\force(p\forcec A\to B)$ \ that is \
$\pt q(\chi'_F\xi\force(q\forcec A\to p\et q\forcec B))$.\\
Thus, let \ $\eta\force q\forcec A$ and $\pi\in\|p\et q\forcec B\|$~; we must show \
$\chi'_F\xi\star\eta\ps\pi\in\bbot$.\\
By the  recurrence hypothesis, we have \ $(\chi_A\eta,q)\fforce A$, therefore
$(\xi,p)(\chi_A\eta,q)\fforce B$ \ or else, by definition of the algebra ${\mathcal B}$: \
$((\ov{\alpha}_0\xi)(\chi_A)\eta,p\et q)\fforce B$.\\
Applying again the recurrence hypothesis, we have \
$(\chi'_B)(\ov{\alpha}_0\xi)(\chi_A)\eta\force(p\et q\forcec B)$ \ and therefore:\\
$(\chi'_B)(\ov{\alpha}_0\xi)(\chi_A)\eta\star\pi\in\bbot$. But we have:\\ $\chi'_F\xi\star\eta\ps\pi\succ\chi'_F\star\xi\ps\eta\ps\pi\succ
(\chi'_B)(\ov{\alpha}_0\xi)(\chi_A)\eta\star\pi$ \ from theorem~\ref{beta_red_gauche}~;
the desired result follows.
\qed

\smallskip\noindent
A formula $F$ is said to be \emph{first order} if it is obtained by the following rules:\\
$\bullet$~~$\bot$ is first order.\\
$\bullet$~~If $A,B$ are first order, then $A\to B$ is first order.\\
$\bullet$~~If $B$ is first order, {\sf R} is a predicate symbol and $t_1,\ldots,t_k$ are terms with 
parameters, then ${\sf R}(t_1,\ldots,t_k)\to B$, $t_1=t_2\mapsto B$ are first order.\\
$\bullet$~~If $A$ is first order, then \ $\pt x\,A$ is first order ($x$ is an individual variable).

\smallskip\noindent
{\small{\bfseries Remarks.}\\
\phantom ii)~If \ $A$ \ is a first order formula, it is the same for \ $\pt x\inde\,A$.\\
ii)~This notion will be extended below (see proposition~\ref{n_eps_p_b}).}

\begin{thm}\label{premier_ordre}
Let $F$ be a closed first order formula. There exist two proof-like terms $\delta_F,\delta'_F$, which depend
only on the propositional structure of $F$, such that we have:\\
$\xi\force(\C[p]\to F)$ \ $\Fl$ \ $(\delta_F\xi,p)\fforce F$~;\\
$(\xi,p)\fforce F$ \ $\Fl$ \ $\delta'_F\xi\force(\C[p]\to F)$\\
for every $\xi\in\Lbd$ and $p\in P$.
\end{thm}
\proof The proof is by recurrence on the construction of $F$ following the above rules.

\smallskip\noindent
$\bullet$~~If $F$ is $\bot$, we put:\\
$\delta_\bot=\lbd x(\chi)\lbd y(x)(\alpha)y$ \ with  $\alpha::p\et q\Fl p$ .\\
$\delta'_\bot=\lbd x\lbd y(\chi'x)(\alpha')y$ \ with  $\alpha'::p\Fl p\et\1$ .

\smallskip\noindent
Indeed, suppose that \ $\xi\force\C[p]\to\bot$ and let us show that \ $(\delta_\bot\xi,p)(\pi,q)\in\bbbot$,
that is:\\
$(\delta_\bot\xi\star\pi,p\et q)\in\bbbot$. Thus, let \ $\tau\in\C[p\et q]$, so that
$\alpha\tau\in\C[p]$, so that \ $\xi\star\alpha\tau\ps\pi\in\bbot$, by hypothesis on $\xi$, which gives \
$\delta_\bot\xi\star\pi^\tau\in\bbot$.

\smallskip\noindent
Conversely, if \ $(\xi,p)\fforce\bot$, we have \ $(\xi,p)\star(\pi,\1)\equiv(\xi\star\pi,p\et\1)\in\bbbot$ \
for every $\pi\in\Pi$.\\
But, if $\tau\in\C[p]$, we have \ $\alpha'\tau\in\C[p\et\1]$, therefore \ $\xi\star\pi^{\alpha'\tau}\in\bbot$,
thus \ $\delta'_\bot\xi\star\tau\ps\pi\in\bbot$.\\
Therefore \ $\delta'_\bot\xi\force\C[p]\to\bot$.

\smallskip\noindent
$\bullet$~~If $F$ is $A\to B$, we put:\\
$\delta_{A\to B}=\lbd x\lbd y(\chi)\lbd z((\chi')(\delta_B)\lbd d((x)(\alpha)z)(\delta'_Ay)(\beta)z)(\gamma)z$ \
with\\
$\alpha::p\et(q\et r)\Fl p$; \ $\beta::p\et(q\et r)\Fl q$~; \ $\gamma::p\et(q\et r)\Fl\1\et r$.

\smallskip\noindent
Indeed, suppose that \ $\xi\force\C[p],A\to B$, $(\eta,q)\fforce A$ \ and \ $(\pi,r)\in\vv B\vv$.\\
We must show \ $(\delta_{A\to B}\xi,p)\star(\eta,q)\ps(\pi,r)\in\bbbot$, that is
$(\delta_{A\to B}\xi\star\eta\ps\pi,p\et(q\et r))\in\bbbot$.\\
Thus, let \ $\tau\in\C[p\et(q\et r)]$~; we must show \
$\delta_{A\to B}\xi\star\eta\ps\pi^\tau\in\bbot$.\\
We have \ $\alpha\tau\in\C[p],\beta\tau\in\C[q]$~; but, by the recurrence hypothesis, we have:\\
$\delta'_A\eta\force\C[q]\to A$, therefore \ $(\delta'_A\eta)(\beta)\tau\force A$ \ and \
$((\xi)(\alpha)\tau)(\delta'_A\eta)(\beta)\tau\force B$~;\\
thus \ $\lbd d((\xi)(\alpha)\tau)(\delta'_A\eta)(\beta)\tau\force\C[\1]\to B$.\\
From the recurrence hypothesis, we have \
$((\delta_B)\lbd d((\xi)(\alpha)\tau)(\delta'_A\eta)(\beta)\tau,\1)\fforce B$, thus:\\
$((\delta_B)\lbd d((\xi)(\alpha)\tau)(\delta'_A\eta)(\beta)\tau,\1)\star(\pi,r)\in\bbbot$, that is:\\
$((\delta_B)\lbd d((\xi)(\alpha)\tau)(\delta'_A\eta)(\beta)\tau\star\pi,\1\et r)\in\bbbot$.\\
But, we have $\gamma\tau\in\C[\1\et r]$, therefore \
$(\delta_B)\lbd d((\xi)(\alpha)\tau)(\delta'_A\eta)(\beta)\tau\star\pi^{\gamma\tau}\in\bbot$, and thus:\\
$((\chi')(\delta_B)\lbd d((\xi)(\alpha)\tau)(\delta'_A\eta)(\beta)\tau)(\gamma)\tau\star\pi\in\bbot$. 
It follows that:\\
$(\chi)\lbd z((\chi')(\delta_B)\lbd d((\xi)(\alpha)z)(\delta'_A\eta)(\beta)z)(\gamma)z\star\pi^\tau\in\bbot$ \
so that \ $\delta_{A\to B}\xi\star\eta\ps\pi^\tau\in\bbot$.

\smallskip\noindent
We now put:\\
$\delta'_{A\to B}=\lbd x\lbd y\lbd z((\delta'_B)(\ov{\alpha_0}x)(\delta_A)\lbd d\,z)(\alpha)y$ \
with \ $\alpha::p\Fl p\et\1$.

\smallskip\noindent
Suppose that \ $(\xi,p)\fforce A\to B$~; let $\tau\in\C[p]$, $\eta\force A$ and $\pi\in\|B\|$. We must show:\\
$\delta'_{A\to B}\xi\star\tau\ps\eta\ps\pi\in\bbot$. We have $\lbd d\,\eta\force\C[\1]\to A$~; applying the
recurrence hypothesis, we have $((\delta_A)\lbd d\,\eta,\1)\fforce A$, thus \
$(\xi,p)((\delta_A)\lbd d\,\eta,\1)\fforce B$ \
that is \ $((\ov{\alpha_0}\xi)(\delta_A)\lbd d\,\eta,p\et\1)\fforce B$.\\
Applying again the recurrence hypothesis, we find:\\
$(\delta'_B)(\ov{\alpha_0}\xi)(\delta_A)\lbd d\,\eta\force\C[p\et\1]\to B$.
Since we have $\alpha\tau\in\C[p\et\1]$, we get:\\
$(\delta'_B)(\ov{\alpha_0}\xi)(\delta_A)\lbd d\,\eta\star\alpha\tau\ps\pi\in\bbot$ and finally \
$\delta'_{A\to B}\xi\star\tau\ps\eta\ps\pi\in\bbot$.

\smallskip\noindent
$\bullet$~~If $F\equiv\R(\vec{q})\to B$, where $\R$ is a $k$-ary predicate symbol and $\vec{p}\in P^k$, we put:\\
$\delta_{R\to B}=\lbd x\lbd y(\ov{\alpha})(\delta_B)\lbd z(x)zy$ with  \
$\alpha::p\et(\1\et r)\Fl p\et r$.\\
$\delta'_{R\to B}=\lbd x\lbd y\lbd z((\delta'_B)(\ov{\alpha}_0)xz)(\alpha')y$ \ with  \ $\alpha'::p\Fl p\et\1$.

\smallskip\noindent
Suppose that \ $\xi\force\C[p],\R[\vec{q}]\to B$ and let \ $\eta\in|\R[\vec{q}]|$, $(\pi,r)\in\vv B\vv$.
We must show:\\
$(\delta_{R\to B}\xi,p)\star(\eta,\1)\ps(\pi,r)\in\bbbot$, that is \ 
$(\delta_{R\to B}\xi\star\eta\ps\pi,p\et(\1\et r))\in\bbbot$.\\
Thus, let \ $\tau\in\C[p\et(\1\et r)]$~; we must show \
$\delta_{R\to B}\xi\star\eta\ps\pi^\tau\in\bbot$. But, we have:\\
$\lbd z(\xi)z\eta\force\C[p]\to B$, and thus \ $((\delta_B)\lbd z(\xi)z\eta,p)\fforce B$,
by the recurrence hypothesis.\\
It follows that \ $((\delta_B)\lbd z(\xi)z\eta,p)\star(\pi,r)\in\bbbot$, that is:\\
$((\delta_B)\lbd z(\xi)z\eta\star\pi,p\et r)\in\bbbot$. But we have \ $\alpha\tau\in\C[p\et r]$,
and therefore:\\
$(\delta_B)\lbd z(\xi)z\eta\star\pi^{\alpha\tau}\in\bbot$, thus \ 
$(\ov{\alpha})(\delta_B)\lbd z(\xi)z\eta\star\pi^\tau\in\bbot$, therefore \
$\delta_{R\to B}\xi\star\eta\ps\pi^\tau\in\bbot$.

\smallskip\noindent
Suppose now that \ $(\xi,p)\fforce\R(\vec{q})\to B$~; let \ $\tau\in C[p]$, $\eta\in|\R[\vec{q}]|$
and \ $\pi\in\|B\|$.\\
We must show \ $\delta'_{R\to B}\xi\star\tau\ps\eta\ps\pi\in\bbot$. But, we have \
$(\xi,p)(\eta,\1)\fforce B$, that is:\\
$((\ov{\alpha}_0)\xi\eta,p\et\1)\fforce B$, thus \
$(\delta'_B)(\ov{\alpha}_0)\xi\eta\force\C[p\et\1]\to B$, by recurrence hypothesis.\\
But, we have \ $\alpha'\tau\in\C[p\et\1]$, therefore \
$(\delta'_B)(\ov{\alpha}_0)\xi\eta\star\alpha'\tau\ps\pi\in\bbot$, hence the result.

\smallskip\noindent
$\bullet$~~If $F\equiv(p_1=p_2\mapsto B)$, we put \ $\delta_F=\delta_B$ \ and \ $\delta'_F=\delta'_B$.\\
Indeed, suppose that \ $\xi\force\C[p]\to(p_1=p_2\mapsto B)$ and $(\pi,q)\in\vv p_1=p_2\mapsto B\vv$.
We must show that \ $(\delta_B\xi,p)\star(\pi,q)\in\bbbot$. Since $\vv p_1=p_2\mapsto B\vv\ne\vide$,
we have $p_1=p_2$, thus $(\pi,q)\in\vv B\vv$ and $\xi\force\C[p]\to B$. Hence the result, by the recurrence
hypothesis.

\smallskip\noindent
Suppose now that \ $(\xi,p)\fforce p_1=p_2\mapsto B$, $\tau\force\C[p]$ et
$\pi\in\|p_1=p_2\mapsto B\|$. We must show \ $\delta'_B\star\tau\ps\pi\in\bbot$. Since
$\|p_1=p_2\mapsto B\|\ne\vide$, we have $p_1=p_2$, therefore $\pi\in\|B\|$ and \
$(\xi,p)\fforce B$. Hence the result, by the recurrence hypothesis.

\smallskip\noindent
$\bullet$~~If $F\equiv\pt x\,A$, we put \ $\delta_F=\delta_A$ \ and \ $\delta'_F=\delta'_A$.

\smallskip\noindent
Indeed, if \ $\xi\force\C[p]\to\pt x\,A$, we have $\xi\force\C[p]\to A[a/x]$ for every $a\in P$. By the
recurrence hypothesis, we have $(\delta_A\xi,p)\fforce A[a/x]$~; thus \ $(\delta_A\xi,p)\fforce\pt x\,A$.

\smallskip\noindent
If $(\xi,p)\fforce\pt x\,A$, we have \ $(\xi,p)\fforce A[a/x]$ for every $a\in P$. By the recurrence
hypothesis, we have $\delta'_A\xi\force\C[p]\to A[a/x]$~; thus \ $\delta'_A\xi\force\C[p]\to\pt x\,A$.
\qed

\subsection*{The generic ideal}\noindent
We define a unary predicate ${\mathcal J}:P\to{\mathcal P}(\PPi)$ in the model ${\mathcal N}$ (second order parameter
of arity~1), by putting \ ${\mathcal J}(p)=\Pi\fois\{p\}$~; we call it \emph{the generic ideal}.\\
Thus, the binary predicate ${\mathcal J}^+:P^2\to{\mathcal P}(\Pi)$ which corresponds to it in the model ${\mathcal M}$, is
such that ${\mathcal J}^+(p,q)=\vide$ (resp. $\Pi$) if $p\ne q$ (resp. $p=q$). In other words:\\
\centerline{${\mathcal J}^+(p,q)$ is the predicate $p\ne q$.}

\noindent
The formula \ $p\force{\mathcal J}(q)$ is \ $\pt r(\C[p\et r]\to{\mathcal J}^+(r,q))$. Therefore, we have:\\
$\|p\force{\mathcal J}(q)\|=\|\neg\C[p\et q]\|$~; in other words:\\
\centerline{$p\force{\mathcal J}(q)$ is exactly $\neg\C[p\et q]$.}

\smallskip\noindent
{\bfseries Notations.}\\
$\bullet$~~We denote by \ $p\sqle q$ \ the formula $\pt r(\neg\C[q\et r]\to\neg\C[p\et r])$ \ and \
by \ $p\sim q$ the formula \ $p\sqle q\land q\sqle p$, that is \ $\pt r(\neg\C[q\et r]\dbfl\neg\C[p\et r])$.\\
In the sequel, we shall often write \ $F\to\C[p]$ \ instead of \ $\neg\C[p]\to\neg F$~;\\
Then \ $p\sqle q$ \ is written \ $\pt r(\C[p\et r]\to\C[q\et r])$ and $p\sim q$ is written
$\pt r(\C[p\et r]\dbfl\C[q\et r])$.\\
{\small{\bfseries Remark.} We recall that $\C[p]$ is not a formula, but a subset of $\Lbd$~; in fact, in some
realizability models which will be considered below, there will exist a formula $\CC[p]$ such that:\\ $|\C[p]|=\{\tau\in\Lbd_c;$ $\tau\force\CC[p]\}$.
In such cases, we can identify $\C[p]$ with the formula $\CC[p]$.}\\
$\bullet$~~If $F$ is a closed formula, we shall write \ $\fforce F$ \ to mean that there exists a
proof-like term $\theta$ such that \ $(\theta,\1)\fforce F$. From proposition~\ref{1_to_p}(i),
this is equivalent to say that there exists a proof-like term \ $\theta$ such that \ $(\theta,p)\fforce F$ for
every $p\in P$.

\begin{prop}\label{xi_p_force_Jq}\ \\
\emph{\phantom{ii}i)}~~$\xi\force\neg\C[p\et q]$ $\Fl$ $(\chi\xi,p)\fforce{\mathcal J}(q)$~;\\
\phantom{iii)}~~$(\xi,p)\fforce{\mathcal J}(q)$ $\Fl$ $\chi'\xi\force\neg\C[p\et q]$.\\
\emph{\phantom iii)}~~$\xi\force\pt r(\C[p\et(\1\et r)],\C[q]\to\bot)$ $\Fl$ $(\chi\xi,p)\fforce\neg\C[q]$~;\\
\phantom{iii)}~~$(\xi,p)\fforce\neg\C[q]$ $\Fl$ $\chi'\xi\force\pt r(\C[p\et(\1\et r)],\C[q]\to\bot)$.\\
\emph{iii)}~~If $\xi\force\neg{\sf R}(a_1,\ldots,a_k)$ \ then \ $(\xi,p)\fforce\neg{\sf R}(a_1,\ldots,a_k)$
for all $p$\\
(${\,\sf R}$ is a predicate symbol of arity $k$).
\end{prop}
\proof{\ }\ \\
\phantom{ii}i)~If $\xi\force\neg\C[p\et q]$, then $\xi\star\tau\ps\pi\in\bbot$ and therefore
$\chi\xi\star\pi^\tau\in\bbot$ for all $\tau\in\C[p\et q]$. Thus, we have: \
$(\chi\xi\star\pi,p\et q)\in\bbbot$, that is \ $(\chi\xi,p)\star(\pi,q)\in\bbbot$ for every $\pi\in\Pi$,
i.e. \ $(\chi\xi,p)\fforce{\mathcal J}(q)$.

\smallskip\noindent
If $(\xi,p)\fforce{\mathcal J}[q]$, we have $(\xi,p)\star(\pi,q)\in\bbbot$, thus $(\xi\star\pi,p\et q)\in\bbbot$
for all $\pi\in\Pi$. Therefore, we have \ $\xi\star\pi^\tau\in\bbot$, that is \
$\chi'\xi\star\tau\ps\pi\in\bbot$ \ for each \ $\tau\in\C[p\et q]$. Therefore \ $\chi'\xi\force\neg\C[p\et q]$.

\smallskip\noindent
\phantom iii)~If $\xi\force\pt r(\C[p\et(\1\et r)],\C[q]\to\bot)$, we have $\xi\star\upsilon\ps\tau\ps\pi\in\bbot$
if \ $\upsilon\in\C[p\et(\1\et r)]$ \ and \ $\tau\in\C[q]$. Therefore \ $\chi\xi\star\tau\ps\pi^\upsilon\in\bbot$,
thus \ $(\chi\xi\star\tau\ps\pi,p\et(\1\et r))\in\bbbot$ \ that is:\\ $(\chi\xi,p)\star(\tau,\1)\ps(\pi,r)\in\bbot$.
But \ $(\tau,\1)$ is arbitrary in \ $\C_{\mathcal N}[q]$, and therefore:\\
$(\chi\xi,p)\fforce\C[q]\to\bot$.

\smallskip\noindent
If $(\xi,p)\fforce\neg\C[q]$, we have $(\xi,p)\star(\tau,\1)\ps(\pi,r)\in\bbbot$, and therefore
$(\xi\star\tau\ps\pi,p\et(\1\et r))\in\bbbot$ for each $\tau\in\C[q]$. Thus, we have \
$\xi\star\tau\ps\pi^\upsilon\in\bbot$ \ therefore \ $\chi'\xi\star\upsilon\ps\tau\ps\pi\in\bbot$
for each \ $\upsilon\in\C[p\et(\1\et r)]$.
It follows that \ $\chi'\xi\force\pt r(\C[p\et(\1\et r)],\C[q]\to\bot)$.\\
iii)~~Let $\tau\in|{\sf R}(a_1,\ldots,a_k)|$~; we have $\xi\star\tau\ps\pi\in\bbot$ for all $\pi\in\Pi$,
thus \ $(\xi\star\tau\ps\pi,a)\in\bbbot$ for all $a\in P$, and therefore \ $(\xi,p)\star(\tau,\1)\ps(\pi,q)\in\bbbot$.
 \qed

\begin{thm}[Elementary properties of the generic ideal]\label{elem_gen}\ \\
\emph{\phantom{ii}i)}~~$(\ov{\alpha},\1)\fforce\neg{\mathcal J}(\1)$ \ with  \ $\alpha::\1\et(p\et q)\Fl p\et\1$.\\
\emph{\phantom iii)}~~$(\theta,\1)\fforce\pt x(\neg\C[x]\to{\mathcal J}(x))$ \ where \
$\theta=\lbd x(\chi)\lbd y((\chi'x)(\beta)y)(\alpha)y$\\
with \ $\alpha::\1\et(p\et q)\Fl q$ \ and \
$\beta::\1\et(p\et q)\Fl p\et(\1\et\1)$.\\
\emph{iii)}~~$(\theta,\1)\fforce\pt x\pt y({\mathcal J}(x\et y),\neg{\mathcal J}(x)\to{\mathcal J}(y))$ \
where \ $\theta=\lbd x\lbd y(\ov{\alpha})(y)(\ov{\beta})x$ \
with \ $\alpha::\1\et(p'\et(q'\et q))\Fl q'\et((q\et p')\et\1)$ \ and \
$\beta::(q\et p')\et p\Fl p'\et(p\et q)$.\\
\emph{iv)}~~$(\theta,\1)\fforce\pt x(\pt y(\neg\C[x\et y]\to{\mathcal J}(y))\to\neg{\mathcal J}(x))$ \ where \
$\theta=\lbd x\lbd y(\ov{\gamma})(x)\lbd z(\chi'y)(\beta)z$, with \
$\beta::p\et q\Fl q\et p$ \ and \ $\gamma::\1\et(r\et(q\et r'))\Fl r\et(\1\et p)$.\\
\emph{\phantom iv)}~~$(\theta,\1)\fforce\pt x\pt y({\mathcal J}(x),y\sqle x\to{\mathcal J}(y))$\\
where \ $\theta=\lbd x\lbd y((\chi)\lbd z(((\chi')(\ov{\alpha}_0y)\lbd z'(\chi'x)(\beta)z')(\alpha)z)(\gamma)z$, \ with \\
$\alpha::\1\et(p'\et(r\et q))\Fl(r\et\1)\et(\1\et\1)$~; \
$\alpha'::\1\et(p'\et(q'\et q))\Fl q\et p'$~; \ $\beta::p\et q\Fl q\et p$.
\end{thm}
\proof{\ }\ \\
\phantom{ii}i) Let \ $(\xi,p)\fforce{\mathcal J}(\1)$~; we must show that \ $(\ov{\alpha},\1)\star(\xi,p)\ps(\pi,q)\in\bbbot$, that is
to say:\\
$(\ov{\alpha}\star\xi\ps\pi,\1\et(p\et q))\in\bbbot$. But, from proposition~\ref{gamma::tu}, we have:\\
$(\ov{\alpha}\star\xi\ps\pi,\1\et(p\et q))\succ(\xi\star\pi,p\et\1)\equiv(\xi,p)\star(\pi,\1)$.\\
Now, we have \ $(\xi,p)\star(\pi,\1)\in\bbbot$ by hypothesis on $(\xi,p)$.

\smallskip\noindent
\phantom iii)Let \ $(\eta,p)\fforce\neg\C[q]$ and $(\pi,q)\in\vv{\mathcal J}(q)\vv$. We must show that
$(\theta,\1)\star(\eta,p)\ps(\pi,q)\in\bbbot$, i.e. \ $(\theta\star\eta\ps\pi,\1\et(p\et q))\in\bbbot$.
Thus, let \ $\tau\in\C[\1\et(p\et q)]$~; we must show that \ $\theta\star\eta\ps\pi^\tau\in\bbot$.\\
From proposition~\ref{xi_p_force_Jq}, we have \ $\chi'\eta\force\C[p\et(\1\et\1)],\C[q]\to\bot$.\\
Now, we have $\beta\tau\in\C[p\et(\1\et\1)]$ and $\alpha\tau\in\C[q]$, therefore 
$\chi'\eta\star\beta\tau\ps\alpha\tau\ps\pi\in\bbot$ \ thus\\ 
$(\chi)\lbd y((\chi'\eta)(\beta)y)(\alpha)y\star\pi^\tau\in\bbot$ thus
$\theta\star\eta\ps\pi^\tau\in\bbot$.

\smallskip\noindent
iii) Let $(\xi,p')\fforce{\mathcal J}(p\et q)$, \ $(\eta,q')\fforce\neg{\mathcal J}(p)$ \ and \
$(\pi,q)\in\vv{\mathcal J}(q)\vv$. We must show that:\\
$(\theta,\1)\star(\xi,p')\ps(\eta,q')\ps(\pi,q)\in\bbbot$, i.e. \
$(\theta\star\xi\ps\eta\ps\pi,\1\et(p'\et(q'\et q)))\in\bbbot$.\\
From propositions~\ref{1_to_p}(ii) and~\ref{gamma::tu}, it suffices to show:\\
$((\ov{\alpha})(\eta)(\ov{\beta})\xi\star\pi,\1\et(p'\et(q'\et q)))\in\bbbot$ \ then \
$(\eta\star\ov{\beta}\xi\ps\pi,q'\et((q\et p')\et\1))\in\bbbot$, that is:\\
$(\eta,q')\star(\ov{\beta}\xi,q\et p')\ps(\pi,\1)\in\bbbot$.\\
By hypothesis on $(\eta,q')$, we have now to show that \
$(\ov{\beta}\xi,q\et p')\fforce{\mathcal J}(p)$, i.e.:\\
$(\ov{\beta}\xi,q\et p')\star(\varpi,p)\in\bbbot$, or else \
$(\ov{\beta}\xi\star\varpi,(q\et p')\et p)\in\bbbot$ for all $\varpi\in\Pi$.\\
But, by proposition~\ref{gamma::tu}, we have:\\
$(\ov{\beta}\xi\star\varpi,(q\et p')\et p)\succ(\xi\star\varpi,p'\et(p\et q))
\equiv(\xi,p')\star(\varpi,p\et q)\in\bbbot$ by hypothesis on $(\xi,p')$.

\smallskip\noindent
iv)~~Let $(\xi,q)\fforce{\mathcal J}(p)$ and $(\eta,r)\fforce\pt q(\neg\C[p\et q]\to{\mathcal J}(q))$~;
we must show that:\\ $(\theta,\1)\star(\eta,r)\ps(\xi,q)\ps(\pi,r')\in\bbbot$, that is \
$(\theta\star\eta\ps\xi\ps\pi,\1\et(r\et(q\et r')))\in\bbbot$.\\
From proposition~\ref{xi_p_force_Jq}(i), we have $\chi'\xi\force\neg\C[q\et p]$. Let \ $\tau\in\C[p\et q]$,
thus \ $\beta\tau\in\C[q\et p]$ \ therefore \ $\chi'\xi\star\beta\tau\ps\rho\in\bbot$ for every $\rho\in\Pi$.
Therefore, we have \ $\lbd x(\chi'\xi)(\beta)x\star\tau\ps\rho\in\bbot$, thus
$\lbd z(\chi'\xi)(\beta)z\force\neg\C[p\et q]$. From proposition~\ref{xi_p_force_Jq}(iii), we have \
$(\lbd z(\chi'\xi)(\beta)z,\1)\fforce\neg\C[p\et q]$.\\
By hypothesis on $(\eta,r)$, we thus have \ $(\eta,r)\star(\lbd z(\chi'\xi)(\beta)z,\1)\ps(\pi,q)\in\bbbot$, i.e.:\\
$(\eta\star\lbd z(\chi'\xi)(\beta)z\ps\pi,r\et(\1\et q))\in\bbbot$, thus \
$((\ov{\gamma})(\eta)\lbd z(\chi'\xi)(\beta)z\star\pi,\1\et(r\et(q\et r')))\in\bbbot$\\
(proposition~\ref{gamma::tu}) and therefore \
$(\theta\star\eta\ps\xi\ps\pi,\1\et(r\et(q\et r')))\in\bbbot$.

\smallskip\noindent
\phantom iv) Let $(\xi,p')\fforce{\mathcal J}(p)$ and $(\eta,r)\fforce q\sqle p$~; we must show that:\\
$(\theta,\1)\star(\xi,p')\ps(\eta,r)\ps(\pi,q)\in\bbbot$ for all $\pi\in\Pi$, that is \
$(\theta\star\xi\ps\eta\ps\pi,\1\et(p'\et(r\et q)))\in\bbbot$.\\
From proposition~\ref{xi_p_force_Jq}(i), we have $\chi'\xi\force\neg\C[p'\et p]$, thus \
$\lbd z'(\chi'\xi)(\beta)z'\force\neg\C[p\et p']$: indeed, if $\tau\in\C[p\et p']$ and $\rho\in\Pi$, we have \
$\lbd z'(\chi'\xi)(\beta)z'\star\tau\ps\rho\succ(\chi'\xi)(\beta)\tau\star\rho\in\bbot$
since $\beta\tau\in\C[p'\et p]$.\\
Then, from proposition~\ref{xi_p_force_Jq}(iii), we have \ $(\lbd z'(\chi'\xi)(\beta)z',\1)\fforce\neg\C[p\et p']$.
But, by hypothesis on $(\eta,r)$, we have \ $(\eta,r)\fforce(\neg\C[p\et p']\to\neg\C[q\et p'])$.
It follows that:\\
$(\eta,r)(\lbd z'(\chi'\xi)(\beta)z',\1)\fforce\neg\C[q\et p']$, i.e. \
($(\ov{\alpha}_0\eta)\lbd z'(\chi'\xi)(\beta)z',r\et\1)\fforce\neg\C[q\et p']$.\\
From proposition~\ref{xi_p_force_Jq}(ii), we have \
$(\chi')(\ov{\alpha}_0\eta)\lbd z'(\chi'\xi)(\beta)z'\force\C[(r\et\1)\et(\1\et\1)],\C[q\et p']\to\bot$.\\
Let \ $\tau\in\C[\1\et(p'\et(r\et q))]$, therefore \ $\alpha\tau\in\C[(r\et\1)\et(\1\et\1)]$ and
$\alpha'\tau\in\C[q\et p']$.\\
Thus, we have:\\
$(((\chi')(\ov{\alpha}_0\eta)\lbd z'(\chi'\xi)(\beta)z')(\alpha)\tau)(\gamma)\tau\star\pi\in\bbot$, therefore:\\
$(\chi)\lbd z(((\chi')(\ov{\alpha}_0\eta)\lbd z'(\chi'\xi)(\beta)z')(\alpha)z)(\alpha')z\star\pi^\tau\in\bbot$.
In other words:\\
$((\chi)\lbd z(((\chi')(\ov{\alpha}_0\eta)\lbd z'(\chi'\xi)(\beta)z')(\alpha)z)(\alpha')z\star\pi,
\1\et(p'\et(r\et q)))\in\bbbot$\\
or else, from proposition~\ref{1_to_p}(ii): \
$(\theta\star\xi\ps\eta\ps\pi,\1\et(p'\et(r\et q)))\in\bbbot$.
\qed

\begin{thm}[Density]\label{densite}\ \\
For each function \ $\phi:P\to P$, we have:\\
$(\theta,\1)\fforce\pt x(\neg\C[x\et\phi(x)]\to{\mathcal J}(x)),\pt x\,{\mathcal J}(x\et\phi(x))\to\bot$\\
where \ $\theta=(\ov{\beta})\lbd x\lbd y(x)(\vartheta)y$, \
$\vartheta=(\chi)\lbd d\lbd x\lbd y(\chi'x)(\alpha)y$~;\\
with \ $\alpha::q\et r\Fl q\et(q\et r)$~; \ $\beta::\1\et(p\et(q\et r))\Fl p\et(\1\et q)$.
\end{thm}

\proof
Let $(\xi,p)\fforce\pt x(\neg\C[x\et\phi(x)]\to{\mathcal J}(x))$,
$(\eta,q)\fforce\pt x\,{\mathcal J}(x\et\phi(x))$ and $(\pi,r)\in\PPi$.\\
we must show that \ $(\theta\star\xi\ps\eta\ps\pi,\1\et(p\et(q\et r)))\in\bbbot$~;
thus, let \ $\tau_0\in\C[\1\et(p\et(q\et r))]$. We must show \
$\theta\star\xi\ps\eta\ps\pi^{\tau_0}\in\bbot$.\\
We first show that \ $(\vartheta\eta,\1)\fforce\neg\C[q\et\phi(q)]$.\\
Thus, let \ $(\varpi,r')\in\PPi$ and $\tau\in\C[q\et\phi(q)]$~; we must show \
$(\vartheta\eta,\1)\star(\tau,\1)\ps(\varpi,r')\in\bbbot$\\
i.e. \ $(\vartheta\eta\star\tau\ps\varpi,\1\et(\1\et r'))\in\bbbot$ or else \
$\vartheta\eta\star\tau\ps\varpi^{\tau'}\in\bbot$ \ for each \ $\tau'\in\C[\1\et(\1\et r')])$.\\
Now, $\vartheta\eta\star\tau\ps\varpi^{\tau'}\succ\eta\star\varpi^{\alpha\tau}$ \ and \
$\alpha\tau\in\C[q\et(q\et\phi(q))]$. Thus, it suffices to show:\\
$(\eta\star\varpi,q\et(q\et\phi(q)))\in\bbbot$ or else \
$(\eta,q)\star(\varpi,q\et\phi(q))\in\bbbot$.\\
But this follows from the hypothesis on $(\eta,q)$, which implies \ $(\eta,q)\fforce{\mathcal J}(q\et\phi(q))$.

\smallskip\noindent
By hypothesis on $\xi$, we have $(\xi,p)\fforce\neg\C[q\et\phi(q)]\to{\mathcal J}(q)$. It follows that:\\
$(\xi,p)\star(\vartheta\eta,\1)\ps(\pi,q)\in\bbbot$, that is \
$(\xi\star\vartheta\eta\ps\pi,p\et(\1\et q))\in\bbbot$.\\
But we have \ $\tau_0\in\C[\1\et(p\et(q\et r))])$, thus \ $\beta\tau_0\in\C[p\et(\1\et q)]$.
It follows that \ $\xi\star\vartheta\eta\ps\pi^{\beta\tau_0}\in\bbot$.\\
This gives the desired result, since \
$\theta\star\xi\ps\eta\ps\pi^{\tau_0}\succ\xi\star\vartheta\eta\ps\pi^{\beta\tau_0}$.
\qed

\section{Countable downward chain condition}\noindent
In this section, we consider a standard realizability algebra ${\mathcal A}$ and a
${\mathcal A}$-model ${\mathcal M}$. We suppose that the set $P$ (domain of variation of individual variables) has a power $\ge2^{\aleph_0}$.
We choose a surjection \ $\varepsilon:P\to{\mathcal P}(\Pi)^\NN$ and we define a binary
predicate in the model ${\mathcal M}$, which we denote also by $\varepsilon$, by putting:\\
\centerline{$\|n\eps p\|=\varepsilon(p)(n)$ if $n\in\NN$~; \ $\|n\eps p\|=\vide$ if
$n\notin\NN$}\\
(we use, for the predicate $\varepsilon$, the notation \ $n\eps p$ instead of
$\varepsilon(n,p)$).\\
Therefore, the predicate $\varepsilon$ enables us to associate, with each individual,
a set of integers which are its \emph{elements}. Proposition~\ref{RPN} shows that the following axiom is realized:

\smallskip\noindent
\emph{For every set, there exists an individual which has the same integer elements}.

\smallskip\noindent
This axiom will be called \emph{axiom of representation of predicates on $\NN$} and
denoted by RPN.

\begin{prop}[RPN]\label{RPN}\ \\
$\lbd x(x)\ul{0}\,\ul{0}\force\pt X\ex x\pt n\inde(Xn\dbfl n\eps x)$.
\end{prop}

\proof
This formula is \
$\pt X(\pt x[\pt n($ent$(n),Xn\to n\eps x),\pt n($ent$(n),n\eps x\to Xn)\to\bot]\to\bot)$.\\
Thus, we consider a unary parameter ${\mathcal X}:P\to{\mathcal P}(\Pi)$ and a term $\xi\in\Lbd$
such that:

\smallskip\noindent
$\xi\force\pt x[\pt n($ent$(n),{\mathcal X}n\to n\eps x),\pt n($ent$(n),n\eps x\to{\mathcal X}n)\to\bot]$.

\smallskip\noindent
We must show that \ $\lbd x(x)\ul{0}\,\ul{0}\star\xi\ps\pi\in\bbot$, or else
$\xi\star\ul{0}\ps\ul{0}\ps\pi\in\bbot$ for every stack $\pi\in\Pi$.\\
By definition of $\varepsilon$, there exists $p_0\in P$ such that \
${\mathcal X}n=\|n\eps p_0\|$ for every integer $n$.\\
But, we have: \
$\xi\force\pt n($ent$(n),{\mathcal X}n\to n\eps p_0),\pt n($ent$(n),n\eps p_0\to{\mathcal X}n)\to\bot$.\\
Thus, it suffices to show that \ $\ul{0}\force\pt n($ent$(n),{\mathcal X}n\to n\eps p_0)$\\
and \ $\ul{0}\force\pt n($ent$(n),n\eps p_0\to{\mathcal X}n)$.\\
Recall that the predicate ent$(x)$ is defined as follows:

\smallskip\noindent
$|\,$ent$(n)|=\{\ul{n}\}$ \ if $n\in\NN$ \ and \ $|\,$ent$(n)|=\vide$ if $n\notin\NN$.

\smallskip\noindent
Therefore, we have to show:\\
$\ul{0}\star\ul{n}\ps\eta\ps\rho\in\bbot$ \ for all $n\in\NN$, $\eta\force{\mathcal X}(n)$ \
and \ $\rho\in\|n\eps p_0\|$~;\\
$\ul{0}\star\ul{n}\ps\eta'\ps\rho'\in\bbot$ \ for all $n\in\NN$, $\eta'\force n\eps p_0$ \ and \ $\rho'\in{\mathcal X}(n)$.\\
But this follows from \ $\eta\star\rho\in\bbot$ \ and \ $\eta'\star\rho'\in\bbot$, which is trivially true, since ${\mathcal X}n=\|n\eps p_0\|$.
\qed

\smallskip\noindent
We suppose now that $\{\C,\et,\1\}$ is a forcing structure in ${\mathcal M}$.
Then we define also the symbol $\varepsilon$ in the ${\mathcal B}$-model ${\mathcal N}$ by
putting:\\
$\vv n\eps p\vv=\|n\eps p\|\fois\{\1\}$ for $n,p\in P$. In other words\\
$\vv n\eps p\vv=\{(\pi,\1);\;\pi\in\eps(p)(n)\}$ if $n\in\NN$~; $\vv n\eps p\vv=\vide$ if $n\notin\NN$.

\begin{prop}\label{n_eps_p}
The predicate \ $\varepsilon^+(q,n,p)$ is \ $q=\1\mapsto n\eps p$.\\
The formula \ $q\forcec n\eps p$ \ is \ $\C[q\et\1]\to n\eps p$.
\end{prop}\noindent
\proof Immediate, by definition of $\vv n\eps p\vv$.
\qed

\begin{prop}\label{n_eps_p_b}\ \\
\emph{\phantom ii)}~~$\xi\force(\C[p]\to n\eps q)$ $\Fl$ $(\delta\xi,p)\fforce n\eps q$ \
where \ $\delta=\lbd x(\chi)\lbd y(x)(\alpha)y$ and $\alpha::p\et\1$ $\Fl$ $p$.\\
\emph{ii)~~}$(\xi,p)\fforce n\eps q$ $\Fl$ $\delta'\xi\force(\C[p]\to n\eps q)$ \
where \ $\delta'=\lbd x\lbd y(\chi'x)(\alpha')y$ and $\alpha'::p$ $\Fl$ $p\et\1$.
\end{prop}
\proof{\ }\ \\
We have \ $(\xi,p)\fforce n\eps p$ $\Dbfl$ $(\xi,p)\star(\pi,\1)\in\bbbot$ \ for all $\pi\in\|n\eps p\|$,
or else:\\
$(\xi,p)\fforce n\eps p$ \ $\Dbfl$ \ $\xi\star\pi^\tau\in\bbot$ \ for each $\tau\in\C[p\et\1]$ \ and \ $\pi\in\|n\eps p\|$.\\
\phantom ii)~Suppose that \ $\xi\force(\C[p]\to n\eps q)$, $\tau\in\C[p\et\1]$ \ and \ $\pi\in\|n\eps p\|$. Then,we have:\\
$\delta\xi\star\pi^\tau\succ\xi\star\alpha\tau\ps\pi\in\bbot$, since $\alpha\tau\in\C[p]$.\\
ii)~Suppose that \ $(\xi,p)\force n\eps q$, $\tau\in\C[p]$ \ and \ $\pi\in\|n\eps p\|$. Then,we have:\\
$\delta'\xi\star\tau\ps\pi\succ\xi\star\pi^{\alpha'\tau}\in\bbot$, since \ $\alpha'\tau\in\C[p\et\1]$.
\qed

\smallskip\noindent
The notion of \emph{first order formula} has been defined previously (see theorem~\ref{premier_ordre}).
We extend this definition with the following clause:

\smallskip\noindent
$\bullet$~~$t\eps u$ is first order, for all terms $t,u$.

\smallskip\noindent
Proposition~\ref{n_eps_p_b} shows that theorem~\ref{premier_ordre} remains true for
this extended notion.

\smallskip\noindent
We say that the forcing structure $\{\C,\et,\1\}$ satisfies the
\emph{countable downward chain condition} (in abridged form \emph{c.d.c.}) if there exists a
proof-like term {\sf cdc} such that:

\smallskip\noindent
${\sf cdc}\force\pt X[
\pt n\inde\ex p\,X(n,p),\pt n\inde\pt p\pt q(X(n,p),X(n,q)\to p=q),\\
\hspace*{2.5em}\pt n\inde\pt p\pt q(X(n,p),X(sn,q)\to q\sqle p)\to\\
\hspace*{2.5em}\ex p'\{\pt n\inde\pt p(X(n,p)\to p'\sqle p),
(\pt n\inde\pt p(X(n,p)\to\C[p])\to\C[p'])\}]$.

\smallskip\noindent
{\small The intuitive meaning of this formula is:\\
If $X(n,p)$ is a decreasing sequence of conditions, then there exists a condition $p'$
which is less than all of them~; moreover, if all these conditions are non trivial,
then $p'$ is non trivial.}

\smallskip\noindent
We intend, in this section to prove the:

\begin{thm}[Conservation of reals]\label{cons_reels}\ \\
If the c.d.c. is verified, then there exists a proof-like term \ {\sf crl} \ such that:\\
({\sf crl}$,\1)\fforce\pt X\ex x\pt n\inde(Xn\dbfl n\eps x)$.
\end{thm}

\proof
This means that the axiom RPN, which is realized in the ${\mathcal A}$-model ${\mathcal M}$
(see proposition~\ref{RPN}) is also realized in the generic ${\mathcal B}$-model ${\mathcal N}$.

\smallskip\noindent
{\bfseries Notation.}\\
The formula $\pt q(\C[p\et q],q\forcec Xn\to p\forcec Xn)$  reads as
``~$p$ decides $Xn$~'', and is denoted by\\
$p\forcec\pm Xn$.\\
It can also be written as \ $\pt q\pt r(\C[p\et q],q\forcec Xn,\C[p\et r]\to X^+(r,n))$.\\
If ${\mathcal X}:P\to{\mathcal P}(\Pi\fois P)$ is a unary predicate in the ${\mathcal B}$-model
${\mathcal N}$,\\
and ${\mathcal X}^+:P^2\to{\mathcal P}(\Pi)$ is the corresponding binary predicate
in the standard ${\mathcal A}$-model ${\mathcal M}$,\\
the formula $\pt q(\C[p\et q],q\forcec{\mathcal X}n\to p\forcec{\mathcal X}n)$ is thus also denoted by \ $p\forcec\pm{\mathcal X}n$.

\begin{thm}\label{pX_decide}
If the c.d.c. is verified, there exists a proof-like term \ {\sf dec} \ such that:\\
{\sf dec}$\force\pt X\pt p_0\ex p'\{(\C[p_0]\to\C[p']),p'\sqle p_0,\pt n\inde(p'\forcec\pm Xn)\}$.
\end{thm}\noindent
{\small{\bfseries Remark.} This formula means that, for any predicate $X$, the set of
conditions which decide $Xn$ for all integers $n$ is dense.}

\smallskip\noindent
We first show how theorem~\ref{cons_reels} can be deduced from this
theorem~\ref{pX_decide}.\\
From theorem~\ref{forcec-fforce}, it is sufficient to find a proof-like term {\sf crl0} such that:\\
{\sf crl0}$\force\1\forcec\pt X\ex x\pt n\inde(Xn\dbfl n\eps x)$\\
or else, since \ $\1\forcec\neg A\equiv\pt p_0((p_0\forcec A),\C[\1\et p_0]\to\bot)$:\\
{\sf crl0}$\force\pt X\pt p_0[(p_0\forcec\pt q\{\pt n\inde(Xn\dbfl n\eps q)\to\bot\}),
\C[\1\et p_0]\to\bot]$.\\
From theorem~\ref{pX_decide}, it is sufficient to find a proof-like term {\sf crl1} such that:\\
\nopagebreak
{\sf crl1}$\force\pt X\pt p_0\pt p'\{(\C[p_0]\to\C[p']),p'\sqle p_0,
\pt n\inde(p'\forcec\pm Xn),\\
\hspace*{\fill}(p_0\forcec\pt q(\pt n\inde(Xn\dbfl n\eps q)\to\bot)),\C[\1\et p_0]\to\bot\}$.\\
It is sufficient to find a proof-like term {\sf crl2} such that:\\
{\sf crl2}$\force\pt X\pt p_0\pt p'\{(p_0\forcec\pt q(\pt n\inde(Xn\dbfl n\eps q)\to\bot)),p'\sqle p_0,\\
\hspace*{\fill}\pt n\inde(p'\forcec\pm Xn),\C[p']\to\bot\}$.\\
Indeed, we take then
{\sf crl1}$=\lbd x\lbd y\lbd z\lbd u\lbd v((x)(${\sf crl2}$)uyz)(\delta)v$
with \ $\delta::\1\et p\Fl p$~;\\
(recall that the formula $\C[p_0]\to\C[p']$ is written, in fact,
as $\neg\C[p']\to\neg\C[p_0]$).

\smallskip\noindent
We fix \ ${\mathcal X}^+:P^2\to{\mathcal P}(\Pi)$, $p_0,p'\in P$,
$\xi\force(p_0\forcec\pt q(\pt n\inde({\mathcal X}n\dbfl n\eps q)\to\bot))$, $\eta\force p'\sqle p_0$,
$\zeta\force\pt n\inde(p'\forcec\pm{\mathcal X}n)$ and $\tau\in\C[p']$. We must have \
({\sf crl2})$\xi\eta\zeta\tau\force\bot$.\\
We choose \ $q_0\in P$ such that we have $\|n\eps q_0\|=\|p'\forcec{\mathcal X}n\|$ for all $n\in\NN$, which is possible, by definition of $\varepsilon$.\\
We trivially have \
$\xi\force(p_0\forcec(\pt n\inde(n\eps q_0\to{\mathcal X}n),\pt n\inde({\mathcal X}n\to n\eps q_0)\to\bot))$.\\
But, the formula \
$p_0\forcec(\pt n\inde(n\eps q_0\to{\mathcal X}n),\,\pt n\inde({\mathcal X}n\to n\eps q_0)\to\bot)$ is written as:\\
$\pt r\pt r'(r\forcec\pt n\inde(n\eps q_0\to{\mathcal X}n),\;r'\forcec\pt n\inde({\mathcal X}n\to n\eps q_0),\;
\C[(p_0\et r)\et r']\to\bot)$.\\
Replacing $r$ and $r'$ with $p'$, we obtain:\\
$\xi\force(p'\forcec\pt n\inde(n\eps q_0\to{\mathcal X}n),\;p'\forcec\pt n\inde({\mathcal X}n\to n\eps q_0),\;\C[(p_0\et p')\et p']\to\bot)$.\\
From \ $\tau\in\C[p']$ and $\eta\force\pt r(\neg\C[p_0\et r]\to\neg\C[p'\et r])$,
we deduce that:\\
$\lbd h((\eta)\lbd x(h)(\beta)x)(\alpha)\tau\force\neg\neg\C[(p_0\et p')\et p']$\\
where $\alpha,\beta$ are \ $\C$-expressions such that $\alpha:p\Fl p\et p$~; \
$\beta::p\et q\Fl(p\et q)\et q$.\\
Thus, we have:\\
(1)~~$\lbd y\lbd z((\eta)\lbd x(\xi yz)(\beta)x)(\alpha)\tau\force\\
\hspace*{\fill}(p'\forcec\pt n\inde(n\eps q_0\to{\mathcal X}n)),(p'\forcec\pt n\inde({\mathcal X}n\to n\eps q_0))\to\bot$.

\smallskip\noindent
$\bullet$~~The formula \ $p'\forcec\pt n\inde(n\eps q_0\to{\mathcal X}n)$ is written as \
$\pt n\inde\pt r(r\forcec n\eps q_0\to p'\et r\forcec{\mathcal X}n)$.\\
But \ $r\forcec n\eps q_0\equiv\C[r\et\1]\to n\eps q_0$ (proposition~\ref{n_eps_p}) \
$\equiv\C[r\et\1]\to p'\forcec{\mathcal X}(n)$ \ by definition of~$q_0$.
Therefore \ $p'\forcec\pt n\inde(n\eps q_0\to{\mathcal X}n)\equiv
\pt n\inde\pt r((\C[r\et\1]\to p'\forcec{\mathcal X}(n))\to p'\et r\forcec{\mathcal X}n)\equiv$\\
$\pt n\inde\pt r\pt q'[\pt q(\C[r\et\1],\C[p'\et q]\to{\mathcal X}^+(q,n)),\C[(p'\et r)\et q']\to{\mathcal X}^+(q',n)]$.\\
Thus, we have:\\
(2)~~$\lbd d\lbd x\lbd y((x)(\alpha')y)(\beta')y\force(p'\forcec\pt n\inde(n\eps q_0\to{\mathcal X}n))$\\
with \ $\alpha'::(p\et r)\et q\Fl r\et\1$ \ and \ $\beta'::(p\et r)\et q\Fl p\et q$.

\smallskip\noindent
$\bullet$~~The formula \ $p'\forcec\pt n\inde({\mathcal X}n\to n\eps q_0)$ is written as \
$\pt n\inde\pt r(r\forcec{\mathcal X}n\to p'\et r\forcec n\eps q_0)$,\\
or else: \ $\pt n\inde\pt r(r\forcec{\mathcal X}n,\C[(p'\et r)\et\1]\to n\eps q_0)$,
that is, by definition of $q_0$:\\
$\pt n\inde\pt r(r\forcec{\mathcal X}n,\C[(p'\et r)\et\1]\to p'\forcec{\mathcal X}n)$. \
But, we have:\\
$\zeta\force\pt n\inde(p'\forcec\pm{\mathcal X}n)$, in other words \
$\zeta\force\pt n\inde\pt r(r\forcec{\mathcal X}n,\C[p'\et r]\to p'\forcec{\mathcal X}n)$.
Therefore:\\
(3)~~$\lbd n\lbd x\lbd y(\zeta nx)(\alpha'')y\force p'\forcec
\pt n\inde({\mathcal X}n\to n\eps q_0)$ \ with \ $\alpha''::(p\et r)\et\1\Fl p\et r$.

\smallskip\noindent
It follows from (1,2,3) that:\\
$((\lbd y\lbd z((\eta)\lbd x(\xi yz)(\beta)x)(\alpha)\tau)\;\lbd d\lbd x\lbd y((x)(\alpha')y)(\beta')y)
\;\lbd n\lbd x\lbd y(\zeta nx)(\alpha'')y\force\bot$.\\
Therefore, we can put \ \ {\sf crl2} $=\\
\lbd x_0\lbd y_0\lbd z_0\lbd u((\lbd y\lbd z((y_0)\lbd x(x_0yz)(\beta)x)(\alpha)u)
\lbd d\lbd x\lbd y((x)(\alpha')y)(\beta')y)\lbd n\lbd x\lbd y(z_0nx)(\alpha'')y$.
\qed

\smallskip\noindent
The remaining of this section is devoted to the proof of theorem~\ref{pX_decide}.

\subsection*{Definition of a sequence by dependent choices}\noindent
In this section, we are given a fixed element $p_0\in P$ \ and a finite sequence of
formulas with  parameters $\vec{F}(n,p,p')$. We are also given a proof-like term \
{\sf dse} \ such that:\\
{\sf dse}$\force\pt n\pt p\ex p'\,\vec{F}(n,p,p')$.

\smallskip\noindent
{\small{\bfseries Remark.} The aim of this section is to write down a formula $\Phi(x,y)$
which represents the graph of a function $\phi:\NN\to P$ such that the formulas $\phi(0)=p_0$
and $\pt n\inde\vec{F}(n,\phi(n),\phi(n+1))$ are realized by proof-like terms.
We shall only apply the results of this section to a particular sequence $\vec{F}$
of length 3.}

\smallskip\noindent
From theorem~\ref{ACI}(i) (axiom of choice for individuals), there exists a function \
$f:P^3\to P$ \ such that: \ \
$\vsig\force\pt n\pt p(\pt k\inde(\vec{F}(n,p,f(n,p,k))\to\bot)\to
\pt p'(\vec{F}(n,p,p')\to\bot))$.

\smallskip\noindent
It follows that \ $\lbd x(${\sf dse}$)(\vsig)x\force
\pt n\pt p(\pt k\inde(\vec{F}(n,p,f(n,p,k))\to\bot)\to\bot)$.

\smallskip\noindent
We define a function denoted by \ $(m\ppt n)$, from $P^2$ into $P$, by putting, for $m,n\in P$:\\
$(m\ppt n)=1$ if $m,n\in\NN$ and $m<n$~; $(m\ppt n)=0$ otherwise.

\smallskip\noindent
Obviously, the relation $(m\ppt n)=1$ is well founded on $P$.\\
Thus, from theorem~\ref{bien_fonde}(ii), we have:\\
$\Y\force\pt k(\pt l($ent$(l),\vec{F}(n,p,f(n,p,l))\to(l\ppt k)\ne1),$ent$(k),\vec{F}(n,p,f(n,p,k))\to\bot)\\
\hspace*{\fill}\to\pt k($ent$(k),\vec{F}(n,p,f(n,p,k))\to\bot)$.\\
Therefore, if we set \ $\Y'=\lbd x(\Y)\lbd y\lbd z(x)zy$, we have:\\
$\Y'\force\pt k\inde\{\pt l\inde(\vec{F}[n,p,f(n,p,l)]\to(l\ppt k)\ne1),\vec{F}[n,p,f(n,p,k)]\to\bot\}\\
\hspace*{\fill}\to\pt k\inde(\vec{F}[n,p,f(n,p,k)]\to\bot)$.\\
Thus, we have:\\
$\lbd x(${\sf dse}$)(\vsig)(\Y')x\force\pt k\inde\{\pt l\inde(\vec{F}[n,p,f(n,p,l)]\to(l\ppt k)\ne1),\vec{F}[n,p,f(n,p,k)]\to\bot\}\to\bot$.

\smallskip\noindent
We define the formula \ $G(n,p,k)\equiv\pt l\inde(\vec{F}(n,p,f(n,p,l))\to(l\ppt k)\ne1)$ and the finite sequence of formulas \ $\vec{H}(n,p,k)\equiv\{G(n,p,k),\vec{F}(n,p,f(n,p,k))\}$. Then, we have shown:

\begin{lem}\label{H_existe}
{\sf dse0}$\force\pt n\pt p\ex k\inde\{\vec{H}(n,p,k)\}$, with \
{\sf dse0} $=\lbd x(${\sf dse}$)(\vsig)(\Y')x$.
\end{lem}\noindent
{\small{\bfseries Remark.} The meaning of $\vec{H}(n,p,k)$ is ``$k$ is the least integer
such that $\vec{F}(n,p,f(n,p,k))$''.}

\begin{lem}\label{H_unicite}
Let \ \cp \ be a proof-like term such that, for every $m,n\in\NN$, we have:\\
\cp$\star\ul{m}\ps\ul{n}\ps\xi\ps\eta\ps\zeta\ps\pi\succ\xi\star\pi$ (resp. \ $\eta\star\pi$, \ $\zeta\star\pi$)
if \ $m<n$ \ (resp. \ $n<m$, $m=n$). \ Then:\\
\emph{\phantom ii)}~\cp$\force\pt m\inde\pt n\inde((m\ppt n)\ne1,(n\ppt m)\ne1,m\ne n\to\bot)$.\\
\emph{ii)}~{\sf dse1}$\force\pt n\pt p\pt k\inde\pt{k'}\,\inde(\vec{H}(n,p,k),\vec{H}(n,p,k'),k\ne k'\to\bot)$\\
with \ {\sf dse1}$\,=\lbd k\lbd k'\lbd x\lbd\vec{y}
\lbd x'\lbd\vec{y}'(($\cp$\,k'k)(x)k'\vec{y}')(x')k\vec{y}$,
where \ $\vec{y},\vec{y}'$ are two sequences of distinct variables of the same length as
the sequence $\vec{F}$.
\end{lem}
\proof{\ }\ \\
\phantom ii)~Trivial.\\
ii)~Let \ $\xi\force G(n,p,k)$, $\vec{\eta}\force\vec{F}(n,p,f(n,p,k))$, \
$\xi'\force G(n,p,k')$, $\vec{\eta}'\force\vec{F}(n,p,f(n,p,k'))$\\
and \ $\zeta\force k\ne k'$.
We must show \
$\cp\star\ul{k}'\ps\ul{k}\ps(\xi)\ul{k}'\vec{\eta}'\ps(\xi')\ul{k}\vec{\eta}\ps\zeta\ps\pi\in\bbot$.\\
If $k=k'$, it remains to prove \ $\zeta\star\pi\in\bbot$~; but this is true because
we then have $\zeta\force\bot$.\\
If $k'<k$, it remains to prove \ $\xi\star\ul{k}'\ps\vec{\eta}'\ps\pi\in\bbot$. This
results immediately from:\\
$\xi\force\pt{k'}\,\inde(\vec{F}(n,p,f(n,p,k'))\to (k'\ppt k)\ne1)$ \ and \ thus:\\
$\xi\force$ent$(k'),\vec{F}(n,p,f(n,p,k'))\to\bot$, \ since $k'<k$.
\qed

\smallskip\noindent
We now define the binary predicate:\\
$\Phi(x,y)\equiv\pt X(\pt n\pt p\pt k\inde(\vec{H}(n,p,k),X(n,p)\to X(sn,f(n,p,k))),X(0,p_0)\to X(x,y))$\\
and we show that $\Phi(x,y)$ is a sequence of conditions (functional relation on $\NN$)
and also some other properties of $\Phi$.

\smallskip\noindent
{\small{\bfseries Remark.} Intuitively, the predicate $\Phi$ is the graph of the function
$\phi$ of domain $\ennl$, recursively defined by the conditions: \ $\phi(0)=p_0$~; \ $\phi(n+1)=f'(n,\phi(n))$\\
where $f'(n,p)$ is $f(n,p,k)$ for the least $k$ such that $F(n,p,f(n,p,k))$.
Unfortunately, we cannot introduce $f'$ as a function symbol because, unlike $f$, it is not defined in
the ground model.}

\begin{lem}\label{rec_Phi}\ \\
\emph{\phantom{ii}i)}~~$\lbd x\lbd y\,y\force\Phi(0,p_0)$.\\
\emph{\phantom iii)}~~$\lbd x(x)II\force\pt y(\Phi(0,y)\to y=p_0)$.\\
\emph{iii)}~~{\sf rec}$\force\pt x\pt y
\pt k\inde(\vec{H}(x,y,k),\Phi(x,y)\to\Phi(sx,f(x,y,k)))$\\
where \ {\sf rec} $=\lbd k\lbd x\lbd\vec{y}\lbd x'\lbd z\lbd u(zkx\vec{y})(x')zu$\\
and \ $\vec{y}$ is a sequence of distinct variables of the same length as \ $\vec{F}$.
\end{lem}
\proof{\ }\ \\
\phantom{ii}i)~Trivial.

\smallskip\noindent
\phantom iii)~We define the  binary predicate ${\mathcal X}:P^2\to{\mathcal P}(\Pi)$ by putting:\\
${\mathcal X}(0,q)=\|q=p_0\|$ \ and \ ${\mathcal X}(p,q)=\vide$ for $p\ne0$.\\
We replace $X$ with ${\mathcal X}$ in the definition of $\Phi(0,y)$. Since we have \ $sn\ne0$ for all $n\in P$,
we obtain \ $\|\Phi(0,y)\|\supset\|\top,p_0=p_0\to y=p_0\|$~; hence the result.

\smallskip\noindent
iii)~Let $\xi\force G(x,y,k)$, $\vec{\eta}\force\vec{F}(x,y,f(x,y,k))$, \ $\xi'\force\Phi(x,y)$,\\
$\zeta\force\pt n\pt p\pt k\inde(\vec{H}(n,p,k),X(n,p)\to X(sn,f(n,p,k)))$,\\
$\upsilon\force X(0,p_0)$ \ and \ $\pi\in\|X(sx,f(x,y,k))\|$.\\
Then \ $\xi'\zeta\upsilon\force X(x,y)$, therefore \
$\zeta\star\ul{k}\ps\xi\ps\vec{\eta}\ps\xi'\zeta\upsilon\ps\pi\in\bbot$ \
i.e. \ ({\sf rec}$)\ul{k}\xi\vec{\eta}\xi'\zeta\upsilon\star\pi\in\bbot$.
\qed

\begin{lem}\label{ccd1}
{\sf cdc1}$\force\pt n\inde\ex p\,\Phi(n,p)$ \ where:\\
{\sf cdc1}$\,=\lbd n((n)\lbd x\lbd y(x)\lbd z(${\sf cd1}$)zy)\lbd x(x)\lbd x\lbd y\,y$\\
with {\sf cd1}$\,=\lbd x\lbd y(${\sf dse0}$)\lbd l\lbd\vec{z}(y)
(${\sf rec}$)l\vec{z}x$~;\\
$\vec{z}$ is a sequence of distinct variables of the same length as $\vec{H}$.
\end{lem}\noindent
Proof by recurrence on $n$~; we have $\lbd x\lbd y\,y\force\Phi(0,p_0)$, therefore
$\lbd x(x)\lbd x\lbd y\,y\force\ex y\,\Phi(0,y)$.\\
We now show that \ {\sf cd1}$\force\Phi(x,y)\to\ex y\Phi(sx,y)$.\\
Thus, we consider \ $\xi\force\Phi(x,y)$, \ $\eta\force\pt y(\Phi(sx,y)\to\bot)$.\\
We have \ {\sf rec}$\force\pt l\inde(\vec{H}(x,y,l),\Phi(x,y)\to\Phi(sx,f(x,y,l)))$
(lemma~\ref{rec_Phi}iii),\\
$\eta\force(\Phi(sx,f(x,y,l))\to\bot)$, and therefore:\\
$\lbd l\lbd\vec{z}(\eta)(${\sf rec}$)l\vec{z}\xi\force
\pt l\inde(\vec{H}(x,y,l)\to\bot)$, where $\vec{z}$ has the same length as
$\vec{H}$.\\
Now, we have \ {\sf dse0}$\force\ex k\inde\{\vec{H}(x,y,k)\}$ (lemma~\ref{H_existe})~; therefore:\\
({\sf dse0})$\lbd l\lbd\vec{z}(\eta)(${\sf rec}$)l\vec{z}\xi\force\bot$, that is \
({\sf cd1})$\xi\eta\force\bot$.

\smallskip\noindent
Thus, we have shown that \ {\sf cd1}$\force\pt y(\Phi(x,y)\to\ex y\Phi(sx,y))$,
and it follows that:\\
$\lbd x\lbd y(x)\lbd z(${\sf cd1}$)zy\force\ex y\Phi(x,y)\to\ex y\Phi(sx,y)$.
\qed

\begin{lem}\label{ccd2}
There exists a proof-like term \ {\sf cdc2} such that:\\
{\sf cdc2}$\force\pt n\inde\pt p\pt q(\Phi(n,p),\Phi(n,q)\to p=q)$.
\end{lem}

\proof
We give a detailed proof, by recurrence on $n$. It enables us to write explicitly
the proof-like term \ {\sf cdc2}.

\smallskip\noindent
For $n=0$, the lemma~\ref{rec_Phi}(ii) gives the result: \ $\Phi(0,p),\Phi(0,q)\to p=q$.\\
Let us fix $m$ and suppose that \ $\pt p\pt q(\Phi(m,p),\Phi(m,q)\to p=q)$.\\
We define the binary predicate:\\
$\Psi(n,q)\equiv\pt p\pt k\inde(n=sm,\vec{H}(m,p,k),\Phi(m,p)\to q=f(m,p,k))$.

\smallskip\noindent
We show that \ $\force\pt p\pt k\inde(\vec{H}(n,p,k),\Phi(n,p)\to\Psi(sn,f(n,p,k)))$, that is to say:\\
\nopagebreak
$\force\pt p\pt q\pt k\inde\pt l\inde\\
\hspace*{\fill}
\{\vec{H}(n,p,k),\Phi(n,p),sn=sm,\vec{H}(m,q,l),\Phi(m,q)\to f(n,p,k)=f(m,q,l)\}$.\\
But we have $\|sn=sm\|=\|n=m\|$, $\Phi(m,p),\Phi(m,q)\to p=q$ by hypothesis of
recurrence~; $\vec{H}(m,p,k),\vec{H}(m,p,l)\to k=l$ (lemma~\ref{H_unicite}(ii)), and
it follows that \ $f(n,p,k)=f(m,q,l)$.

\smallskip\noindent
If we put \ $\Psi'(x,y)\equiv\Phi(x,y)\land\Psi(x,y)$, we have:\\
$\force\pt p\pt k\inde(\vec{H}(n,p,k),\Psi'(n,p)\to\Psi'(sn,f(n,p,k)))$~; we have also
$\force\Psi'(0,p_0)$. This shows that \ $\force(\Phi(x,y)\to\Psi'(x,y))$ by making $X\equiv\Psi'$ in the definition of~$\Phi$.\\
Thus, we have \ $\force\Phi(sm,q)\to\pt p\pt k\inde(\vec{H}(m,p,k),\Phi(m,p)
\to q=f(m,p,k))$.\\
It follows that:\\
$\force\Phi(sm,q),\Phi(sm,q')\to\\
\hspace*{\fill}
\pt p\pt k\inde(\vec{H}(m,p,k),\Phi(m,p)\to(q=f(m,p,k))\land(q'=f(m,p,k)))$\\
and therefore \
$\force\Phi(sm,q),\Phi(sm,q')\to\pt p\pt k\inde(\vec{H}(m,p,k),\Phi(m,p)\to q=q')$.\\
Thus, we obtain \ $\force\Phi(sm,q),\Phi(sm,q')\to q=q'$, \ since we have \
{\sf cdc1}$\force\ex p\,\Phi(m,p)$ \ by lemma~\ref{ccd1} and \
{\sf dse0}$\force\pt p\ex k\inde\{\vec{H}(m,p,k)\}$  by lemma~\ref{H_existe}.
\qed

\subsubsection*{Resumption of the proof of theorem~\ref{pX_decide}}\noindent
In order to show theorem~\ref{pX_decide}, we fix \ $p_0\in P$ and a binary predicate
${\mathcal X}:P^2\to{\mathcal P}(\Pi)$.\\
We have to find a proof-like term \ {\sf dec} \ such that:\\
{\sf dec}$\force\ex p'\{(\C[p_0]\to\C[p']),p'\sqle p_0,\pt n\inde(p'\forcec\pm{\mathcal X}n)\}$.

\smallskip\noindent
We apply the above results, taking for \ $\vec{F}(n,p,p')$ the sequence of three
formulas:\\
$\{(\C[p]\to\C[p']),\,(p'\sqle p),\,p'\forcec\pm{\mathcal X}n\}$.\\
Lemma~\ref{densite_decide} below gives a proof-like term \ {\sf dse} \ such that
{\sf dse}$\force\pt n\pt p\ex p'\{\vec{F}(n,p,p')\}$.

\begin{lem}\label{densite_decide}
{\sf dse}$\force\pt p\ex p'\{\vec{F}(n,p,p')\}$\\
where \ {\sf dse}$\,=
\lbd a(\lbd h(aII)\lbd x\lbd y\,h)\lbd z(\Ccc)\lbd k((a\lbd x\,xz)\beta')\lbd x\lbd y(k)(y)(\alpha)x$\\
with \ $\beta'=\lbd x\lbd y(x)(\beta)y$, \ $\alpha::(p\et q)\et r\Fl r\et q$ \ and \
$\beta::(p\et q)\et r\Fl p\et r$.
\end{lem}

\proof
The formula we consider is written as \
$\pt p'[(\C[p]\to\C[p']),p'\sqle p\,,(p'\forcec\pm{\mathcal X}n)\to\bot]\to\bot$.\\
Thus, let \ $\xi\force\pt p'[(\C[p]\to\C[p']),p'\sqle p\,,
(p'\forcec\pm{\mathcal X}n)\to\bot]$. We must show \ ({\sf dse}$)\xi\force\bot$.

\smallskip\noindent
$\bullet$~~We show that \ $\lbd h(\xi II)\lbd x\lbd y\,h\force\neg(p\forcec{\mathcal X}n)$:\\
Let \ $\zeta\force(p\forcec{\mathcal X}n)$~; therefore, we have \
$\lbd x\lbd y\,\zeta\force(p\forcec\pm{\mathcal X}n)$~; indeed:\\
$p\forcec\pm{\mathcal X}n\equiv\pt q(\C[p\et q],q\forcec{\mathcal X}n\to p\forcec{\mathcal X}n)$.\\
But, we have \ $\xi\force\;(\C[p]\to\C[p]),p\sqle p\,,(p\forcec\pm{\mathcal X}n)\to\bot$~;\\
we have \ $I\force\C[p]\to\C[p]$ and $I\force p\sqle p$ \
(since $p'\sqle p\equiv\pt q(\neg\C[p\et q]\to\neg\C[p'\et q])$).\\
Thus \ $(\xi II)\lbd x\lbd y\,\zeta\force\bot$, hence the result.

\smallskip\noindent
$\bullet$~~We now show \
$\lbd z(\Ccc)\lbd k((\xi\lbd x\,xz)\beta')\lbd x\lbd y(k)(y)(\alpha)x
\force(p\forcec{\mathcal X}n)$.\\
Thus, let \ $\tau\in\C[p\et q]$ and $\pi\in{\mathcal X}^+(q,n)$. We must show:\\
$((\xi\lbd x\,x\tau)\beta')\lbd x\lbd y(\kk_\pi)(y)(\alpha)x\star\pi\in\bbot$.
But, we have \
$\lbd x\,x\tau\force\neg\neg\C[p\et q]$,\\
$\beta'\force p\et q\sqle p$ (lemma~\ref{p_et_q_le_p}) and
$\xi\force(\neg\C[p\et q]\to\neg\C[p]),p\et q\sqle p\,,
(p\et q\forcec\pm{\mathcal X}n)\to\bot$~; thus:\\
$(\xi\lbd x\,x\tau)\beta'\force((p\et q\forcec\pm{\mathcal X}n)\to\bot)$.
Therefore, it is sufficient to show:\\
$\lbd x\lbd y(\kk_\pi)(y)(\alpha)x\force(p\et q\forcec\pm{\mathcal X}n)$, i.e.:\\
$\lbd x\lbd y(\kk_\pi)(y)(\alpha)x\force\pt r(\C[(p\et q)\et r],\,r\forcec{\mathcal X}n\to p\et q\forcec{\mathcal X}n)$.
In fact, we show:\\
$\lbd x\lbd y(\kk_\pi)(y)(\alpha)x\force\pt r(\C[(p\et q)\et r],\,r\forcec
{\mathcal X}n\to\bot)$.\\
Thus, let \ $\upsilon\in\C[(p\et q)\et r]$ \ and \ $\eta\force(r\forcec{\mathcal X}n)$. We must show:\\
$(\kk_\pi)(\eta)(\alpha)\upsilon\star\rho\in\bbot$ for all $\rho\in\Pi$, i.e. \
$(\eta)(\alpha)\upsilon\star\pi\in\bbot$. But, we have  \ $(\alpha)\upsilon\in\C[r\et q]$,\\
therefore \ $(\eta)(\alpha)\upsilon\force{\mathcal X}^+(q,n)$, hence the result, since $\pi\in{\mathcal X}^+(q,n)$.

\smallskip\noindent
$\bullet$~~It follows that \
$(\lbd h(\xi II)\lbd x\lbd y\,h)\lbd z(\Ccc)\lbd k((\xi\lbd x\,xz)\beta')\lbd x\lbd y(k)(y)(\alpha)x\force\bot$\\
i.e. \ ({\sf dse})$\xi\force\bot$, which completes the proof.
\qed

\begin{lem}\label{p_et_q_le_p}
Let $\beta::(p\et q)\et r\Fl p\et r$. Then $\lbd x\lbd y(x)(\beta)y\force\pt p\pt q((p\et q)\sqle p)$.
\end{lem}

\proof
This formula is written \ $\pt p\pt q\pt r(\neg\C[p\et r],\C[(p\et q)\et r]\to\bot)$.\\
Therefore, let \ $\xi\force\neg\C[p\et r],\tau\in\C[(p\et q)\et r]$, thus $\beta\tau\in\C[p\et r]$ \ and \
$(\xi)(\beta)\tau\force\bot$.\\
Thus, we obtain \ $\lbd x\lbd y(x)(\beta)y\star\xi\ps\tau\ps\pi\in\bbot$ for every $\pi\in\Pi$.
\qed

\smallskip\noindent
We propose now to apply the countable downward chain condition to the binary predicate $\Phi(x,y)$.
Lemmas~\ref{ccd1} and~\ref{ccd2} show that the first two hypothesis of the c.d.c.
are realized by {\sf cdc1} and {\sf cdc2}. The third one is given by lemma~\ref{ccd3}
below.

\begin{lem}\label{ccd3}
There exist two proof-like terms \ {\sf cdc3} \ and \ {\sf for}  such that:\\
\emph{\phantom ii)}~~{\sf cdc3}$\force\pt n\inde\pt p\pt q(\Phi(n,p),\Phi(sn,q)\to q\sqle p)$.\\
\emph{ii)}~~{\sf for}$\force\pt n\inde\pt q(\Phi(sn,q)\to q\forcec\pm{\mathcal X}n)$.
\end{lem}

\proof
By lemma~\ref{rec_Phi}(iii), we have:\\
{\sf rec}$\force\pt k\inde(\vec{H}(n,p,k),\Phi(n,p)\to\Phi(sn,f(n,p,k)))$.
Using {\sf cdc2} (lemma~\ref{ccd2}), we get:\\
$\force\pt k\inde(\vec{H}(n,p,k),\Phi(n,p),\Phi(sn,q)\to q=f(n,p,k))$.\\
Now, $\vec{H}(n,p,k)$ is a sequence of four formulas, the last two of which are:\\
$f(n,p,k)\sqle p$ \ and \ $f(n,p,k)\forcec\pm{\mathcal X}n$.\\
\phantom ii)~It follows first that \ $\force\pt k\inde(\vec{H}(n,p,k),\Phi(n,p),\Phi(sn,q)\to q\sqle p)$.\\
Hence the result, since we have \
{\sf dse0}$\force\ex k\inde\{\vec{H}(n,p,k)\}$ (lemma~\ref{H_existe}).\\
ii)~It follows also that \ $\force\pt k\inde(\vec{H}(n,p,k),\Phi(n,p),\Phi(sn,q)\to q\forcec
\pm{\mathcal X}n)$.\\
Thus, we obtain \ $\force\pt n\inde\pt q(\Phi(sn,q)\to q\forcec\pm{\mathcal X}n)$ since we have \
{\sf cdc1}$\force\pt n\inde\ex p\,\Phi(n,p)$ (lemma~\ref{ccd1}) and \
{\sf dse0}$\force\pt n\pt p\ex k\inde\{\vec{H}(n,p,k)\}$ (lemma~\ref{H_existe}).
\qed

\smallskip\noindent
We can now apply the c.d.c. to the predicate $\Phi(x,y)$, which gives a proof-like term
{\sf cdc0} such that \ {\sf cdc0}$\force\ex p'\{\vec{\Omega}(n,p,p')\}$ \ with :\\
$\vec{\Omega}(n,p,p')\equiv\{\pt n\inde\pt p(\Phi(n,p)\to p'\sqle p),\;
\pt n\inde\pt p(\Phi(n,p),\neg\C[p]\to\bot),\neg\C[p']\to\bot\}$.

\smallskip\noindent
Therefore, in order to complete the proof of theorem~\ref{pX_decide}, it is sufficient to
find proof-like terms \ {\sf dec0,dec1,dec2} \ such that:

\smallskip\noindent
{\sf dec0}$\force\pt p'(\vec{\Omega}(n,p,p'),\neg\C[p_0],\C[p']\to\bot)$~;\\
{\sf dec1}$\force\pt p'(\vec{\Omega}(n,p,p')\to p'\sqle p_0)$~;\\
{\sf dec2}$\force\pt p'(\vec{\Omega}(n,p,p')\to\pt n\inde(p'\forcec\pm{\mathcal X}n))$.

\smallskip\noindent
Thus, let \ $\omega_0,\omega_1\in\Lbd$ \ be such that:\\
$\omega_0\force\pt n\inde\pt p(\Phi(n,p)\to p'\sqle p)$ \ and \
$\omega_1\force\pt n\inde\pt p(\Phi(n,p),\neg\C[p]\to\bot),\neg\C[p']\to\bot$

\smallskip\noindent
Applying lemma~\ref{rec_Phi}(i) with  $n=0,p=p_0$, we obtain \
$(\omega_0)\lbd x\lbd y\,y\force p'\sqle p_0$.\\
Therefore, we can take \ {\sf dec1} $=\lbd a\lbd b(a)\lbd x\lbd y\,y$.

\begin{lem}\label{ccd4}
{\sf cdc4}$\force(\C[p_0]\to\pt n\inde\pt p(\Phi(n,p),\neg\C[p]\to\bot))$\\
where \ {\sf cdc4}$\,=
\lbd a\lbd b\lbd c((b\lbd x_0\lbd x_1\lbd x_2\lbd x_3\lbd x\lbd y(x)(x_1)y)\lbd x\,xa)c$.
\end{lem}

\proof
Let \ $\tau\in\C[p_0]$, $\xi\force\Phi(n,p)$ and $\eta\force\neg\C[p]$.\\
Making \ $X(x,y)\equiv\neg\neg\C[y]$ in the definition de $\Phi$, we get:\\
%\nopagebreak
$\xi\force\pt n'\pt p'\pt k\inde
(G[n',p',k],\vec{F}[n',p',f(n',p',k)],\neg\neg\C[p']\to\neg\neg\C[f(n',p',k)]),\\
\hspace*{\fill}\neg\neg\C[p_0],\neg\C[p]\to\bot$.\\
We have \ $\lbd x(x)\tau\force\neg\neg\C[p_0]$.\\
Moreover, since \ $\vec{F}[n',p',q]\equiv
\{(\neg\C[q]\to\neg\C[p']),(q\sqle p'),q\forcec\pm{\mathcal X}n\}$, we easily get:\\
$\lbd x_0\lbd x_1\lbd x_2\lbd x_3\lbd x\lbd y(x)(x_1)y\force\\
\hspace*{\fill}\pt n'\pt p'\pt k\inde(G[n',p',k],\vec{F}[n',p',f(n',p',k)],
\neg\neg\C[p']\to\neg\neg\C[f(n',p',k)])$.\\
It follows that \
$((\xi\lbd x_0\lbd x_1\lbd x_2\lbd x_3\lbd x\lbd y(x)(x_1)y)\lbd x(x)\tau)\eta\force\bot$,
i.e. \ ({\sf cdc4})$\tau\xi\eta\force\bot$.
\qed

\smallskip\noindent
From lemma~\ref{ccd4}, we immediately deduce \
$\lbd x(\omega_1)(${\sf cdc4}$)x\force\C[p_0],\neg\C[p']\to\bot$.\\
Therefore, we can put \ {\sf dec0} $=\lbd a\lbd b\lbd x(b)(${\sf cdc4}$)x$.

\begin{lem}\label{force_inf}\ \\
\emph{\phantom ii)}~~{\sf lef0}$\force\pt p\pt q(p\forcec{\mathcal X}n,\,q\sqle p\to q\forcec{\mathcal X}n)$ \ with  \
{\sf lef0}$\,=\lbd x\lbd y\lbd z(\Ccc)\lbd k((y)\lbd u(k)(x)u)z$.\\
\emph{ii)}~~{\sf lef1}$\force\pt p\pt q(p\forcec\pm{\mathcal X}n,\,q\sqle p\to q\forcec\pm{\mathcal X}n)$ \ with \\
{\sf lef1}$\,=\lbd x\lbd y\lbd z\lbd u((${\sf lef0}$)(\Ccc)\lbd h((y)\lbd v(h)(x)vu)z$.
\end{lem}
\proof{\ }\ \\
\phantom ii) This is immediate, if we write explicitly the formulas:\\
$p\forcec{\mathcal X}n\equiv\pt r(\C[p\et r]\to{\mathcal X}^+(r,n))$~;\\
$q\sqle p\equiv\pt r(\neg\C[p\et r]\to\neg\C[q\et r])$~;\\
$q\forcec{\mathcal X}n\equiv\pt r(\C[q\et r]\to{\mathcal X}^+(r,n))$.\\
We declare \ $x:p\forcec{\mathcal X}n$, \ $y:q\sqle p$, \ $z:\C[q\et r]$, \ $k:\neg{\mathcal X}^+n$.\\
ii)~~We write down the formulas:\\
$p\forcec\pm{\mathcal X}n\equiv\pt r(\C[p\et r],r\forcec{\mathcal X}n\to p\forcec{\mathcal X}n)$~;\\
$q\sqle p\equiv\pt r(\neg\C[p\et r]\to\neg\C[q\et r])$~;\\
$q\forcec\pm{\mathcal X}n\equiv\pt r(\C[q\et r],r\forcec{\mathcal X}n\to q\forcec{\mathcal X}n)$.\\
We declare \ $x:p\forcec\pm{\mathcal X}n$, \ $y:q\sqle p$, \ $z:\C[q\et r]$, \ $u:r\forcec{\mathcal X}n$, \
$v:\C[p\et r]$, \ $h:\neg(p\force{\mathcal X}n)$.
\qed

\smallskip\noindent
By means of lemmas~\ref{ccd3}(ii) and~\ref{force_inf} and also \
$\omega_0\force\pt n\inde\pt p(\Phi(n,p)\to p'\sqle p)$, we obtain:\\
$\lbd n\lbd x((${\sf lef1}$)(${\sf for}$)nx)(\omega_0)nx
\force\pt n\inde\pt q(\Phi(sn,q)\to p'\forcec\pm{\mathcal X}n)$.\\
But, we have \ {\sf cdc1}$\force\pt n\inde\ex p\,\Phi(n,p)$ (lemma~\ref{ccd1})~; it follows that:\\
$\lbd n(\Ccc)\lbd k((${\sf cdc1}$)(s)n)\lbd x(k)((${\sf lef1}$)(${\sf for}$)nx)(\omega_0)nx
\force\pt n\inde(p_{\mathcal X}\forcec\pm{\mathcal X}n)$.\\
Thus, we can put \
{\sf dec2} $=\lbd a\lbd b\lbd n(\Ccc)\lbd k((${\sf cdc1}$)(s)n)\lbd x(k)((${\sf lef1}$)(${\sf for}$)nx)(a)nx$.

\smallskip\noindent
This completes the proof of theorem~\ref{pX_decide}.
\qed

\section{The ultrafilter axiom on \texorpdfstring{$\NN$}{N}}\noindent
Let us consider a standard realizability algebra ${\mathcal A}$ and a ${\mathcal A}$-model ${\mathcal M}$ in
which the individual set (which is also the set of conditions) is $P={\mathcal P}(\Pi)^{\NN}$.\\
The binary relation $\eps$ is defined by \ $\|n\eps p\|=p(n)$ if $n\in\NN$~;
otherwise \ $\|n\eps p\|=\vide$.\\
$\1$ is defined by \ $\1(n)=\vide$ for every $n\in\NN$~;\\
$\et$ is defined by \ $\|n\eps(p\et q)\|=\|n\eps p\land n\eps q\|$ for every $n\in\NN$.

\subsubsection*{The axiom of representation of predicates on $\NN$ (RPN)}\noindent
We define the following recursive function of arity $k$, denoted by $(n_1,\ldots,n_k)$ (coding of $k$-uples): \
$(n_1,n_2)=n_1+(n_1+n_2)(n_1+n_2+1)/2$~; $(n_1,\ldots,n_{k+1})=((n_1,\ldots,n_k),n_{k+1})$.

\begin{prop}\label{rep_pred}
$\force\pt X\ex x\pt y_1\indi\ldots\pt y_k\indi((y_1,\ldots,y_k)\eps x\dbfl X(y_1,\ldots,y_k))$ where $X$
is a predicate variable of arity $k$.
\end{prop}

\proof
Let \ ${\mathcal X}:P^k\to{\mathcal P}(\Pi)$ be a predicate of arity $k$. We define \ $a\in P$ \ by putting:\\
$a(n)={\mathcal X}(n_1,\ldots,n_k)$ for $n\in\NN$, $n=(n_1,\ldots,n_k)$. Then, we have immediately:\\
$I\force\pt y_1\inde\ldots\pt y_k\inde((y_1,\ldots,y_k)\eps a\to{\mathcal X}(y_1,\ldots,y_k))$ \ and\\
$I\force\pt y_1\inde\ldots\pt y_k\inde({\mathcal X}(y_1,\ldots,y_k)\to(y_1,\ldots,y_k)\eps a)$.\\
It follows that:\\
$\lbd x(x)I\force\pt X\ex x\pt y_1\inde\ldots\pt y_k\inde
((y_1,\ldots,y_k)\eps x\to X(y_1,\ldots,y_k))$ \ and\\
$\lbd x(x)I\force\pt X\ex x\pt y_1\inde\ldots\pt y_k\inde
(X(y_1,\ldots,y_k)\to(y_1,\ldots,y_k)\eps x)$.

\smallskip\noindent
Then, it suffices to apply theorem~\ref{mm1}.
\qed

\subsubsection*{The comprehension scheme for $\NN$ (CSN)}\noindent
Let \ $F[y,x_1,\ldots,x_k]$ \ be a formula the free variables of which are taken among
$y,x_1,\ldots,x_k$. We define a $k$-ary function \ $g_F:P^k\to P$, in other words \
$g_F:P^k\fois\NN\to{\mathcal P}(\Pi)$ \ by putting \
$g_F(p_1,\ldots,p_k)(n)=\|F[n,p_1,\ldots,p_k]\|$ for every $n\in\NN$.

\begin{prop}\label{comp_ent}
We have \ $\force\pt x_1\ldots\pt x_k\pt y\indi(y\eps g_F(x_1,\ldots,x_k)\dbfl F[y,x_1,\ldots,x_k])$ for
every formula \ $F[y,x_1,\ldots,x_k]$.
\end{prop}

\proof
Indeed, we have trivially:\\
$I\force \pt x_1\ldots\pt x_k\pt y\inde(y\eps g_F(x_1,\ldots,x_k)\to F[y,x_1,\ldots,x_k])$ \ and\\
$I\force \pt x_1\ldots\pt x_k\pt y\inde(F[y,x_1,\ldots,x_k]\to y\eps g_F(x_1,\ldots,x_k))$.

\smallskip\noindent
Then, it suffices to apply theorem~\ref{mm1}.
\qed

\smallskip\noindent
{\small{\bfseries Remark.}\\
The binary function symbol $\et$ is obtained by applying CSN to the formula $y\eps x_1\land y\eps x_2$.} 

\subsection*{The generic model}\noindent
We denote by $\C[x]$ the formula $\pt m\indi\ex n\indi(m+n)\eps x$, which says that the set~$x$ of
integers is infinite. The predicate $\C$ is defined by this formula: for every $p\in P$,
$|\C[p]|$ is, by definition, the set \ $\{\tau\in\Lbd;\;\tau\force\C[p]\}$.\\
It follows that the condition \ $\gamma::t(p_1,\ldots,p_n)$ $\Fl$ $u(p_1,\ldots,p_n)$ is written as:\\
$\lbd x\,\gamma x\force\pt p_1\ldots\pt p_n(\C[t(p_1,\ldots,p_n)]\to\C[u(p_1,\ldots,p_n)])$.

\smallskip\noindent
Therefore, in order to complete the definition of the algebra ${\mathcal B}$ (and of the
${\mathcal B}$-model ${\mathcal N}$), it remains to find proof-like terms $\alpha_0,\alpha_1,\alpha_2,\beta_0,\beta_1,\beta_2$ such that:

\smallskip\noindent
$\alpha_0\force\pt p\pt q\pt r(\C[(p\et q)\et r]\to\C[p\et(q\et r)])$~; \ $\alpha_1\force\pt p(\C[p]\to\C[p\et\1])$~;\\
$\alpha_2\force\pt p\pt q(\C[p\et q]\to\C[q])$~; \ $\beta_0\force\pt p(\C[p]\to\C[p\et p])$~; \
$\beta_1\force\pt p\pt q(\C[p\et q]\to\C[q\et p])$~;\\
$\beta_2\force\pt p\pt q\pt r\pt s(\C[((p\et q)\et r)\et s]\to\C[(p\et(q\et r))\et s])$.

\smallskip\noindent
Now, we easily have, in  natural deduction:\\
$\vdash\theta:\pt n(n\eps x\to n\eps x')\to(\C[x]\to\C[x'])$ \ with  \
$\theta=\lbd f\lbd u\lbd m\lbd h(um)\lbd n\lbd x(hn)(f)x$.\\
Therefore, by theorem~\ref{adequat} (adequacy lemma), we can put \ $\alpha_i=\theta\alpha^{*}_i$ \
and \ $\beta_i=\theta\beta_i^*$, with proof-like terms $\alpha_i^*,\beta_i^*(0\le i\le 2)$ such that:\\
$\vdash\alpha_0^*:\pt X\pt Y\pt Z\{(X\land Y)\land Z\to X\land(Y\land Z)\}$~; \
$\vdash\alpha_1^*:\pt X\{X\to X\land\top\}$~; \ $\vdash\alpha_2^*:\pt X\pt Y\{X\land Y\to Y\}$~; \
$\vdash\beta_0^*:\pt X\{X\to X\land X\}$~; \ $\vdash\beta_1^*:\pt X\pt Y\{X\land Y\to Y\land X\}$~;\\
$\vdash\beta_2^*:\pt X\pt Y\pt Z\pt U\{((X\land Y)\land Z)\land U\to(X\land(Y\land Z))\land U\}$.

\subsubsection*{The countable downward chain condition}\noindent
In this section, we show the:

\begin{thm}\label{ccd_ver1}\ \\
The forcing structure $\{\C,\et,\1\}$ satisfies the countable downward chain condition in ${\mathcal M}$.
\end{thm}\noindent
{\small{\bfseries Remark.} The proof of this theorem is a formalization of the following simple result:\\
The set of infinite subsets of $\NN$ with the preorder ``$p\sqle q$ $\Dbfl$ $p\setminus q$ is finite'',
satisfies the countable downward chain condition.\\
The proof is as follows: let $p_n$ be a decreasing sequence for this preorder~; put $h_n=\bigcap_{i\le n}p_i$,
$k_n=$ the first element of $h_n$ which is \ $\ge n$, and consider $\{k_n\;;\,n\in\NN\}$ which is an infinite subset
of $\NN$.}

\smallskip
\proof
We have to find a proof-like term {\sf cdc} such that:\\
{\sf cdc}$\force\pt X\ex x\{\pt n\inde\ex p\,X(n,p),\pt n\inde\pt p\pt q(X(n,p), X(n,q)\to p=q),\\
\hspace*{4.2em}\pt n\inde\pt p\pt q(X(n,p),X(sn,q)\to q\sqle p)\to\\
\hspace*{4.2em}\pt n\inde\pt p(X(n,p)\to x\sqle p)
\land(\pt n\inde\pt p(X(n,p)\to\C[p])\to\C[x])\}$\\
where $p\sqle q$ is the formula $\pt r(\C[p\et r]\to\C[q\et r])$.

\smallskip\noindent
By theorem~\ref{mm1}, this amounts to find a proof-like term {\sf cdc'} such that:

\smallskip\noindent
{\sf cdc'}$\force\pt X\ex x\{\pt n\indi\ex p\,X(n,p),\pt n\indi\pt p\pt q(X(n,p), X(n,q)\to p=q),\\
\hspace*{4.2em}\pt n\indi\pt p\pt q(X(n,p),X(sn,q)\to q\sqle p)\to\\
\hspace*{4.2em}\pt n\indi\pt p(X(n,p)\to x\sqle p)
\land(\pt n\indi\pt p(X(n,p)\to\C[p])\to\C[x])\}$.

\smallskip\noindent
By theorem~\ref{adequat} (adequacy lemma), given a formula $F$, we can use the following method
to show $\force F$:\\
First, show $\force A_1,\ldots,\force A_k$, then
show \ $A_1,\ldots,A_k\vdash F$ \ by means of the rules of classical second order natural deduction
(which contains the comprehension scheme), and of the following axioms which are realized by proof-like terms in the ${\mathcal A}$-model ${\mathcal M}$:

\smallskip\noindent
$\bullet$~~$t\ne u$ for all closed terms $t,u$ \ which take distinct values in ${\mathcal M}$.\\
$\bullet$~~$\pt x_1\indi\ldots\pt x_k\indi(t(x_1,\ldots,x_k)=u(x_1,\ldots,x_k))$
for all the equations between terms which are true in $\NN$.\\
$\bullet$~~The foundation scheme (SCF\/, see theorem~\ref{bien_fonde}ii) which consists of the formulas:\\
$\pt X_1\ldots\pt X_k\{\pt x\indi[\pt y\indi(X_1y,\ldots,X_ky\to f(y,x)\ne1),X_1x,\ldots,X_kx\to\bot]\\
\hspace*{\fill}\to\pt x\indi(X_1x,\ldots,X_kx\to\bot)\}$\\
where \ $f:P^2\to P$ \ is such that the relation \ $f(y,x)=1$ \ is well founded on $\NN$.\\
$\bullet$~~The axiom of choice scheme for individuals (ACI, see theorem~\ref{ACI}) which consists of
the formulas \ \ $\pt\vec{x}(\pt y\indi F(\vec{x},f_F(\vec{x},y))\to\pt y\,F(\vec{x},y))$~;\\
$\vec{x}=(x_1,\ldots,x_k)$ is a finite sequence of variables, $\pt\vec{x}\pt y\indi F$ is an
arbitrary closed formula, and $f_F$ is a function symbol of arity $k+1$.\\
$\bullet$~~The axiom of representation of predicates on $\NN$ (RPN, see proposition~\ref{rep_pred})
which consists of the formulas \ \ $\pt X\ex x\pt\vec{y}\indi((y_1,\ldots,y_k)\eps x\dbfl X\vec{y})$~;\\
$\vec{y}=(y_1,\ldots,y_k)$ is a sequence of $k$ variables and $X$ is a predicate variable of arity $k$.\\
$\bullet$~~The comprehension scheme for integers (CSN, see proposition~\ref{comp_ent}),
which consists of the formulas \ \ $\pt\vec{x}\pt y\indi(y\eps g_F(\vec{x})\dbfl F[y,\vec{x}])$~;\\
$\vec{x}=(x_1,\ldots,x_k)$ is a sequence of $k$ variables, $\pt\vec{x}\pt y\indi F$ is an arbitrary
closed formula, and $g_F$ is a function symbol of arity $k$.

\begin{lem}
$\vdash\pt p\pt q(p\sqle q\dbfl\ex m\indi\pt n\indi(n+m\eps p\to n+m\eps q))$.
\end{lem}

\proof
We apply the CSN to the formula $F[y,x]\equiv y\neps x$~; thus, we obtain:

\smallskip\noindent
\centerline{$\vdash\pt x\pt y\indi(y\eps\neg x\dbfl y\neps x)$}

\noindent
using the notation $\neg x$ for $g_F(x)$.

\smallskip\noindent
We have \ $p\sqle q\equiv\pt r(\C[p\et r]\to \C[q\et r])$ \ and therefore \
$p\sqle q\;\vdash\C[p\et\neg q]\to \C[q\et\neg q]$.\\
But, we have \ $\C[q\et\neg q]\vdash\pt m\indi\ex n\indi(m+n\eps q\land m+n\neps q)\;\vdash\bot$, \
and thus:\\
$p\sqle q\;\vdash\neg\C[p\et\neg q]$, \ that is \
$\vdash p\sqle q\to\ex m\indi\pt n\indi\neg(m+n\eps p\land\neg(m+n\eps q))$.

\smallskip\noindent
Conversely, from the hypothesis:\\
$\pt n{'}\,\indi(m'+n'\eps p\to m'+n'\eps q),\pt m\indi\ex n\indi(m+n\eps p\land m+n\eps r)$, \
we deduce:\\
$\pt m\indi\ex n\indi((m'+m)+n\eps p\land(m'+m)+n\eps r)$, \ then:\\
$\pt m\indi\ex n\indi(m+(m'+n)\eps q\land m+(m'+n)\eps r)$ \ then:\\
$\pt m\indi\ex n\indi(m+n\eps q\land m+n\eps r)$. \ Therefore:\\
$\pt n{'}\,\indi(m'+n'\eps p\to m'+n'\eps q)\;\vdash\;\C[p\et r]\to\C[q\et r]$ \ and thus:\\
$\ex m'\pt n{'}\,\indi(m'+n'\eps p\to m'+n'\eps q)\;\vdash\;\C[p\et r]\to\C[q\et r]$.
\qed

\smallskip\noindent
Applying RPN and the comprehension scheme, we obtain \ $\force\pt X\ex h\,D(h,X)$ \
with:\\
$D(h,X)\equiv\pt k\indi\pt n\indi((k,n)\eps h\dbfl\pt q\pt i\indi(i\le n,X(i,q)\to k\eps q))$.\\
{\small{\bfseries Remark.} The intuitive meaning of \ $D(h,X)$ is:	~$h$ is the individual associated
with the decreasing sequence of conditions $X'$, the $n$-th term of which is the intersection of the
$n$ first terms of the sequence~$X$.}

\smallskip\noindent
We apply CSN to the formula $F(k,n,h)\equiv(k,n)\eps h$. Thus, we obtain:\\
$\vdash\pt n\pt h\pt k\indi\pt n(k\eps g_F(n,h)\dbfl(k,n)\eps h)$.\\
We shall use the notation $h_n$ for $g_F(n,h)$. Therefore, we have:

\smallskip\noindent
\centerline{$\vdash\pt n\pt h\pt k\indi(k\eps h_n\dbfl (k,n)\eps h)$.}

\smallskip\noindent
and it follows that:\\
\centerline{$D(h,X)\vdash
\pt k\indi\pt n\indi(k\eps h_n\dbfl\pt q\pt i\indi(i\le n,X(i,q)\to k\eps q))$}

\smallskip\noindent
We put \ $\Phi(k,h,n)\equiv\ex i\indi\{\pt j\indi(j+n\eps h_n\to(j<i)\ne1),\,i+n\eps h_n,\,k=i+n\}$.\\
{\small{\bfseries Remark.}
The intuitive meaning of $\Phi(k,h,n)$ is: ``~$k$ is the first element of $h_n$ which is $\ge n$~''.}\\
We apply CSN to the formula $F(k,h)\equiv\ex n\indi\,\Phi(k,h,n)$. Thus, we obtain:\\
$\vdash\pt h\pt k\indi(k\eps g_F(h)\dbfl\ex n\indi\,\Phi(k,h,n))$.\\
We shall use the notation \ $\inf(h)$ for $g_F(h)$. Therefore, we have:

\smallskip\noindent
\centerline{$\vdash\pt h\pt k\indi(k\eps\inf(h)\dbfl\ex n\indi\,\Phi(k,h,n))$.}

\smallskip\noindent
The hypothesis of the c.d.c. are:

\smallskip\noindent
$H_0[X]\equiv\pt n\indi\ex p\,X(n,p)$~;\\
$H_1[X]\equiv\pt n\indi\pt p\pt q(X(n,p), X(n,q)\to p=q)$~;\\
$H_2[X]\equiv\pt n\indi\pt p\pt q(X(n,p),X(sn,q)\to q\sqle p)$~;\\
$H_3[X]\equiv\pt n\indi\pt p(X(n,p)\to\C[p])$.

\smallskip\noindent
We put \ $\vec{H}[X]\equiv\{H_0[X],H_1[X],H_2[X],H_3[X]\}$ \ and \
$\vec{H}_*\![X]=\{H_0[X],H_1[X],H_2[X]\}$.

\smallskip\noindent
Thus, it is sufficient to show:\\
$D(h,X),\vec{H}_*\![X]\,\vdash\pt n\indi\pt p(X(n,p)\to\inf(h)\sqle p)$ \ and\\
$D(h,X),\vec{H}[X]\,\vdash\,\C[\inf(h)]$.

\smallskip\noindent
{\bfseries Notation.} The formula \ $\pt n\indi(n\eps p\to n\eps q)$ \ is denoted by \ $p\subseteq q$.

\begin{lem}\label{DhH1}
$D(h,X)\vdash\pt m\indi\pt n\indi(h_{n+m}\subseteq h_n)$.
\end{lem}

\proof
This formula is written \ $\pt m\indi\pt n\indi\pt k\indi(k\eps h_{n+m}\to k\eps h_n)$. Now, we have:

\smallskip\noindent
$D(h,X)\vdash\pt m\indi\pt n\indi\pt k\indi(k\eps h_{n+m}\to\pt q\pt i\indi(i\le n+m,X(i,q)\to
k\eps q))$~;\\
$\vdash\pt m\indi\pt n\indi\pt k\indi[\pt q\pt i\indi(i\le n+m,X(i,q)\to k\eps q)\to
\pt q\pt i\indi(i\le n,X(i,q)\to k\eps q)]$:\\
$D(h,X)\vdash\pt m\indi\pt n\indi\pt k\indi(\pt q\pt i\indi(i\le n,X(i,q)\to k\eps q)\to k\eps h_n)$.
\qed

\begin{lem}\label{DhH2_0}
$D(h,X),H_0[X],H_1[X]\vdash\pt n\indi\pt k\indi\pt p(X(sn,p),\,k\eps p,\,k\eps h_n\to k\eps h_{sn})$.
\end{lem}

\proof
We have \ $D(h,X),$ int$(k),$ int$(n)\vdash\pt p\pt i\indi(i\le sn,X(i,p)\to k\eps p)\to k\eps h_{sn}$.\\
But, we have \ \ int$(n),$ int$(i),\,i\le sn\,\vdash\,i\le n\lor i=sn$, and therefore:\\
int$(n),\,\pt p\pt i\indi(i\le n,X(i,p)\to k\eps p),\,\pt p(X(sn,p)\to k\eps p)\;\vdash\;\\
\hspace*{\fill}\pt p\pt i\indi(i\le sn,X(i,p)\to k\eps p)$.\\
It follows that:\\
$D(h,X),$ int$(k),$ int$(n)\vdash
\pt p\pt i\indi(i\le n,X(i,p)\to k\eps p),\pt p(X(sn,p)\to k\eps p)\to k\eps h_{sn}$, \ i.e.:\\
$D(h,X),$ int$(k),$ int$(n)\vdash k\eps h_n,\pt p(X(sn,p)\to k\eps p)\to k\eps h_{sn}$. Therefore:\\
$D(h,X),$ int$(k),$ int$(n),H_0[X],H_1[X]\vdash\pt p(k\eps h_n,X(sn,p),k\eps p\to k\eps h_{sn})$.
\qed

\begin{lem}\label{DhH2}
$D(h,X),\vec{H}_*\![X]\vdash\pt n\indi\pt p(X(n,p)\to p\sqle h_n)$.
\end{lem}

\proof By recurrence on $n$. We must show:\\
$D(h,X),\vec{H}_*\![X],$ int$(n)\,\vdash\,
\pt p\ex m\indi\pt l\indi(X(n,p),l+m\eps p\to l+m\eps h_n)$.\\
For $n=0$, we have \ $D(h,X)\vdash\pt k\indi(\pt q(X(0,q)\to k\eps q)\to k\eps h_0)$.
Thus, it suffices to show:\\
$D(h,X),\vec{H}_*\![X]\,\vdash\,
\pt p\ex m\indi\pt l\indi\pt q(X(0,p),l+m\eps p,X(0,q)\to l+m\eps q)$,\\
which follows, in fact, from $H_1[X]$, that is \ $X(0,p),X(0,q)\to p=q$.\\
The recurrence hypothesis is $\pt p(X(n,p)\to p\sqle h_n)$~;\\
$H_2[X]$ is $\pt p\pt q(X(n,p),X(sn,q)\to q\sqle p)$~; $H_0[X]$ is $\ex p\,X(n,p)$.\\
Moreover, we have easily \ $q\sqle p,p\sqle r\vdash q\sqle r$. Thus, it follows that:\\
$\pt p(X(sn,p)\to p\sqle h_n)$, i.e. \ $\pt p\ex m\indi\pt l\indi(X(sn,p),l+m\eps p\to l+m\eps h_n)$.\\
Now, we have, by lemma~\ref{DhH2_0}:\\
$D(h,X),H_0[X],H_1[X]\,\vdash X(sn,p),\,l+m\eps p,\,l+m\eps h_n\to l+m\eps h_{sn}$.\\
Therefore, we have \ $\pt p\ex m\indi\pt l\indi(X(sn,p),l+m\eps p\to l+m\eps h_{sn})$ \ that is:\\
$\pt p(X(sn,p)\to p\sqle h_{sn})$, which is the desired result.
\qed

\begin{lem}\label{DhH2_1}
$D(h,X),\vec{H}(X)\vdash\pt n\indi\C[h_n]$.
\end{lem}

\proof
We have \ $\pt n\indi\pt p(X(n,p)\to\C[p])$ \ from $H_3$. Moreover, we have easily:\\
$\vdash\pt p\pt q(\C[p],p\sqle q\to\C[q])$. Thus, applying lemma~\ref{DhH2}, we obtain:\\
$D(h,X),\vec{H}(X)\vdash\pt n\indi\pt p(X(n,p)\to\C[h_n])$. Hence the result, from $H_0[X]$.
\qed

\begin{lem}\label{DhH3}
$D(h,X),\vec{H}[X]\vdash\pt n\indi\ex k\indi\Phi(k,h,n)$.
\end{lem}

\proof
By the foundation scheme (SCF), we have:\\
$\vdash\,\pt i\indi\{\pt j\indi(j+n\eps h_n\to(j\ppt i)\ne1),i+n\eps h_n\to\bot\}
\to\pt i\indi(i+n\eps h_n\to\bot)$.\\
But, we have \ $D(h,X),\vec{H}[X]\vdash\pt n\indi\C[h_n]$ (lemma~\ref{DhH2_1}), therefore:\\
$D(h,X),\vec{H}[X]\vdash\pt n\indi\ex i\indi i+n\eps h_n$. It follows that:\\
$D(h,X),\vec{H}[X]\vdash\pt n\indi\ex i\indi\{\pt j\indi(j+n\eps h_n\to(j\ppt i)\ne1),i+n\eps h_n\}$.
\qed

\begin{lem}\label{DhH4}
$D(h,X),\vec{H}[X]\;\vdash\;\C[\inf(h)]$.
\end{lem}

\proof
We have \ $\C[\inf(h)]\equiv\pt m\indi\ex i\indi(i+m\eps\inf(h))$.\\
Now, by definition of the function symbol \ $\inf$, we have:\\
$\vdash\pt h\pt k\indi(k\eps\inf(h)\dbfl\ex n\indi\Phi(k,h,n))$.\\
Therefore \ $\vdash\C[\inf(h)]\dbfl\pt m\indi\ex i\indi\ex n\indi\Phi(i+m,h,n)$.\\
By definition de $\Phi$, we have trivially \
$\vdash\pt n\indi\pt k\indi(\Phi(k,h,n)\to\ex i\indi(k=i+n))$.\\
Moreover, we have \ $D(h,X),\vec{H}[X]\;\vdash\;\pt n\indi\ex k\indi\Phi(k,h,n)$ (lemma~\ref{DhH3}).\\
Therefore \ $D(h,X),\vec{H}[X]\;\vdash\;\pt n\indi\ex i\indi\,\Phi(i+n,h,n)$, thus \
$D(h,X),\vec{H}[X]\;\vdash\;\C[\inf(h)]$.
\qed

\begin{lem}\label{DhH5}\ \\
$D(h,X),\vec{H}_*\![X]\vdash\pt h\pt k\indi\pt k{'}\indi
\pt n\indi\pt n{'}\indi(\Phi(k,h,n),\Phi(k',h,n'),k'>k\to n'>n)$.
\end{lem}
\proof
We have \ $\Phi(k,h,n)\equiv\ex i\indi\vec{\Psi}(k,h,n,i)$, with :\\
$\vec{\Psi}(k,h,n,i)\equiv\{\pt j\indi(j+n\eps h_n\to(j\ppt i)\ne1),\; i+n\eps h_n,\; k=i+n\}$.\\
Thus, we have to show:\\
$D(h,X),\vec{H}_*\![X],$ int$(k),$ int$(k'),$ int$(n),$ int$(n'),$ int$(i),$ int$(i')\,\vdash\,
\vec{\Xi}(h,k,n,i,k',n',i')\to\bot$\\
with \ $\vec{\Xi}(h,k,n,i,k',n',i')\equiv
\{\vec{\Psi}(k,h,n,i),\,\vec{\Psi}(k',h,n',i'),\,k'>k,\,n'\le n\}$ \ that is:\\
$\vec{\Xi}(h,k,n,i,k',n',i')\equiv\\
\{\pt j\indi(j+n\eps h_n\to(j\ppt i)\ne1),\; i+n\eps h_n,\; k=i+n,\\
\pt j{'}\,\indi(j'+n'\eps h_{n'}\to(j'\ppt i')\ne1),\; i'+n'\eps h_{n'},\; k'=i'+n',\\
k'>k,\,n'\le n\}$.\\
From \ $n'\le n$ \ and \ $k=i+n$, we deduce \ $n'\le k$, thus \ $k=j'+n'$.\\
From \ $k'>k$, we deduce \ $i'+n'>k$, and thus \ $j'<i'$.\\
Therefore, we have \ $j'+n'\neps h_{n'}$, i.e. \ $k\neps h_{n'}$. But, from \ $n'\le n$, we deduce \ $h_n\subseteq h_{n'}$
(lemma~\ref{DhH1}), thus \ $k\neps h_n$, which contradicts \ $i+n\eps h_n,\; k=i+n$.
\qed

\smallskip\noindent
By definition of $\Phi$, we have trivially \
$\vdash\pt n\indi\pt k\indi(\Phi(k,h,n)\to k\eps h_n)$.

\smallskip\noindent
By lemmas~\ref{DhH1} and~\ref{DhH5}, we get:\\
$D(h,X),\vec{H}_*\![X]\vdash\pt h\pt k\indi\pt k{'}\indi
\pt n\indi\pt n{'}\indi(\Phi(k,h,n),\Phi(k',h,n'),k'>k\to k'\eps h_n)$.\\
Lemma~\ref{DhH3} gives \ $\pt n\indi\ex k\indi\Phi(k,h,n)$. It follows that:\\
$D(h,X),\vec{H}_*\![X]\vdash\pt n\indi\ex k\indi\pt n{'}\,\indi\pt k{'}\,\indi
(\Phi(k',h,n'),k'>k\to k'\eps h_n)$,\\
and therefore \ $D(h,X),\vec{H}_*\![X]\vdash\pt n\indi(\inf(h)\sqle h_n)$.

\smallskip\noindent
But, we have trivially \ $D(h,X)\vdash\pt n\indi\pt k\indi\pt p(k\eps h_n,X(n,p)\to k\eps p)$.
Therefore, finally:\\
$D(h,X),\vec{H}_*\![X]\vdash\pt n\indi\pt p(X(n,p)\to\inf(h)\sqle p)$.

\smallskip\noindent
We have eventually obtained the desired proof-like term {\sf cdc'}, which completes the proof of theorem~\ref{ccd_ver1}.
\qed

\subsection*{The ultrafilter}\noindent
In the model ${\mathcal N}$, we have defined the \emph{generic ideal} ${\mathcal J}$, which is a
unary predicate, by putting: \ ${\mathcal J}(p)=\Pi\fois\{p\}$ \ for every $p\in P$.

\smallskip\noindent
By theorem~\ref{elem_gen}, we have:

\smallskip\noindent
\phantom{ii}i) $\fforce\neg{\mathcal J}(\1)$\\
\phantom iii) $\fforce\pt x(\neg\C[x]\to{\mathcal J}(x))$\\
iii) $\fforce\pt x\pt y({\mathcal J}(x\et y)\to{\mathcal J}(x)\lor{\mathcal J}(y))$\\
iv) $\fforce\pt x(\pt y(\neg\C[x\et y]\to{\mathcal J}(y))\to\neg{\mathcal J}(x))$\\
\phantom{i}v) $\fforce\pt x\pt y({\mathcal J}(x),y\sqle x\to{\mathcal J}(y))$

\smallskip\noindent
By theorem~\ref{premier_ordre}, we have \ $\force F$ $\Dbfl$ $\fforce F$ \ for every closed
first order formula $F$.

\smallskip\noindent
{\small{\bfseries Remark.} A ``first order'' formula contains quantifiers on the individuals
which, by means of the symbol $\eps$, represent the subsets of $\NN$. Therefore, it is a
\emph{second order} formula from the point of view of Arithmetic. But it contains no quantifier on sets of
individuals.}

\smallskip\noindent
By theorems~\ref{mm1} and~\ref{entMN}, we can use, in $F$, the quantifier $\pt x\indi$, since the quantifier $\pt x\inde$ is first order.

\smallskip\noindent
Therefore, we have:

\smallskip\noindent
\phantom{ii}vi)~$\fforce\C[x]\dbfl\pt m\indi\ex n\indi(m+n\eps x)$\\
\phantom ivii)~$\fforce y\sqle x\dbfl\ex m\indi\pt n\indi(m+n\eps y\to m+n\eps x)$\\
viii)~$\fforce\pt n\indi n\eps\1$~; $\fforce\pt x\pt y\pt n\indi(n\eps x\et y\dbfl n\eps x\land n\eps y)$

\smallskip\noindent
since all these formulas are first order.
Properties (i) to (viii) show that, in the ${\mathcal B}$-model ${\mathcal N}$, the following formula is
realized:\\
${\mathcal J}$ is a maximal non trivial ideal on the Boolean algebra of the subsets of $\NN$
\emph{which are represented by individuals.}

\smallskip\noindent
Now, by theorems~\ref{cons_reels} and~\ref{ccd_ver1}, the following formula is
realized in ${\mathcal N}$:\\
\emph{Every subset of $\NN$ is represented by an individual.}

\smallskip\noindent
Thus the following formula is realized in ${\mathcal N}$:\\
\emph{${\mathcal J}$ is a maximal non trivial ideal on the Boolean algebra of the subsets of $\NN$.}

\subsection*{Programs obtained from proofs}\noindent
Let $F$ be a formula of \emph{second order arithmetic}, that is to say a second order formula
every individual quantifier of which is restricted to $\NN$ and every second order quantifier
of which is restricted to ${\mathcal P}(\NN)$.\\
We associate with $F$, a \emph{first order} formula  $F^\dag$, defined by recurrence on $F$:

\smallskip\noindent
$\bullet$~~If $F$ is $t=u$, $F^\dag\equiv F$.\\
$\bullet$~~If $F$ is $Xt$, $F^\dag$ is $t\eps X^-$, where $X^-$ is an \emph{individual} variable
associated with the unary predicate variable $X$.\\
$\bullet$~~If $F$ is $A\to B$, $F^\dag$ is $A^\dag\to B^\dag$.\\
$\bullet$~~If $F$ is $\pt x\,A$, $F^\dag$ is $\pt x\indi\,A^\dag$.\\
$\bullet$~~If $F$ is $\pt X\,A$, $F^\dag$ is $\pt X^-\,A^\dag$.

\smallskip\noindent
We note that, if $F$ is a formula of \emph{first order arithmetic}, then $F^\dag$ is simply
the restriction $F^{\mbox{\small int}}$ of $F$ to the predicate int$(x)$. 

\smallskip\noindent
Let $F$ be a closed formula of second order arithmetic and let us consider a proof of~$F$, which
uses the axiom of dependent choice DC and the axiom UA of ultrafilter on $\NN$, written in the
following form, with a constant ${\mathcal J}$ of predicate:
``${\mathcal J}$ is a maximal non trivial ideal on ${\mathcal P}(\NN)$~''.\\
We can transform it immediately into a proof of $F^\dag$ if we add the axiom RPN of representation of predicates on $\NN$: \ $\pt X\ex x\pt y(y\eps x\dbfl Xy)$. Thus, we obtain:\\
$x:$ UA, $y:$ RPN, $z:$ DC$^\dag\vdash t[x,y,z]:F^\dag$.\\
Therefore, we have \ $\vdash u:\,$UA, RPN $\to G$ \ with  $u=\lbd x\lbd y\lbd z\,t[x,y,z]$ and $G\equiv$ DC$^\dag\to F^\dag$.\\
Thus, $G$ is a \emph{first order formula}.\\
In the previous section, we obtained proof-like terms \ $\theta,\theta'$ \ such that \
$(\theta,\1)\fforce UA$ \ and \ $(\theta',\1)\fforce$ RPN (theorems~\ref{cons_reels} and~\ref{ccd_ver1}).\\
Therefore, theorem~\ref{adequat_B} (adequacy lemma) gives
$(u^*,\1_u)(\theta,\1)(\theta',\1)\fforce G$, that is to say:\\
$(v,(\1_u\et\1)\et\1)\fforce G$ \ with  $v=((\ov{\alpha}_0)(\ov{\alpha}_0)u^*\theta)\theta'$.\\
By theorem~\ref{premier_ordre}, we thus have \
$\delta'_Gv\force\C[(\1_u\et\1)\et\1]\to G$, that is:\\
$\delta'_Gv\force\C[(\1_u\et\1)\et\1],$ DC$^\dag\to F$.\\
The axiom DC$^\dag$ is consequence of ACI (axiom of choice for individuals). Therefore,
by theorem~\ref{ACI}, we have a proof-like term \ $\eta_0\force$ DC$^\dag$.\\
Moreover, we have obviously a proof-like term $\xi_0\force\C[(\1_u\et\1)\et\1]$.\\
Thus, finally, we have $\delta'_Gv\xi_0\eta_0\force F$.\\
Then, we can apply to the program \ $\zeta=\delta'_Gv\xi_0\eta_0$ \ all the results obtained in the
framework of usual classical realizability. The case when $F$ is an arithmetical (resp. $\Pi_1^1$)
formula is considered in~\cite{krivine3}
(resp.~\cite{krivine4}).\\
Let us take two very simple examples:

\smallskip\noindent
If $F\equiv\pt X(X1,X0\to X1)$, we have \ $\zeta\star\kappa\ps\kappa'\ps\pi\succ\kappa\star\pi$
for all terms $\kappa,\kappa'\in\Lbd$ and every stack $\pi\in\Pi$.

\smallskip\noindent
If $F\equiv\pt m\indi\ex n\indi(\phi(m,n)=0)$, where $\phi$ is a function symbol, then for every $m\in\NN$, \ there exists $n\in\NN$ such that \ $\phi(m,n)=0$ \ and \
$\zeta\star\ul{m}\ps T\kappa\ps\pi\succ\kappa\star\ul{n}\ps\pi'$.\\
$T$ is the proof-like term for integer storage, given in theorem~\ref{mm1}(i).\\
$\pi,\kappa$ are arbitrary~; therefore, by taking a constant for $\kappa$, we obtain a program
which computes $n$ from $m$.

\section{Well ordering on \texorpdfstring{$\mathbb{R}$}{R}}\noindent
The ${\mathcal A}$-model ${\mathcal M}$ is the same as in the previous section: the set of individuals is
$P={\mathcal P}(\Pi)^{\NN}$. Recall that an element of $P$ is called sometimes an \emph{individual},
sometimes a \emph{condition}, depending on the context.

\smallskip\noindent
We put $(m,n)=m+(m+n)(m+n+1)/2$ (bijection of \ $\NN^2$ onto \ $\NN$). We define a \ binary function
$\gamma:P^2\to P$ by putting:\\
$\gamma(n,p)(i)=p(i,n)$ if $n\in\NN$~; $\gamma(n,p)$ is arbitrary (for instance $0$) if $n\notin\NN$.

\smallskip\noindent
{\bfseries Notation.} In the sequel, we shall write $p_n$ instead of $\gamma(n,p)$. Thus, it is the same
to give an individual \ $p$ \ or a sequence of individuals \ $p_n (n\in\NN$).\\
If $i,n\in\NN$, we have \ $\|(i,n)\eps p\|=\|i\eps p_n\|$.

\smallskip\noindent
We fix a well ordering \ $\trl$ \ on $P={\mathcal P}(\Pi)^{\NN}$, which is strict (i.e. $\pt x\neg(x\trl x)$)
and isomorphic to the cardinal $2^{\aleph_0}$:
every proper initial segment of $\trl$ is therefore of power $<2^{\aleph_0}$. We define a binary function,
denoted by \ $(p\trl q)$ \ by putting \ $(p\trl q)=1$ if $p\trl q$~; \ $(p\trl q)=0$ otherwise.\\
Since the relation $(p\trl q)=1$ is well founded on $P$, we have (theorem~\ref{bien_fonde}):\\
$\Y\force\pt X[\pt x(\pt y((y\trl x)=1\mapsto Xy)\to Xx)\to\pt x\,Xx]$\\
in the ${\mathcal A}$-model ${\mathcal M}$, but also in every ${\mathcal B}$-model ${\mathcal N}$.\\
We shall write, in abridged form, $y\trl x$ for $(y\trl x)=1$.\\
Thus, in ${\mathcal M}$ and ${\mathcal N}$, the relation $\trl$ is well founded but, in general, not total.\\
It is a strict order relation, in both models~; indeed we have immediately, in the model~${\mathcal M}$:
$I\force\pt x((x\trl x)\ne1)$~; \
$I\force\pt x\pt y\pt z((x\trl y)=1\mapsto((y\trl z)=1\mapsto(x\trl z)=1))$.\\
Since all these formulas are first order, by theorem~\ref{premier_ordre}, we have also,
in the model~${\mathcal N}$:\\
$\fforce\pt x((x\trl x)\ne1)$~; \
$\fforce\pt x\pt y\pt z((x\trl y)=1\mapsto((y\trl z)=1\mapsto(x\trl z)=1))$.

\smallskip\noindent
A condition $p\in P$ is also a sequence of individuals $p_k$. Intuitively, we shall consider it, as
``~the set of individuals $p_{k+1}$ for $k\eps p_0$~''~; we define accordingly the condition $\1$,
the formula $\C[p]$ which says that $p$ is a non trivial condition, and the binary operation \ $\et$.

\smallskip\noindent
$\1$ is the empty set, in other words \ $i\eps\1_0$ (i.e. $(i,0)\eps\1$) must be false. Therefore,
we put:\\
$\1(n)=\Pi$ for every $n\in\NN$.

\smallskip\noindent
A condition is non trivial if the set of individuals, which is associated with it, is totally ordered
by \ $\trl$. Therefore, we put:\\
$\C[p]\equiv\pt i\inde\pt j\inde(i\eps p_0,j\eps p_0\to E[p_{i+1},p_{j+1}])$ \ with :\\
$E[x,y]\equiv(x=y\lor x\trl y\lor y\trl x)$ \ that is \
$E[x,y]\equiv(x\ne y,(x\trl y)\ne1,(y\trl x)\ne1\to\bot)$.

\smallskip\noindent
The set associated with $p\et q$ is the union of the sets associated with $p$ and with $q$~;
therefore, we put:\\
$p\et q=r$ \ where $r_0$ is defined by: \ $\|2i\eps r_0\|=\|i\eps p_0\|$~; $\|2i+1\eps r_0\|=\|i\eps q_0\|$~;\\
$r_{j+1}$ is defined by: \ $r_{2i+1}=p_{i+1}$~; \ $r_{2i+2}=q_{i+1}$.

\smallskip\noindent
The notation \ $p\subset q$ \ means that the set associated with $q$ contains the one associated
with~$p$.\\
Therefore, we put:\\
$p\subset q\equiv\pt i\inde(i\eps p_0\to\ex j\inde\{j\eps q_0,p_{i+1}=q_{j+1}\})$.

\begin{lem}\label{subset_trans}\ \\
\emph{\phantom ii)}~~$\theta\force\pt p\pt q\pt r(p\subset q,q\subset r\to p\subset r)$ \ with \
$\theta=\lbd f\lbd g\lbd i\lbd x\lbd h(fix)\lbd j\lbd y(g)jyh$.\\
\emph{ii)}~~$\theta'\force\pt p\pt q\pt r(p\subset q\to p\et r\subset q\et r)$ \ with:\\
$\theta'=\lbd f\lbd i\lbd y\lbd u((ei)(u)iy)(((f)(d_2)iy)\lbd j(u)(d_0)j$\\
where $d_0,d_1,d_2,e$ are proof-like terms representing respectively the recursive functions:\\
$n\mapsto 2n$, $n\mapsto 2n+1$, $n\mapsto[n/2]$, $n\mapsto$ parity of $n$ ($e$ returns boolean values).
\end{lem}
\proof{\ }\ \\
\phantom ii)~~We suppose:\\
$f\force\pt i($ent$(i),i\eps p_0,\pt j($ent$(j),j\eps q_0\to p_{i+1}\ne q_{j+1})\to\bot)$~;\\
$g\force\pt j($ent$(j),j\eps q_0,\pt k($ent$(k),k\eps r_0\to q_{j+1}\ne r_{k+1})\to\bot)$~;\\
$x\force i\eps p_0$~; $h\force\pt k($ent$(k),k\eps r_0\to p_{i+1}\ne r_{k+1})$~; and we have \ $\ul{i}\in|$ent$(i)|$.\\
It follows that \ $f\ul{i}x\force\pt j($ent$(j),j\eps q_0\to p_{i+1}\ne q_{j+1})\to\bot$.\\
Suppose that \ $y\force j\eps q_0$ \ and let \ $\ul{j}\in|$ent$(j)|$.

\smallskip\noindent
If $p_{i+1}=q_{j+1}$, then \ $g\ul{j}yh\force\bot$~; \ therefore \ $g\ul{j}yh\force p_{i+1}\ne q_{j+1}$.
We have shown:\\
$\lbd j\lbd y(g)jyh\force\pt j($ent$(j),j\eps q_0\to p_{i+1}\ne q_{j+1})$. Therefore \
$(f\ul{i}x)\lbd j\lbd y(g)jyh\force\bot$.

\smallskip\noindent
ii)~~We suppose:\\
$f\force\pt i($ent$(i),i\eps p_0,\pt j($ent$(j),j\eps q_0\to p_{i+1}\ne q_{j+1})\to\bot)$~;\\
$y\force i'\eps(p\et r)_0$~; \
$u\force\pt j'($ent$(j'),j'\eps(q\et r)_0\to(p\et r)_{i'+1}\ne(q\et r)_{j'+1})$.\\
If we replace $j'$ with \ $2j''$, and then with \ $2j''+1$, we obtain, by definition of \ $\et$:

\smallskip\noindent
(1) \ \ $(u)(d_0)\ul{j}''\force j''\eps q_0\to(p\et r)_{i'+1}\ne q_{j''+1}$~;\\
(2) \ \ $(u)(d_1)\ul{j}''\force j''\eps r_0\to(p\et r)_{i'+1}\ne r_{j''+1}$.

\smallskip\noindent
Then, there are two cases:

\smallskip\noindent
$\bullet$~~If $i'=2i''$, we have \ $y\force i''\eps p_0$ \ and, by~(1), \
$(u)(d_0)\ul{j}''\force j''\eps q_0\to p_{i''+1}\ne q_{j''+1}$. \ Therefore:\\
$\lbd j(u)(d_0)j\force\pt j($ent$(j),j\eps q_0\to p_{i''+1}\ne q_{j+1})$ and it follows that:\\
\centerline{$(((f)(d_2)\ul{i}')y)\lbd j(u)(d_0)j\force\bot$.}

\smallskip\noindent
$\bullet$~~If $i'=2i''+1$, we have \ $y\force i''\eps r_0$ \ and, by~(2), \
$(u)(d_1)\ul{j}''\force j''\eps r_0\to r_{i''+1}\ne r_{j''+1}$.\\
By making $j''=i''$, we obtain \ $(u)(d_1)\ul{i}''\force i''\eps r_0\to\bot$ \ and therefore:\\
\centerline{$(u)\ul{i}'y\force\bot$.}

\smallskip\noindent
Thus, in both cases, we get: \ \
$((e\ul{i}')(u)\ul{i}'y)(((f)(d_2)\ul{i}')y)\lbd j(u)(d_0)j\force\bot$.
\qed

\begin{lem}\label{p_sub_q_Cq_Cp}\ \\
\emph{\phantom ii)}~~$\theta\force\pt p\pt q(p\subset q,\C[q]\to\C[p])$ \ with \\
$\theta=\lbd f\lbd g\lbd i\lbd i'\lbd x\lbd x'\lbd u\lbd v\lbd w(fi'x')\lbd j'\lbd y'(fix)\lbd j\lbd y(g)jj'yy'uvw$.\\
\emph{ii)}~~$\force\pt p\pt q\pt r(p\subset q,\C[q\et r]\to\C[p\et r])$ \ in other words \
$\force\pt p\pt q(p\subset q\to q\sqle p)$.
\end{lem}
\proof{\ }\ \\
\phantom ii)~~Let \ $f\force p\subset q,g\force\C[q]$, \ that is:\\
$f\force\pt i($ent$(i),i\eps p_0,\pt j($ent$(j),j\eps q_0\to p_{i+1}\ne q_{j+1})\to\bot)$~;\\
$g\force\pt j\pt j'($ent$(j),$ ent$(j'),j\eps q_0,j'\eps q_0\to E[q_{j+1},q_{j'+1}])$ \ with :\\
$E[x,y]\equiv(x\ne y,(x\trl y)\ne1,(y\trl x)\ne1\to\bot)$.\\
Let\ $x\force  i\eps p_0,x'\force i'\eps p_0,u\force p_{i+1}\ne p_{i'+1},
v\force(p_{i+1}\trl p_{i'+1})\ne1,w\force(p_{i'+1}\trl p_{i+1})\ne1$.\\
Let \ $y\force j\eps q_0$, $y'\force j'\eps q_0$.\\
We have \ $g\ul{j}\,\ul{j}'yy'\force E[q_{j+1},q_{j'+1}]$~; if $p_{i+1}=q_{j+1}$ and $p_{i'+1}=q_{j'+1}$, then:\\
$g\ul{j}\,\ul{j}'yy'\force E[p_{i+1},p_{i'+1}]$, and therefore \ $g\ul{j}\,\ul{j}'yy'uvw\force\bot$.\\
Thus, we have \ $\lbd j\lbd y(g)jj'yy'uvw\force$ent$(j),j\eps q_0\to\bot$ if \
$p_{i+1}=q_{j+1}$ and $p_{i'+1}=q_{j'+1}$.\\
Therefore, \
$\lbd j\lbd y(g)jj'yy'uvw\force\pt j($ent$(j),j\eps q_0\to p_{i+1}\ne q_{j+1})$ \
if \ $p_{i'+1}=q_{j'+1}$, thus:\\
$(f\ul{i}x)\lbd j\lbd y(g)jj'yy'uvw\force\bot$ \ if \ $p_{i'+1}=q_{j'+1}$, thus:\\
$\lbd j'\lbd y'(f\ul{i}x)\lbd j\lbd y(g)jj'yy'uvw\force
\pt j'($ent$(j'),j'\eps q_0\to p_{i'+1}\ne q_{j'+1})$. Therefore:\\
$(f\ul{i}'x')\lbd j'\lbd y'(f\ul{i}x)\lbd j\lbd y(g)jj'yy'uvw\force\bot$.

\smallskip\noindent
ii)~~Follows immediately from (i) and \ $\force\pt p\pt q\pt r(p\subset q\to p\et r\subset q\et r)$ \
(lemma~\ref{subset_trans}).
\qed

\smallskip\noindent
The following lemma shows that we can build the algebra ${\mathcal B}$ and the ${\mathcal B}$-model ${\mathcal N}$.

\begin{lem}\label{six_qp}
There exist six proof-like terms $\alpha_0,\alpha_1,\alpha_2,\beta_0,\beta_1,\beta_2$ such that:\\
$\alpha_0\force\pt p\pt q\pt r(\C[(p\et q)\et r]\to\C[p\et(q\et r)])$~; \
$\alpha_1\force\pt p(\C[p]\to\C[p\et\1])$~;\\
$\alpha_2\force\pt p\pt q(\C[p\et q]\to\C[q])$~; \
$\beta_0\force\pt p(\C[p]\to\C[p\et p])$~; \
$\beta_1\force\pt p\pt q(\C[p\et q]\to\C[q\et p])$~;\\
$\beta_2\force\pt p\pt q\pt r\pt s(\C[((p\et q)\et r)\et s]\to\C[(p\et(q\et r))\et s])$.
\end{lem}

\proof \noindent
We only show the first case. By lemma~\ref{p_sub_q_Cq_Cp}(i), it suffices to find
a proof-like term:\\
$\theta\force\pt p\pt q\pt r(p\et(q\et r)\subset (p\et q)\et r)$.
Thus, we suppose:\\
$y\force i\eps(p\et(q\et r))_0$~;
$u\force\pt j($ent$(j),j\eps((p\et q)\et r)_0\to(p\et(q\et r))_{i+1}\ne((p\et q)\et r)_{j+1})$.\\
There are three cases:\\
$\bullet$~~$i=2i'$~; then, we have \ $y\force i'\eps p_0$. We make $j=2i=4i'$,
therefore:\\
$u\force$ ent$(2i),i'\eps p_0\to p_{i'+1}\ne p_{i'+1}$.
Thus, we have: \ $(u)(d_0)\ul{i}y\force\bot$.\\
$\bullet$~~$i=4i'+1$~; then, we have \ $y\force i'\eps q_0$. We make $j=i+2=4i'+3$, thus:\\
$u\force$ ent$(i+2),i'\eps q_0\to q_{i'+1}\ne q_{i'+1}$. Thus, we have: \ $((u)(\sig)^2\ul{i})y\force\bot$.\\
$\bullet$~~$i=4i'+3$~; then, we have \ $y\force i'\eps r_0$. We make $j=i-3=4i'$, thus:\\
$u\force$ ent$(i-3),i'\eps r_0\to r_{i'+1}\ne r_{i'+1}$. Therefore, we have: \
$((u)(\p)^3\ul{i})y\force\bot$\\
($\p$ is the program for the predecessor).\\
Thus, we put \ $\theta=\lbd i\lbd y\lbd u(((e_4i)(u)(d_0)iy)((u)(\sig)^2i)y)((u)(\p)^3i)y$, \ where $e_4$
is defined by its execution rule: \ $e_4\star\ul{i}\ps\xi\ps\eta\ps\zeta\ps\pi\succ\xi\ps\pi$
(resp. $\eta\ps\pi$, $\zeta\ps\pi$) if $i=4i'$ (resp. $4i'+1,4i'+3$).
\qed

\smallskip\noindent
We now show the:

\begin{thm}\label{ccd_ver2}\ \\
The forcing structure $\{\C,\et,\1\}$ satisfies the countable downward chain condition in ${\mathcal M}$.
\end{thm}
\proof
The hypothesis of the c.d.c. are:

\smallskip\noindent
$H_0\equiv\pt n\ex p\,{\mathcal X}(n,p)$~;\\
$H_1\equiv\pt n\inde\pt p\pt q\{{\mathcal X}(n,p),{\mathcal X}(n,q)\to p=q\}$~;\\
$H_2\equiv\pt n\inde\pt p\pt q({\mathcal X}(n,p),{\mathcal X}(sn,q)\to q\sqle p)$~;\\
$H_3\equiv\pt n\inde\pt p({\mathcal X}(n,p)\to\C[p])$.

\smallskip\noindent
Moreover, by theorem~\ref{ACI}, we have a binary function $f:P^2\to P$ such that:\\
$\vsig\force\pt n\inde(\ex p\,{\mathcal X}(n,p)\to\ex k\inde{\mathcal X}(n,f(n,k)))$.\\
Therefore, by $H_0$, we can also use the hypothesis:

\smallskip\noindent
$H'_0\equiv\pt n\inde\ex k\inde\,{\mathcal X}(n,f(n,k))$.

\smallskip\noindent
Let us put \ $\vec{H}=\{H_0,H'_0,H_1,H_2,H_3\}$ \ and \ $\vec{H}_*=\{H_0,H'_0,H_1,H_2\}$.

\begin{lem}\label{Cp_et_q}
$\vec{H}\vdash\pt p\pt q\pt m\inde\pt n\inde({\mathcal X}(m,p),{\mathcal X}(n,q)\to\C[p\et q])$.
\end{lem}
\proof
We show \ $\pt m\indi\pt n\indi({\mathcal X}(m,p),{\mathcal X}(m+n,q)\to q\sqle p)$ \ by recurrence on $n$.\\
For $n=0$, this follows from \ $H_1,H_3$. For the recurrence step, we use $H_2$.

\smallskip\noindent
Thus, we have \ $\pt p\pt q\pt m\inde\pt n\inde({\mathcal X}(m,p),{\mathcal X}(n,q)\to p\sqle q\lor q\sqle p)$.\\
From $p\sqle q$, we deduce \ $\C[p\et p]\to\C[q\et p]$, \ and the result follows, by $H_3$ \ and \
$\C[p]\to\C[p\et p]$.
\qed

\smallskip\noindent
We define the wanted limit $h$ by defining $h_0$ and $h_{m+1}$ for each $m\in\NN$.\\
For  $m=(i,n,k)$ (that is \ $(i,(n,k))$\,), we put \
$\|m\eps h_0\|=\|{\mathcal X}(n,f(n,k))\land i\eps(f(n,k))_0\|$~;\\
then \ $h_{m+1}=(f(n,k))_{i+1}$.\\
Intuitively, ${\mathcal X}$ defines a sequence of countable sets, and $h$ is the union of these sets.

\smallskip\noindent
$\bullet$~~Proof of \ $\vec{H}_*\vdash{\mathcal X}(n,p)\to h\sqle p$.\\
By lemma~\ref{p_sub_q_Cq_Cp}(ii), it suffices to show \
${\mathcal X}(n,p)\to p\subset h$, that is:\\
${\mathcal X}(n,p),i\eps p_0,\pt m\inde(m\eps h_0,\to h_{m+1}\ne p_{i+1})\to\bot$, for $n,i\in\NN$.\\
We fix $k\in\NN$ \ and we put \ $m=(i,n,k)$. By definition of $h$, it suffices to show:\\
${\mathcal X}(n,p),i\eps p_0,\pt k\inde({\mathcal X}(n,f(n,k)),i\eps(f(n,k))_0,
\to (f(n,k))_{i+1}\ne p_{i+1})\to\bot$.\\
Now, from \ $H_1,{\mathcal X}(n,p),{\mathcal X}(n,f(n,k))$, \ we deduce \ $f(n,k)=p$ \ and therefore:\\
$(f(n,k))_0=p_0$ \ and $(f(n,k))_{i+1}=p_{i+1}$. Thus, it remains to show:\\
${\mathcal X}(n,p),i\eps p_0,\pt k\inde({\mathcal X}(n,f(n,k)),i\eps p_0\to p_{i+1}\ne p_{i+1})\to\bot$.\\
But this formula follows immediately from $H'_0$.

\smallskip\noindent
$\bullet$~~Proof of \ $\vec{H}\vdash\C[h]$.\\
We must show $\C[h]$, that is \ $m\eps h_0,m'\eps h_0\to E[h_{m+1},h_{m'+1}]$. \ Now, we have:

\smallskip\noindent
$m=(i,n,k)$~; \ $\|m\eps h_0\|=\|{\mathcal X}(n,f(n,k))\land i\eps(f(n,k))_0\|$~; \
$h_{m+1}=(f(n,k))_{i+1}$~;\\
$m'=(i',n',k')$~; \ $\|m'\eps h_0\|=\|{\mathcal X}(n',f(n',k'))\land i'\eps(f(n',k'))_0\|$~; \
$h_{m'+1}=(f(n',k'))_{i'+1}$.

\smallskip\noindent
From \ ${\mathcal X}(n,f(n,k)),{\mathcal X}(n',f(n',k'))$, it follows that:\\
$\C[u]$ with  $u=f(n,k)\et f(n',k')$ (lemma~\ref{Cp_et_q}).
Therefore, we have:\\
$\|i\eps(f(n,k))_0\|=\|2i\eps u\|$~; \ $\|i'\eps(f(n',k'))_0\|=\|2i'+1\eps u\|$~;\\
$h_{m+1}=u_{2i+1}$~; \ $h_{m'+1}=u_{2i'+2}$.\\
From $\C[u]$, we deduce \ $E[u_{2i+1},u_{2i'+2}]$, that is \ $E[h_{m+1},h_{m'+1}]$.

\smallskip\noindent
This completes the proof of theorem~\ref{ccd_ver2}.
\qed

\subsection*{The well ordering on \texorpdfstring{${\mathcal P}(\NN)$}{P(N)}}\noindent
In the model ${\mathcal N}$, we define the unary predicate:\\
${\mathcal G}(x)\equiv\ex p\ex i\inde\{\neg{\mathcal J}(p),i\eps p_0,x=p_{i+1}\}$.

\begin{lem}\label{G_total}
$\fforce{\mathcal G}(x),{\mathcal G}(y)\to E[x,y]$.
\end{lem}\noindent
We must show \
$\fforce\neg{\mathcal J}(p),\neg{\mathcal J}(q),i\eps p_0,x=p_{i+1},j\eps q_0,y=q_{j+1}\to E[x,y]$,
that is:\\
$\fforce\neg{\mathcal J}(p),\neg{\mathcal J}(q),i\eps p_0,j\eps q_0\to E[p_{i+1},q_{j+1}]$.\\
By theorem~\ref{elem_gen}(ii) and (iii), we have \ $\fforce\neg{\mathcal J}(p),\neg{\mathcal J}(q)\to\C[p\et q]$.\\
Therefore, it is sufficient to show that \ $\fforce\C[p\et q],i\eps p_0,j\eps q_0\to E[p_{i+1},q_{j+1}]$.\\
We show below that we have \
$I\force\C[p\et q],i\eps p_0,j\eps q_0\to E[p_{i+1},q_{j+1}]$.
Since this is a first order formula, this gives the desired result, by theorem~\ref{premier_ordre}.\\
Indeed, we have: \ $p_{i+1}=(p\et q)_{2i+1}$~; $q_{j+1}=(p\et q)_{2j+2}$~;\\
$\|i\eps p_0\|=\|2i\eps(p\et q)_0\|$~; \ $\|j\eps q_0\|=\|2j+1\eps(p\et q)_0\|$.\\
Therefore, it remains to show:\\
$I\force\C[p\et q],2i\eps(p\et q)_0,2j+1\eps(p\et q)_0\to E[(p\et q)_{2i+1},(p\et q)_{2j+2}]$\\
which is obvious, by definition of $\C[p\et q]$.
\qed

\smallskip\noindent
Lemma~\ref{G_total} shows that $\trl$ is a \emph{total} relation on ${\mathcal G}$. But, moreover,
$\trl$ is a well founded relation in ${\mathcal N}$. Therefore, we have:

\smallskip\noindent
\centerline{$\fforce\,{\mathcal G}$ is \emph{well ordered} by $\trl$.}

\smallskip\noindent
We define now two functions on $P$:

\smallskip\noindent
$\bullet$~~a unary function \ $\delta:P\to P$ \ by putting \
$\|i\eps\delta(p)_0\|=\|i+1\eps p_0\|$~; \ $\delta(p)_{i+1}=p_{i+2}$.\\
$\bullet$~~a binary function \ $\phi:P^2\to P$ \ by putting:\\
$\|0\eps\phi(p,q)_0\|=\vide$~; $\|i+1\eps\phi(p,q)_0\|=\|i\eps p_0\|$~;\\
$\phi(p,q)_1=q$~; \ $\phi(p,q)_{i+2}=p_{i+1}$ for every $i\in\NN$.\\
Therefore, we have \ $\delta(\phi(p,q))=p$ \ and \ $\phi(p,q)_1=q$ \ for all $p,q\in P$ \ and thus:\\
$I\force\pt p\pt q(\delta(\phi(p,q))=p)$~; \ $\mathbf{I}\fforce\pt p\pt q(\delta(\phi(p,q))=p)$~;\\
$I\force\pt p\pt q(\phi(p,q)_1=q)$~; \ $\mathbf{I}\fforce\pt p\pt q(\phi(p,q)_1=q)$.

\smallskip\noindent
Intuitively, $\delta(p)$ defines the set we obtain by removing $p_1$ from the set associated
with $p$~;\\
$\phi(p,q)$ defines the set we obtain by  adding $q$ to the set associated with $p$.

\begin{lem}\label{ajout}
If $p,q\in P$, there exists $q'\in P$ such that \ $\delta(q')=q$ \ and \ $p_i\trl q'$ for every $i\in\NN$.
\end{lem}\noindent
For each $a\in P$, we have $\delta(\phi(q,a))=q$. But the application \ $a\mapsto\phi(q,a)$ \ is
obviously injective, since \ $\phi(q,a)_1=a$. Thus, the set $\{\phi(q,a);\;a\in P\}$ is of cardinal $2^{\aleph_0}$. Now, by hypothesis on $\trl$, every proper initial segment of $P$, for the well ordering
$\trl$, is of cardinal $<2^{\aleph_0}$. Thus, there exists some $a_0\in P$ such that $p_i\trl\phi(q,a_0)$ \ for every $i\in\NN$. Then, it suffices to put $q'=\phi(q,a_0)$.
\qed

\smallskip\noindent
Therefore, we can define a binary function \ $\psi:P^2\to P$ such that we have:\\
$\delta(\psi(p,q))=q$ \ and \ $(p_i\trl\psi(p,q))=1$ \ for all $p,q\in P$ and $i\in\NN$.
Thus, we have:

\smallskip\noindent
$I\force\pt p\pt q(\delta(\psi(p,q))=q)$~; \ $\mathbf{I}\fforce\pt p\pt q(\delta(\psi(p,q))=q)$.\\
$KI\force\pt p\pt q\pt i\inde(p_i\trl\psi(p,q))$~; \
$\mathbf{KI}\fforce\pt p\pt q\pt i\inde(p_i\trl\psi(p,q))$.

\begin{lem}\label{G_atteint}
We have \ $\fforce\pt q\ex x\{{\mathcal G}(x),\delta(x)=q\}$.
\end{lem}
\proof
This is written as \ $\fforce\pt q[\pt x\pt p\pt i\inde(\delta(x)=q,\,i\eps p_0,\,x=p_{i+1}
\to{\mathcal J}(p))\to\bot]$ \
or else:\\
$\fforce\pt q[\pt p\pt i\inde(i\eps p_0,\,\delta(p_{i+1})=q\to{\mathcal J}(p))\to\bot]$.\\
By making $i=0$, it is sufficient to show:\\
(1)\hspace{5em}$\fforce\pt q[\pt p(0\eps p_0,\delta(p_1)=q\to{\mathcal J}(p))\to\bot]$.

\smallskip\noindent
By replacing \ $p$ \ with \ $\phi(p,\psi(p,q))$ \ in (1), we see that it remains to show:\\
\centerline{$\fforce\pt q\neg\pt p\,{\mathcal J}(\phi(p,\psi(p,q)))$.}

\begin{lem}\label{CpC_phi_ap}
$\force\pt p\pt q(\C[p]\to\C[\phi(p,\psi(p,q))])$.
\end{lem}

\proof
We have \ $\C[r]\equiv\pt i\inde\pt j\inde(i\eps r_0,j\eps r_0\to E[r_{i+1},r_{j+1}])$.
Therefore, in order to show that \ $\force\C[p]\to\C[r]$, it suffices to show:\\
(1) \ \ $\force\C[p]\to\pt i\inde\pt j\inde(i+1\eps r_0,j+1\eps r_0\to E[r_{i+2},r_{j+2}])$ \ and\\
(2) \ \ $\force\C[p]\to\pt j\inde(0\eps r_0,j+1\eps r_0\to E[r_1,r_{j+2}])$.\\
We apply this remark by putting \ $r=\phi(p,\psi(p,q))$. Then \ (1) is written as \
$\force\C[p]\to\C[p]$ \ since \ $\|i+1\eps r_0\|=\|i\eps p_0\|$ and $r_{i+2}=p_{i+1}$ \
and the same for $j$.\\
Thus, it suffices to show (2), that is:

\smallskip\noindent
$\force\C[p]\to\\
\hspace*{\fill}\pt j\inde(0\eps\phi(p,\psi(p,q))_0,j+1\eps\phi(p,\psi(p,q))_0
\to E[\phi(p,\psi(p,q))_1,\phi(p,\psi(p,q))_{j+2}])$.

\smallskip\noindent
But, we have \ $I\force\pt p\pt q(0\eps\phi(p,q)_0)$~; \
$I\force\pt p\pt q(j\eps p_0\to j+1\eps\phi(p,\psi(p,q))_0)$~;\\
$I\force\pt p\pt q(\phi(p,\psi(p,q))_1=\psi(p,q))$~; \
$I\force\pt p\pt q(\phi(p,\psi(p,q))_{j+2}=p_{j+1})$.\\
Therefore, it remains to show:\\
$\force\C[p]\to\pt j\inde(j\eps p_0\to E[\psi(p,q),p_{j+1}])$\\
which is trivial, since we have \ $KI\force\pt p\pt q\pt j\inde(p_{j+1}\trl\psi(p,q))$.
\qed

\begin{lem}\label{p_sub_phi_pq}
$\lbd i\lbd x\lbd y((y)(\sig)i)x\force\pt p\pt q(p\subset\phi(p,q))$.
\end{lem}

\proof
This is written as:\\
$\lbd i\lbd x\lbd y((y)(\sig)i)x\force\pt i($ent$(i),i\eps p_0,\pt j($ent$(j),j\eps\phi(p,q)_0
\to\phi(p,q)_{j+1}\ne p_{i+1})\to\bot)$\\
which is immediate, by making $j=i+1$.
\qed

\smallskip\noindent
We have \ $\force p\subset\phi(p,\psi(p,q))$ (lemma~\ref{p_sub_phi_pq}), and it follows that:\\
$\force\phi(p,\psi(p,q))\sqle p$ (lemma~\ref{p_sub_q_Cq_Cp}ii), \ and thus \
$\force\C[\phi(p,\psi(p,q))]\to\C[p\et\phi(p,\psi(p,q))]$.\\
Therefore, by lemma~\ref{CpC_phi_ap}, we have:\\
$\force\pt p\pt q(\C[p]\to\C[p\et\phi(p,\psi(p,q))])$. Since this is a first order formula,
we have, by theorem~\ref{premier_ordre}: \
$\fforce\pt p\pt q(\C[p]\to\C[p\et\phi(p,\psi(p,q))])$\\
and therefore, by theorem~\ref{elem_gen}(ii): \
$\fforce\pt p\pt q(\neg\C[p\et\phi(p,\psi(p,q))]\to{\mathcal J}(p))$.\\
Then, we apply theorem~\ref{densite}, which gives: \
$\fforce\pt q\neg\pt p\,{\mathcal J}(\phi(p,\psi(p,q)))$\\
which is the desired result.
\qed

\begin{thm}\label{bon_ordre}
The following formulas are realized in ${\mathcal N}$:\\
\emph{\phantom ii)}~~There exists a well ordering on the set of individuals.\\
\emph{ii)}~~There exists a well ordering on the power set of \ $\NN$.
\end{thm}
\proof{\ }\ \\
\phantom ii)~~Lemma~\ref{G_atteint} shows that, in ${\mathcal N}$, the function $\delta$ is a surjection
from ${\mathcal G}$ onto the set $P$ of individuals. But, we have seen that the formula: \
``~${\mathcal G}$ is well ordered by $\trl$~'' \ is realized in ${\mathcal N}$.\\
ii)~~By theorems~\ref{cons_reels} and~\ref{ccd_ver2}, the following formula is realized in
${\mathcal N}$: \ ``~Every subset of \ $\NN$ is represented by an individual~''.
Hence the result, by (i).
\qed

\smallskip\noindent
Theorem~\ref{bon_ordre}(ii) enables us to transform into a program any proof of a formula of
second order arithmetic, which uses the existence of a well ordering on $\mathbb{R}$. The method
is the same as the one explained above for the ultrafilter axiom.

\bigskip
\centerline{***********}

\end{document}